%% file: main.tex
\documentclass[preprint]{elsarticle}
\pdfoutput = 1

\usepackage{floatrow}
\usepackage{subfig}

\usepackage[utf8]{inputenc}
\usepackage[T1]{fontenc}
\usepackage{lipsum}
\usepackage{xargs}  
\usepackage[pdftex,dvipsnames,table]{xcolor}  
\usepackage{parskip} 
\usepackage{booktabs,arydshln}
\usepackage{algpseudocode} 
\usepackage{algorithm} 
\usepackage{soul} 

\soulregister\cite7 
\soulregister\ref7
\soulregister\textsubscript7
\soulregister\textsuperscript7

\usepackage[section]{placeins} 
\makeatletter
\AtBeginDocument{%
  \expandafter\renewcommand\expandafter\subsection\expandafter{%
    \expandafter\@fb@secFB\subsection
  }%
}
\makeatother

\usepackage{amssymb}
\usepackage{mathtools}
\usepackage{mycommands}
\usepackage{lipsum}
\usepackage{boldline}
\usepackage{multirow}
\usepackage{booktabs,siunitx}
\usepackage{nicematrix}
\usepackage{caption}
\usepackage[shortcuts]{extdash}
\usepackage{nameref}
\usepackage{geometry}
\usepackage{appendix}

\usepackage[colorinlistoftodos,prependcaption,textsize=tiny]{todonotes}

\newwrite\figuresusedout
\immediate\openout\figuresusedout=\jobname.figs%
  
\NewDocumentCommand{\FigureUsedOut}{m}{%
  \immediate\write\figuresusedout{#1}
}

\NewDocumentCommand{\IncludeGraphics}{ O{} m }{%
  \includegraphics[#1]{#2}\FigureUsedOut{#2}%
}

\usepackage{fancyhdr}
\journal{Int. J. Mechanical Sciences}
\journal{arXiv}
\begin{document}

\begin{frontmatter}

\title{Bayesian Optimal Experimental Design for Constitutive Model Calibration}
\author[1]{D. E. Ricciardi\corref{cor1}}
\ead{dericci@sandia.gov}

\author[1]{D. T. Seidl}
\author[1]{B. T. Lester}
\author[1]{A. R. Jones}
\author[1]{E. M. C. Jones}

\cortext[cor1]{Corresponding author}
\affiliation[1]{organization={Sandia National Laboratories},
    addressline={PO Box 5800},
    postcode={87185},
    city={Albuquerque, NM},
    country={USA}}

\begin{abstract}

Computational simulation is increasingly relied upon for high\-/consequence engineering decisions, and a foundational element to solid mechanics simulations, such as finite element analysis (FEA), is a credible constitutive or material model. Calibration of these complex material models is an essential step; however, the selection, calibration and validation of material models is often a discrete, multi-stage process that is decoupled from material characterization activities, which means the data collected does not always align with the data that is needed. To address this issue, an integrated workflow for delivering an enhanced characterization and calibration procedure---Interlaced Characterization and Calibration (ICC)---is introduced and demonstrated. This framework leverages Bayesian optimal experimental design (BOED) to select the optimal load path for a cruciform specimen in order to collect the most informative data for model calibration. Eventually, the ICC framework will be used for quasi real-time, actively controlled experiments of complex specimens and the calibration of an FEA model. The critical first piece of algorithm development is to demonstrate the active experimental design within a Bayesian framework for a fast model with simulated data. For this demonstration, a material point simulator that models a plane stress elastoplastic material subject to bi-axial loading was chosen. 

The ICC framework is demonstrated on two exemplar problems in which BOED is used to determine which load step to take, e.g., in which direction to increment the strain, at each iteration of the characterization and calibration cycle. Calibration results from data obtained by adaptively selecting the load path within the ICC algorithm are compared to results from data generated under two naive static load paths that were chosen \textit{a priori} based on human intuition. These results are communicated with posterior summaries, and parameter uncertainties are propagated to the model output space. In these exemplar problems, data generated in an adaptive setting resulted in calibrated model parameters with reduced measures of uncertainty compared to the static settings.

\end{abstract}


\end{frontmatter}

\includegraphics[width=\textwidth]{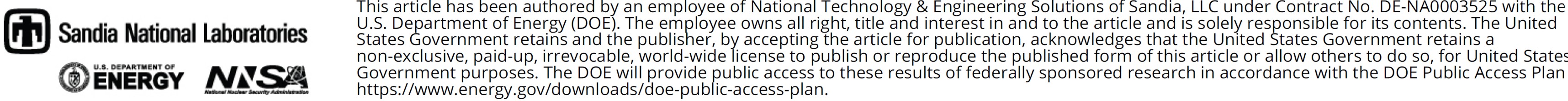}
\clearpage

\floatsetup[figure]{style=plain,subcapbesideposition=top,font=large}
\input{Introduction.tex}
\input{Framework.tex}
\input{Methods}

\input{Exemplar_1.tex}

\input{Exemplar_2.tex}

\input{Conclusions}
\input{Acknowledgements}

\biboptions{sort&compress}
\bibliographystyle{unsrt}
\bibliography{refs}

\appendix
\include{app_A}

\include{app_B}
\include{app_C}

\end{document}

%% file: Introduction.tex
\section{Introduction}\label{sec: introduction}

Material constitutive models---mathematical representations of the complex mechanical behavior of materials under various loading conditions---are critical to solid mechanics simulations like finite element analysis (FEA). These models contain material-dependent parameters that must be determined in order to reliably simulate material behavior. These unknown parameters are obtained by first collecting experimental data, i.e., material characterization, and then using the data to drive an inverse problem for parameter estimation, i.e., calibration.

The process of material characterization and model calibration has evolved over the years. Tensile dog bone samples, which develop a homogeneous state of stress, allow constitutive models to be calibrated analytically. Alternative specimens and test methods, such as notched tension, shear, torsion, interrupted or reversed loading, etc., can be employed to access other stress states, but require advanced calibration methods such as finite element model updating (FEMU) \cite{friswell1995finite}. Additionally, Digital Image Correlation (DIC) delivers full-field kinematic data (e.g., strain on the surface of a test specimen \cite{schreier2009image}) and can be used with FEMU and other advanced calibration methods like the Virtual Fields Method (VFM) \cite{jones2018high, jones2019investigation, jones2018parameter, kramer2014implementation, avril2008overview, pierron2012virtual, friswell1995finite, grediac2006virtual, martins2021calibration}. 

More often than not, material characterization and model calibration are performed deterministically, which provides a single set of optimal parameters. Alternatively, a stochastic approach to model calibration, such as Bayesian inference \cite{gelman1995bayesian}, produces a probability distribution over the unknown model parameters. Under this paradigm, posterior summary metrics such as parameter expected values, variances and credible intervals are useful in characterizing parameter uncertainty, which can then be propagated to the model output. A description of uncertainty is valuable in engineering settings as it can be used to make confident design decisions, especially in high-consequence scenarios. There is a growing body of work in fields of science and engineering that use Bayesian inference for model calibration \cite{honarmandi2019uncertainty,paranjape2021probabilistic,ricciardi2019uncertainty,ricciardi2020uncertainty, choo2023distribution, daghia2007estimation, koutsourelakis2012novel, gogu2010introduction, rappel2018bayesian}. 

Despite these advances in material characterization and model calibration, the two processes are typically decoupled from each other, with each being performed sequentially. Only after a separate validation effort is performed can calibration results be evaluated. If the results are unsatisfactory, a new model must be selected, potentially necessitating another experimental campaign \cite{corona2021anisotropic, karlson2018revisiting, karlson2019hierarchical, corona2021thermal, corona2021thermal2, corona2014evaluation, jones2021anisotropic, zhang2014calibration, martins2021calibration, martins2018comparison}, which is a time-consuming and costly process. 

The expense of this decoupled approach to characterization and calibration may be alleviated with a smart selection, or \emph{design}, of experiments. The overarching goal when selecting an experimental design is to maximize the information content in the data with respect to the goals of the experiment. If the goal is to estimate model parameters (as in this work), then information content of the data may be measured by the resulting uncertainty of the inferred model parameters.

The experimental design may be constructed under several different settings. A \emph{static} experimental design is one such that the characterization experiments are determined prior to seeing any data and that does not adapt as data is collected. At the most fundamental level, human intuition is used to determine the experimental design ahead of data collection, which is the approach traditionally taken and is prone to result in data not optimal for the model calibration problem at hand. Static designs may also be obtained under more structured guidelines. The full factorial design first introduced by Fisher \cite{fisher1966design} constructs a design matrix containing all possible combinations of discrete design settings. Although this design approach is comprehensive, the cost of executing the design matrix grows exponentially as the number of design variables and possible values they can assume increases. A reduced design matrix may be obtained with a fractional factorial design \cite{gunst2009fractional}, which uses a selective subset from the full factorial design matrix while limiting the loss in critical information. These combinatorial approaches are in general not feasible for calibration of constitutive models due to the number of design settings involved and the time and cost requirements associated with running each experiment. Thus, when the experimental design is constructed in a static manner, it is susceptible to result in the inefficient use of limited resources and/or result in sub-optimal data for calibration.

In lieu of determining the experimental design matrix up front, an \emph{adaptive} design---which re-evaluates the design periodically based on the observed data thus far---is appealing for engineering applications in order to maximize the information content in the data while making the best use of limited resources. Bayesian optimal experimental design (BOED), first introduced by Lindley \cite{lindley1956measure}, provides a principled way to adaptively choose an experimental design under uncertainty as it is based on Bayesian inference, which naturally incorporates any \emph{a priori} knowledge of the parameters into the decision-making process as well as uncertainties in the data and model form and any numerical error.\footnote{BOED may likewise be implemented in a static setting in which the optimal design is determined prior to seeing any data.} An \emph{optimal} design of experiments is one that provides the most informative data, and in an ideal scenario, only data that is both necessary and sufficient (i.e., all the data and only the data that is needed) for calibration is collected. In this work, BOED is the paradigm used to adaptively determine the optimal experimental design.

There are several different ways an optimal experimental design (OED) can be specified for constitutive model calibration. Two contrasting approaches involve either determining the optimal load path for a specimen of a given geometry \cite{villarreal2022design, michopoulos2008towards} (the focus of this work), or optimizing the specimen geometry for a given load path \cite{oliveira2022evaluation, thoby2022robustness, barroqueiro2020design}. Both objectives of optimizing the load path and specimen geometry were explored in \cite{souto2016numerical}. Additionally, OED is often used for sensor placement when there are limited physical sensors (such as accelerometers, strain gauges or thermocouples) \cite{papadimitriou2004optimal, chen2022bayesian}. All approaches, however, have the same goal of determining the OED for the purposes of the experiment.

This work utilizes adaptive BOED in a new Interlaced Characterization and Calibration (ICC) framework for material characterization and constitutive model calibration that interlaces the two processes in a quasi real-time feedback loop. In this adaptive experimental design framework, the load path placed on a specimen is determined \emph{in situ} with model calibration in order to leverage information from previous load steps when determining future load steps. To the authors' knowledge, this is the first realization of ICC for load path optimization of a specimen. With the ICC approach, both the outcome (i.e., parameter uncertainty) as well as the process (i.e., time and monetary efficiency) of characterization and calibration are improved.

The remainder of this paper is organized as follows: First, the components of the characterization and calibration framework are presented (Sec.~\ref{sec: framework}), and then a background of the constitutive model, statistical methods, experimental design tools and other details of the framework are described in detail (Sec.~\ref{sec: methods}). Next, the framework is demonstrated on two exemplar problems: The first exemplar focuses on yield anisotropy (Sec.~\ref{sec: exemplar 1}), while the second exemplar additionally considers isotropic hardening (Sec.~\ref{sec: exemplar 2}). In summary, this work addresses issues with the current decoupled approach to material characterization and model calibration by interlacing the two processes, resulting in improved calibration with reduced uncertainty.

%% file: Framework.tex
\section{Framework}\label{sec: framework}

\noindent The overarching goal of the ICC framework is to optimize the load path a specimen of a given geometry is subjected to in order to collect the most informative data for model calibration. In the ICC framework, an initial load is placed on the specimen, experimental data is collected and Bayesian calibration is performed to estimate a probability distribution of the unknown model parameters. Then, given a set of possible next load steps, BOED is used to determine which one to take using the expected information gain (EIG) as a measure of how informative each step is predicted to be. The experiment is actively driven through the load path with the highest EIG, and the characterization-calibration process is repeated for a predetermined number of load steps, which is just one of many possible ending criteria.

\begin{figure}
    \centering
    \includegraphics[width=.8\linewidth]{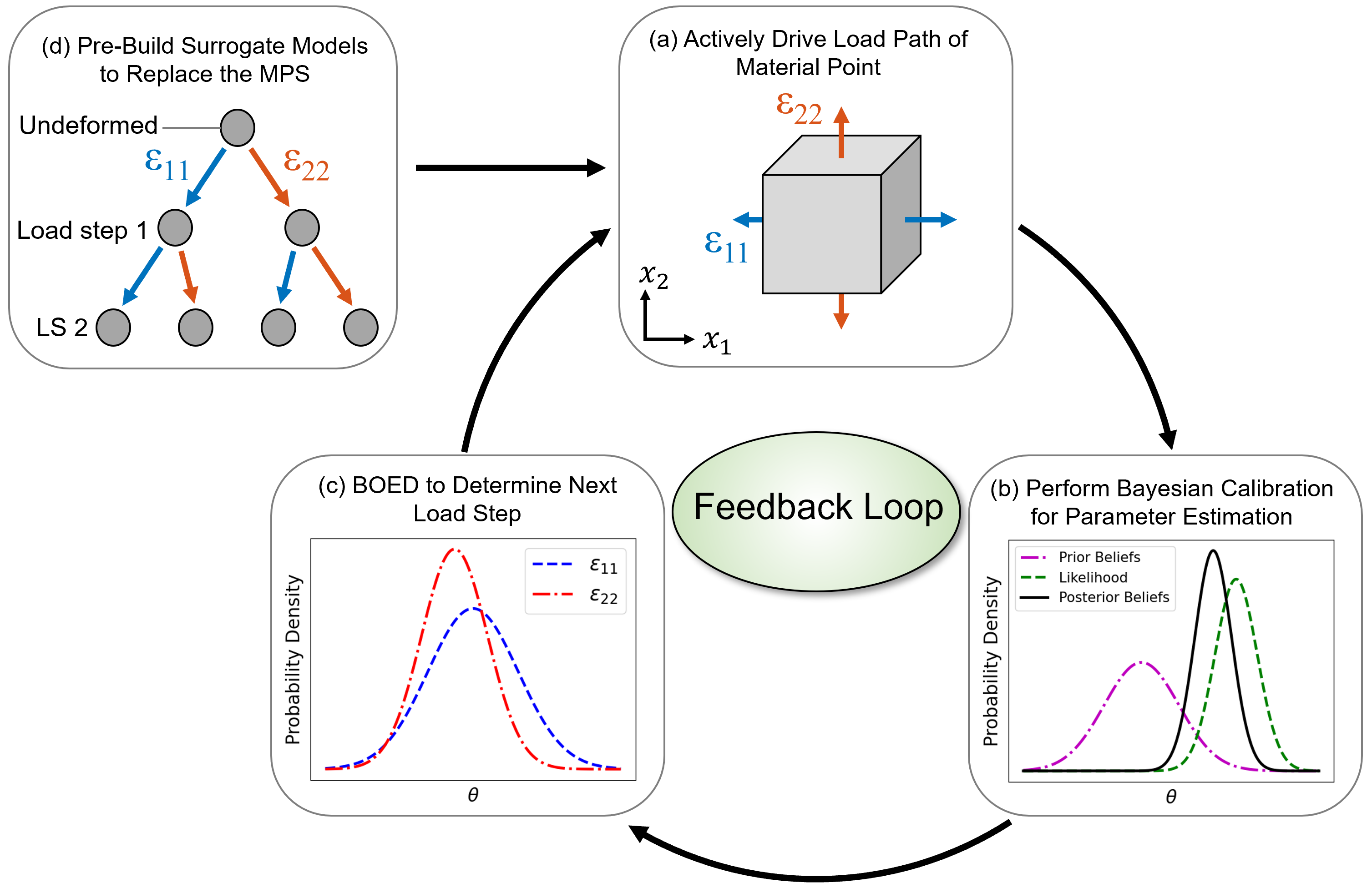}
    \caption{The ICC framework for a material point under bi-axial loading. An initial load step is taken by applying a strain increment along one of the two axes while keeping the other axis fixed (a). Given the collected data from (a), prior knowledge about the parameters $\theta$ is updated via Bayesian inference (b). Given the two possible next load steps (applying a strain increment on either of the two axes), BOED is used to determined the next step (c). Prior to entering the feedback loop, the load path space is reduced to a finite load path tree, and surrogate models are constructed to replace the material point simulator (d). }
    \label{fig: framework}
\end{figure}

\noindent The ICC framework could be applied to numerous experimental configurations, but the target in this work is a cruciform specimen in a bi-axial load frame. As a starting point, a computationally cheap material point simulator (MPS), which represents the center of a cruciform specimen, was used for framework development and testing. The process is illustrated in Fig.~\ref{fig: framework} for the MPS. Starting with plot (a) of Fig.~\ref{fig: framework}, the material point is driven along a load path by applying a strain increment along either of the two axes shown while the other axis is held fixed (displacement = 0), and data is collected; in (b), Bayesian inference is performed, which updates prior knowledge of the unknown parameters by conditioning on the available data to obtain a posterior understanding of the parameters; in (c), BOED is used to determine the load step that will be the most informative for calibration. The process continues in a feedback loop for a given number of load steps. Prior to entering the feedback loop, the load path space is reduced to a finite load path tree in (d), and surrogate models are constructed to replace the MPS, which is discussed in detail in Sec.~\ref{subsec: finite load-path tree}. Since the material under consideration in this work is an aluminum alloy, the load path tree also provides a way for path-dependence to be incorporated into the surrogates, which is important when considering plasticity. The implementation of the ICC framework utilizing the methods described in Sec.~\ref{sec: methods} is detailed in Algorithm~\ref{alg: ICC} in \ref{alg.app}.

\noindent The ICC approach has similar goals as the experimental design protocol proposed by Villarreal et al. \cite{villarreal2022design}, wherein a reinforcement learning algorithm was used for the design of experiments in order to calibrate a history-dependent constitutive model. A key difference is that Villarreal et al. leveraged a pre-trained policy to guide the experimental actions, whereas the ICC framework continuously updates the load path \textit{in situ} during the experiment.  Additionally, Villarreal et al. used a Kalman filter to obtain an estimate of the information gain corresponding to each experimental action, as opposed to the EIG used in the current work. Thus, the tools used within the ICC approach for experimental design are fundamentally different.

%% file: Methods.tex
\section{Methods}\label{sec: methods}

In this section, the different components that make up the framework are described in detail. First, the MPS and elastoplastic material model used in this work---Hill48 anisotropic yield criterion with two different isotropic hardening functions---are detailed along with an introduction to the two exemplar problems (Sec.~\ref{sec: material point simulator}). Next, Bayesian inference for material model calibration (Sec.~\ref{sec: bayesian inference}) as well as the theory of BOED and the EIG calculation (Sec.~\ref{subsec: bayesian optimal experimental design}) are presented. The load path tree that reduces the infinite-dimensional space of possible load paths to a manageable number is described (Sec.~\ref{subsec: finite load-path tree}). Finally, the surrogate models that are constructed at each node of the load path tree are discussed (Sec.~\ref{sec: surrogate model construction}).

\subsection{Material Point Simulator}\label{sec: material point simulator}

To begin, a description of the constitutive model to be calibrated is presented. In the case of the MPS, it is assumed that the material point is subjected to a known total strain history in two directions, $\varepsilon_{11}$ and $\varepsilon_{22}$, under a plane stress assumption. The model is taken to be temperature and rate independent, leaving the out-of-plane strain $\varepsilon_{33}$ and the stresses in the in-plane directions, $\sigma_{11}$ and $\sigma_{22}$, as the unknowns to be calculated, along with any internal state variables.

The material response is taken to be elastoplastic with isotropic hardening and anisotropic yield with the constitutive response given via

\begin{equation}
    \sigma_{ij}=\mathbb{C}_{ijkl}\left(\varepsilon_{kl}-\varepsilon_{kl}^{\text{p}}\right),
\end{equation}

where $\mathbb{C}_{ijkl}$ and $\varepsilon_{kl}^{\text{p}}$ are the elastic stiffness tensor (assumed isotropic) and plastic strain, respectively.  To describe the plastic response, the yield function $g$ is introduced as

\begin{equation}
    g = \phi\left(\sigma_{ij}\right) - \bar{\sigma}\left(\kappa\right),
\end{equation}

where $\phi$ and $\bar{\sigma}$ are the effective and flow stresses describing the shape and size of the yield surface, respectively.  The flow stress is a function of the isotropic hardening variable $\kappa$.  If $g<0$, the material response is elastic while the condition $g=0$ indicates inelastic plastic deformation.

For the hardening behavior, a combined linear and Voce type expression is used such that

\begin{equation}\label{eq: hardening behavior}
    \bar{\sigma}\left(\kappa\right) = \sigma_y + \bar{h}\kappa + A\left(1-\exp\left(-n\kappa\right)\right),
\end{equation}

where $\sigma_y$, $\bar{h}$, $A$ and $n$ are the constant initial yield stress, linear hardening modulus, exponential modulus and the exponential fitting coefficient, respectively. 

With respect to the shape of the yield surface, Hill's effective stress~\cite{hill1948theory} is utilized. This expression, however, is further simplified by two assumptions: the material is in a state of plane stress, and a purely bi-axial stress-state is taken to result from loading, leaving the remaining shear stresses as zero. These assumptions produce only two non-zero stresses, enabling the Hill effective stress to be reduced to

\begin{equation}\label{eq: hill effective stress}
    \phi^2\left(\sigma_{ij}\right) = F\left(\sigma_{22}\right)^2+G\left(\sigma_{11}\right)^2+H\left(\sigma_{11}-\sigma_{22}\right)^2.
\end{equation}

Lastly, using an associative approach, the flow rule is given by 

\begin{eqnarray}
    \dot{\varepsilon}^{\text{p}}_{\gamma\gamma} & = & \dot{\kappa}\frac{\partial\phi}{\partial\sigma_{\gamma\gamma}},~\qquad~\gamma=1,2\\
    \dot{\varepsilon}^{\text{p}}_{33} & = & -\left(\dot{\varepsilon}^{\text{p}}_{11}+\dot{\varepsilon}_{22}^{\text{p}}\right) \nonumber,
\end{eqnarray}

with plasticity incompressibility ($\dot{\varepsilon}^{\text{p}}_{kk}=0$) being used to arrive at the expression for the out-of-plane plastic strain.\footnote{The out-of-plane plastic strain evolution may also be derived by leaving $\sigma_{33}\neq 0$ in the effective-stress expression.}

With a prescribed in-plane strain history, the in-plane stresses, out-of-plane strain, isotropic hardening variable and plastic strains all need to be determined, leading to a solution variable set of

\begin{equation*}
    x := [\sigma_{11}, \, \sigma_{22}, \, \varepsilon_{33}, \, \kappa, \, \varepsilon_{11}^p , \, \varepsilon_{22}^p , \, \varepsilon_{33}^p].
\end{equation*}

The updated state may be found using a classical return mapping algorithm~\cite{simo2006computational} to solve Hooke's Law, yield consistency 
condition and plastic flow rule.  Details of that procedure are left to the cited reference.

\paragraph{Exemplar Problems}\label{sec: exemplar problems}

In this section, two exemplar problems that explore different combinations of material phenomenology are introduced, both of which use the MPS as the forward model, and their results are contained in Sec.~\ref{sec: exemplar 1} and Sec.~\ref{sec: exemplar 2}. In Exemplar~1, the model parameters being inferred from the data, which are referred to as \emph{unknown} parameters, are the anisotropic yield parameters $F$ and $G$ from \eqref{eq: hill effective stress}, and the model parameter vector is $\pars := [F,\, G]^T$ with all other parameters held fixed at the values presented in Table \ref{Tab: parameters in exemplars}. In Exemplar~2, the initial yield stress $\sigma_{y}$ as well as the isotropic hardening variables $A \, \mbox{and} \, n$ from \eqref{eq: hardening behavior} are also inferred so that $\pars := [F,\, G,\, \sigma_{y},\, A,\, n]^T$.

These two exemplars were selected to study the ability of the ICC algorithm to effectively drive calibration under different material responses. Exemplar~1 focuses solely on the anisotropy of the yield surface in order to observe the impact of direction dependence as probed by the bi-axial load frame; Exemplar~2 introduces plastic hardening into the calibration problem. Thus, two distinct material responses may be found with potentially competing calibration needs that pose a challenge for the ICC algorithm. As such, both responses are considered in this work. 

\begin{table}[ht!]
\centering
\begin{tabular}{llll}
\hline
\textbf{Parameter} & \textbf{Exemplar 1} & \textbf{Exemplar 2} & \textbf{Bounds} \\ 
\hlineB{4}
E  & 70 GPa & 70 GPa & -- \\
$\nu$   &  0.3 & 0.3 & -- \\
$F$ & Unknown & Unknown & $[0.3, 0.7]$\\
$G$ & Unknown & Unknown & $[0.3, 0.7]$\\
$H$ & 0.5 & 0.5 & -- \\
$\sigma_{y}$  & 200 MPa & Unknown & $[50, 500]$ MPa \\[0.15 em]
$\bar{h}$   &  200 MPa & 0 MPa & -- \\ 
$A$  & 200 MPa & Unknown &  $[10, 400]$ MPa\\
$n$  & 20 & Unknown & $[1\times 10^{-6}, 100]$ \\
\hline
\end{tabular}
\caption{Parameters in Exemplars~1~and~2.}\label{Tab: parameters in exemplars}
\end{table}

\paragraph{Parameter Cases}

The data used for inference was simulated with the MPS using an assumed true parameter vector $\pars^{true}$ with added Gaussian noise. In Exemplar~1, calibration was performed for four different parameter cases, each having a unique $\pars^{true}$ that is detailed in Table \ref{Tab: exemplar 1 parameter cases}.  These cases were chosen to evaluate the sensitivity of the load path selection algorithm to the level of yield anisotropy in the model. The four cases are labeled according to the $F$ and $G$ values, where $F=G=0.5$ represents an isotropic (J2) yield surface. Also included in the table are the corresponding values of the yield multiplier parameters, $R_{11} = (G + H)^{-1}$ and $R_{22} = (F + H)^{-1}$.\footnote{$R_{ij} = \sigma_{ij}^{y}/\bar{\sigma}$, where $\sigma_{ij}^{y}$ is the stress in that direction at yield.} Parameter H was held fixed at 0.5 in all four cases. Parameters $F$ and $G$ were chosen to have 1) values of equal and opposite distance from 0.5, 2) one value equal to 0.5, 3) both values greater than 0.5 and 4) one value above and one value below 0.5 with unequal distances away from 0.5.  These selections are meant to test algorithmic performance given different parameteric and response contributions.

\begin{table}[ht!]
\centering
\begin{tabular}{lllllll}
\hline
\textbf{Case No.} & \textbf{Case Description} & $\boldsymbol{F}$ & $\boldsymbol{G}$ & $\boldsymbol{R_{11}}$ & $\boldsymbol{R_{22}}$ & $\boldsymbol{R_{33}}$ \\ 
\hlineB{4}
1   & Equal and Opposite    & 0.55    & 0.45   & 1.05   & 0.95    & 1.00  \\
2   & 1 Unchanged           & 0.60    & 0.50   & 1.00   & 0.91    & 0.91  \\
3   & Both \textgreater{} 0.5      & 0.60    & 0.60   & 0.91   & 0.91    & 0.82  \\
4   & Strong Anisotropy     & 0.69    & 0.43  & 1.07   & 0.84    & 0.89  \\ 
\hline
\end{tabular}
\caption{Four different parameter cases studied in Exemplar~1 with varying levels of yield anisotropy.}\label{Tab: exemplar 1 parameter cases}
\end{table}

Two parameter cases were studied in Exemplar~2, and the $\pars^{true}$ used for data generation in each case is detailed in Table \ref{Tab: exemplar 2 parameter cases}. In Case~5, a stronger emphasis was placed on the initial yield stress by setting $\sigma_y$ to a value three times that of the exponential modulus $A$, which describes the Voce component of hardening. In contrast, Case~6 put a stronger emphasis on hardening by setting $A$ to a value three times that of $\sigma_{y}$. The linear component of hardening was removed in this second exemplar by setting $\bar{h} = 0$ so that the effects of the trade-off in the components contributing to the total stress (yield and hardening) could be observed.

\begin{table}[ht!]
\centering
\begin{tabular}{lllllll}
\hline
\textbf{Case No.} & \textbf{Emphasis} & $\boldsymbol{F}$ & $\boldsymbol{G}$ & $\boldsymbol{\sigma_{y}}$ (MPa) & $\boldsymbol{A}$ (MPa) & $\boldsymbol{n}$ \\ 
\hlineB{4}
5   & Yield   & 0.55    & 0.45  & 300  & 100   & 20      \\
6   & Hardening      & 0.55    & 0.45  & 100    & 300   & 20     \\
\hline
\end{tabular}
\caption{Two different parameter cases studied in Exemplar~2, which emphasize the initial yield or Voce component of the hardening behavior.}\label{Tab: exemplar 2 parameter cases}
\end{table}

\subsection{Bayesian Inference}\label{sec: bayesian inference}

Model calibration is performed within the Bayesian paradigm for inference, which provides estimates of parameter uncertainty; the reader is referred to Gelman et al. \cite{gelman1995bayesian} for a thorough introduction to Bayesian inference. The goal in Bayesian inference is to define and update the knowledge of unknown (or uncertain) quantities by conditioning on available information. In the context of model calibration, the unknown quantities are the forward model parameters $\pars \in \Theta \subset \mathbb{R}^{D}$ (see Table~\ref{Tab: parameters in exemplars}), where $D$ is the dimensionality of the parameter space $\Theta$.

Before any data are observed, the uncertainty about the model parameters is modeled through a probability density function (PDF), which is the \emph{prior} distribution (or "prior" for short) with probability density $\pi(\pars)$; prior belief about the parameter ranges and regions of highest probability can be incorporated in this modeling step.

Data for the calibration of constitutive models often takes the form of one-dimensional force-extension curves or field data (e.g., displacement or strain on the surface of a test specimen). The data being used for calibration in this work in the context of the MPS are the in-plane stresses. Once this data $\data \in \mathcal{Y} \subset \mathbb{R}^{N_{obs}}$, where $N_{obs}$ (number of observations) is the dimensionality of the data space $\mathcal{Y}$, becomes available, the prior is updated to a PDF known as the \emph{posterior} distribution (or "posterior") $\pi(\pars \given \data)$ by conditioning on the available information. The probability distribution over the data is $f(\data \given \pars)$, and for fixed $\data$, it is considered as a function of $\pars$ and is called the \textit{likelihood function}, which quantifies the likelihood of observing the data $\data$ given parameters $\pars$. The posterior is obtained via Bayes' rule, which is derived from the definition of conditional probability,

\begin{equation}\label{eq: bayes_rule}
    \pi(\pars \given \data) = \frac{f(\data \given \pars)\pi(\pars)}{\int_\Theta f(\data \given \pars)\pi(\pars)d\pars} \propto f(\data \given \pars)\pi(\pars).
\end{equation}

The term in the denominator of the posterior is called the model \emph{evidence} (also known as the marginal likelihood). Since $\pars$ is integrated out in $\int_\Theta f(\data \given \pars)\pi(\pars)d\pars$, this term is only dependent on the observed data, and for fixed $\data$, it is a constant. Therefore, the posterior can be written as being proportional to the product of the likelihood and prior as in \eqref{eq: bayes_rule}. This form is convenient since direct computation of the marginal likelihood is in general not possible.

An important consideration in Bayesian inference is how to recover uncertainty information when posterior functionals and summaries such as the posterior means and probabilities are not analytically tractable. While the use of numerical integration may be a valid option to obtain these quantities for a parameter space $\Theta$ of low dimension (i.e., $D \leq 4$) \cite{smith2013uncertainty}, its use may lead to large numerical errors in higher dimensions. Thus, the use of alternative methods for approximating these quantities is needed.

A commonly used method for obtaining posterior summaries is Markov Chain Monte Carlo (MCMC) simulation \cite{tierney1994markov,gamerman2006markov,smith1993bayesian,andrieu2003introduction,liu2001monte}. The high accuracy of an MCMC sampling-based approach is achieved at a high computational cost and is too expensive for the quasi real-time characterization and calibration setting in which the ICC framework will eventually operate (additional comments in \ref{sec: MCMC_posterior_approx.app}). Therefore, a highly efficient, non-sampling technique is better suited to recover posterior uncertainties in the ICC framework. One such method is Laplace's approximation, which makes use of the second-order Taylor series expansion of the posterior centered about the maximum \textit{a posteriori} (MAP) probability estimate, providing a Gaussian approximation of the posterior. The MAP estimate is obtained by finding the $\pars$ that maximizes the log of the posterior, or equivalently, minimizes its negative. Define $J(\pars) \coloneqq -\log \pi(\pars \given \data)$, then the MAP estimate is

\begin{equation}\label{eq: MAP}
    \hat{\pars} = \argmin_{\pars \in \Theta} J(\pars).
\end{equation}

The Laplace approximation for the posterior is a normal distribution centered at $\hat{\pars}$ with a covariance matrix determined by the inverse of the Hessian of $J(\pars)$ at the MAP estimate, $\Sigma^{L} = \bs{H}(\hat{\pars})^{-1}$ \cite{penny2011statistical}. The approximation of the posterior is written as

\begin{equation}\label{eq: laplace approx}
    \tilde{\pi}^{L}(\pars \given \data) = \mathcal{N}_{D}(\hat{\pars}, \Sigma^{L}) \approx \pi(\pars \given \data).
\end{equation}

The most straightforward approach to computing the Hessian is through a second-order finite difference approximation about $J(\hat{\pars})$. A primary benefit of using the Laplace approximation for the posterior is the low computational cost compared to MCMC simulation. After the ICC algorithm has completed and all the data has been collected, an MCMC simulation may optionally be used to obtain an approximation to the posterior given all the data---although, this step was not performed in the presented work.

\subsection{Bayesian Optimal Experimental Design}\label{subsec: bayesian optimal experimental design}

The following section discusses BOED modeling steps and considerations within the scope of this work; the reader is referred to Ryan \cite{ryan2016review} for a comprehensive introduction and review of BOED algorithms and concepts. Additionally, the prior works of Long et al.  \cite{long2013fast}, Beck et al. \cite{beck2018fast} and Huan \& Marzouk \cite{huan2013simulation} detail theoretical frameworks in line with what is presented here.

The first step of BOED is to define some metric by which to measure the information content of an experimental design. Define $\xi$ as a single experiment and $\des$ as the experimental design that may be comprised of one or more experiments. Given a design belonging to the space of all possible designs $\des \in \Xi$, experimental data $\data$ and model parameters $\pars$, a utility function $U(\des, \data, \pars)$ is used to quantify the relationship between the design and the resulting information that is obtained from it. Since the outcome $\data$ of performing the experiment(s) in design $\des$ is unknown prior to performing the experiment(s), in practice, the expectation of the utility over the marginal distribution of all possible outcomes and the parameters $\mathbb{E}_{\data,\pars \given \des}[U(\des, \data, \pars)]$ is computed. The optimal design $\des^*$ is the one that maximizes the expected utility,

\begin{equation}\label{eq: optimal design}
  \des^* = \argmax_{\des \in \Xi} \mathbb{E}_{\data,\pars \given \des}[U(\des, \data, \pars)].
\end{equation}

There are a number of ways to formulate the utility function, and the decision should reflect the overarching goal of the experiment and inference. A thorough description of various utility functions is contained in \cite{chaloner1995bayesian}. A common choice for the expected utility function (and the one used in this work) when the interest is in maximizing the information content in the data with respect to parameter uncertainty is the expected Kullback-Leibler (KL) divergence. The KL divergence provides a measure of the statistical distance between one probability distribution $q(\pars)$ and a second reference probability distribution $p(\pars)$,

\begin{equation}\label{eq: kl_divergence}
    \infdiv{p(\pars)}{q(\pars)} = \int_{\Theta} p(\pars)\log\left(\frac{p(\pars)}{q(\pars)}\right)d\pars.
\end{equation}

In the Bayesian setting, the two distributions of interest are the prior $\pi(\pars)$ and the posterior $\pi(\pars \given \data, \des)$ after performing the experiment(s) in $\des$ and observing $\data$. The optimal design is the one that has the greatest expected KL divergence, which means it results in a posterior distribution that has the greatest statistical distance from the prior and therefore contains the most information about the parameters. The expected KL divergence is also commonly referred to as expected information gain (EIG).\footnote{Maximizing the EIG is equivalent to D-optimality when the parameter-to-observable map is linear \cite{chaloner1995bayesian}.} Replacing $p(\pars)$ and $q(\pars)$ in \eqref{eq: kl_divergence} with $\pi(\pars \given \data, \des)$ and $\pi(\pars)$, respectively, and taking the expectation over the marginal distribution of all possible outcomes, the EIG is written as

\begin{subequations}\label{eq: KL divergence EIG}
    \begin{align}
    \mbox{EIG}(\des) &= \mathbb{E}_{\data \given \des} [\infdiv{\pi\left(\pars \given \data, \des\right)}{\pi\left(\pars\right)}] \label{eq: KL 1}\\ 
     &= \int_{Y} \int_{\Theta}\pi(\pars \given \data, \des) f(\data \given \des ) \big[ \log \pi(\pars \given \data, \des) - \log \pi(\pars) \big] d\pars d\data \label{eq: KL 2} \\
     &= \int_{Y} \int_{\Theta} f(\data \given \pars, \des)\pi(\pars) \big[ \log f(\data \given \pars, \des) - \log f(\data \given \des) \big] d\pars d\data \label{eq: KL 3}\\
     &= \mathbb{E}_{\pars} [\infdiv{f(\data \given \pars, \des)}{f(\data \given \des)}] \label{eq: KL 4},
     \end{align}
\end{subequations}   

where $f(\data \given \des)$ is the evidence. 

Using Bayes' rule \eqref{eq: bayes_rule}, the EIG can equivalently be written as the expectation of the KL divergence between the posterior and prior \eqref{eq: KL 1} \& \eqref{eq: KL 2} or between the likelihood and evidence \eqref{eq: KL 3} \& \eqref{eq: KL 4}. In both cases, numerical methods are needed in order to obtain an estimate of the EIG since typically neither formulation can be expressed in a closed form. 

In this work, a nested Monte Carlo estimator is used to obtain an approximation of the EIG using the form found in \eqref{eq: KL 3} since it is straightforward to obtain samples from the prior and data distributions,

\begin{equation}\label{eq: double_nested MC estimator}
    \begin{aligned}
    \widehat{EIG}_{MC}(\des) = \frac{1}{N} \sum_{n=1}^{N} \log \frac{f\left(\data_n \given \pars_{n,0},\des\right)}{\frac{1}{M} \sum_{m=1}^{M} f\left(\data_n \given \pars_{n,m},\des\right)}, \quad \pars_{n,m} \sim \pi(\pars), \quad \data_{n} \sim f(\data \given \pars_{n,0}, \des).
    \end{aligned}
\end{equation}

The approximation in \eqref{eq: double_nested MC estimator} uses samples from the data distribution and prior to evaluate the inner and outer sum of the Monte Carlo estimator. The quality of the estimate is largely controlled by $N$ and $M$, the number of samples used to evaluate the outer and inner sums, which control the estimator variance and bias, respectively \cite{ryan2003estimating}. Additional comments on the nested MC estimator can be found in \ref{sec: EIG_approx.app}.

Recall from Sec.~\ref{sec: introduction} that BOED is especially appealing in an adaptive setting in which previously collected data is incorporated into the design selection process.\footnote{BOED may likewise be used in a static (a.k.a. \emph{batch}) setting in which the EIG is computed once for the optimal batch of size $T$ experiments up-front, EIG($\des$), where $\des = [\xi_{1}, \dots, \xi_{T}]$. However, the static designs that are used for comparison in this work were determined by human intuition and not through static BOED.} In an adaptive setting, the EIG for the $t^{\text{th}}$ experiment in $\des$, $\mbox{EIG}(\xi_{t}), \, t = 1, \ldots, T$, (T is the total number of experiments in $\des$) is computed based on the the data collected from the past $t-1$ experiments. More details of the adaptive BOED approach can be found in \ref{sec: adaptive_boed.app}

When the EIG is calculated for only the next design point $\xi_{t}$ without considering all subsequent design points $\des_{t+1:T}$, it is known as a \emph{myopic} (or greedy) method, where a locally optimal decision is made at each step of the algorithm. While locally optimal decisions may result in the globally optimal experimental design $\des$ (or, in this application, load path from beginning to end), this outcome is not guaranteed. At a minimum, the myopic algorithm is expected to yield improvement over a static design. The decision to use this myopic decision-making process was one of practicality, as the ICC framework will eventually be used in a quasi real-time setting. Since the number of design points grows exponentially with each subsequent load step (illustrated in Fig.~\ref{fig: loadpath_tree} in Sec.~\ref{subsec: finite load-path tree}), so would the computational cost of calculating the EIG for all paths to the leaves of the tree. In addition to this, in reality, it is difficult to hold a sample in an exact position for an indefinite amount of time while these computations are taking place. For these reasons, a myopic approach is utilized.

In summary, an adaptive, myopic BOED algorithm is used within the ICC framework to optimize the load path of a specimen in order to gain the most informative data for calibration. This section is concluded with a summary of the notation used for the experiments in this work: Each load step applied to a specimen corresponds to a single \emph{experiment} $\xi_{t}$, and an \emph{experimental design} consists of the full load path a specimen takes $\boldsymbol{\xi} = \{\xi_{t}\}_{t=1}^{T}$, where $T$ is the number of load steps in the load path. At load step $t$, data $\data_{t}$ is collected. The \emph{optimal experimental design} in this context describes the optimal load path design.

\subsection{Finite Load Path Tree}\label{subsec: finite load-path tree}

\begin{figure}[!ht]
\begin{center}
  \includegraphics[width=0.45\textwidth]{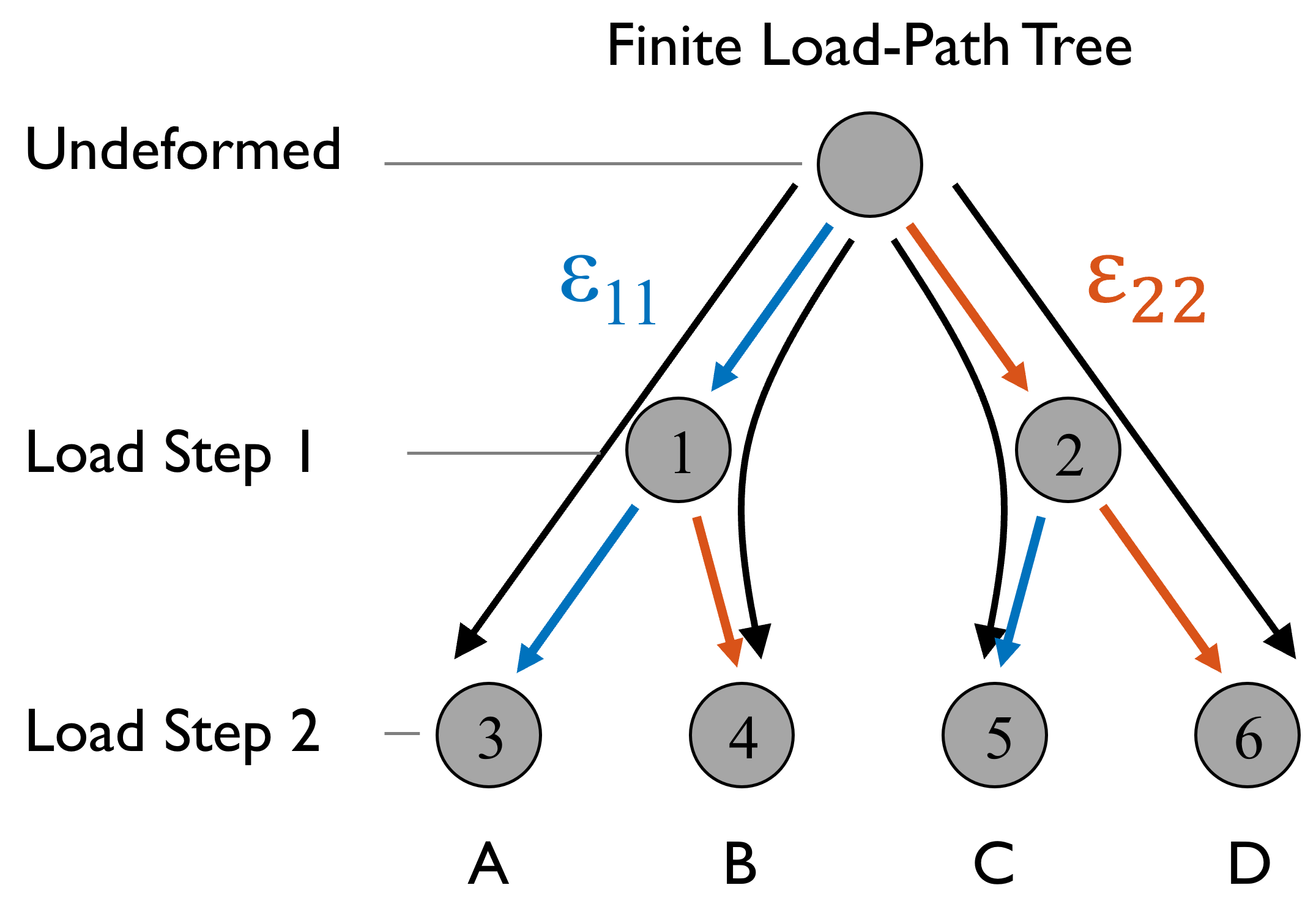}
  \caption{A simplified load path tree showing four possible load paths (A-D), each consisting of two load steps.}
\label{fig: loadpath_tree}
\end{center}
\end{figure}

The finite load path tree that reduces the infinite-dimensional load path space of material deformation to a discrete space is introduced in this section. The graph structure in this network, depicted in Fig.~\ref{fig: loadpath_tree}, is that of a binary tree, as each parent node has two children. This graph stands in for a collection of load steps that involve bi-axial loading of a specimen. The branches could take many forms, but for now, the load path tree is restricted to two children per node, which correspond to applying a strain increment along either the $x_{1}$ or $x_{2}$ direction ($\varepsilon_{11}$ or $\varepsilon_{22}$) of the material point while holding the alternate axis fixed. Other, more complex, branch options may include both positive and negative strains along the $x_{1}$ and $x_{2}$ directions (i.e., node children $\varepsilon_{11}$, $\varepsilon_{22}$, $-\varepsilon_{11}$ and $-\varepsilon_{22}$), as well as combinations of strains applied in both directions simultaneously (i.e., node children $\varepsilon_{11}\varepsilon_{22}$, $-(\varepsilon_{11}\varepsilon_{22})$, $-\varepsilon_{11}\varepsilon_{22}$ and $-\varepsilon_{22}\varepsilon_{11}$) etc. The selection of two children per node is a simplified starting point and is not necessarily guaranteed to be the optimal tree structure. 

As the target application is an aluminum alloy that can fail around 0.15 mm/mm strain (depending on the material orientation) \cite{lu2021solid}, the total strain imposed on the material point throughout the duration of the experiment (from the undeformed state to the end of the final load step) was chosen to be within the plastic regime of deformation away from failure. The total strain was segmented equally across load steps, with the number of steps chosen to keep the computational expense of the EIG \eqref{eq: double_nested MC estimator} reasonable (in combination with the selection of 2 children per node). In Exemplar~1, the total strain was $\varepsilon = 0.05$ mm/mm, segmented into 5 load steps yielding $\Delta \varepsilon = 0.01$ mm/mm strain per load step. In Exemplar~2, a total strain of $\varepsilon = 0.14$ mm/mm was split into 7 load steps with $\Delta \varepsilon = 0.02$ mm/mm strain per load step. In both exemplars, there were 100 simulated pseudotime increments per load step. In Exemplar~1, QoIs $\sigma_{11}$ and $\sigma_{22}$ were stored at the end of each load step (corresponding to each node) and in Exemplar~2, at 3 equally-spaced strain increments between nodes. The strain increment was increased in Exemplar~2 in order to have sufficient data in the plastic region for calibrating the hardening parameters, and the number of measurement points per load step was increased to 3 in order to ensure there was enough data to mitigate issues with identifiabiltiy of the parameters. Note that optimizing the strain increment (as well as the form of the branches) is a subject of future work.

The BOED workflow begins at the root node, where an initial deformation direction may either be chosen by the analyst or the algorithm. If the EIG \eqref{eq: double_nested MC estimator} is used to initiate the design, the EIG estimate is based entirely on the prior modeling and the forward model since there is not yet any data available. Therefore, the estimate is only as good as these two components are. Poor prior modeling and, likewise, a forward model that suffers from a high level of model-form error may lead to a poor EIG estimate. Alternatively, engineering judgement may be used to determine the first step. Both options were explored in this work: The initial step was determined ahead of time in Exemplar~1 and via the EIG estimate in Exemplar~2.

As an example, if the analyst chooses to initially deform the specimen by a set amount of strain along the left branch ($\xi_{1} = \varepsilon_{11}$), it would result in a load path that leads to the state represented by node 1. The experimental data $\data_{1}$ is then collected to produce the observations at load step 1, and Bayesian calibration is performed to compute the posterior distribution of the model parameters conditioned on the available data $\pi(\pars \given \data_{1}, \xi_{1})$. An instance of the BOED step selection problem is then solved: Given the current information about the unknown parameters, which step is the most desirable for calibration, resulting in the greatest reduction in posterior uncertainty---the one leading to node 3 or node 4 ($\xi_{2} = \varepsilon_{11}$ or $\xi_{2} = \varepsilon_{22}$)? If one were to follow the series of load steps leading to node 4, the full load path would consist of $T=2$ load steps---the first leading to node 1 and the second leading to node 4---and would be written as $\des = [\xi_{1}, \xi_{2}] = [\varepsilon_{11}, \varepsilon_{22}]$. This process continues until an exit criterion is met, such as progressing through a predetermined number of load steps $T$.

\begin{figure}%
    \centering
    \sidesubfloat[]{{\includegraphics[width=0.45\textwidth]{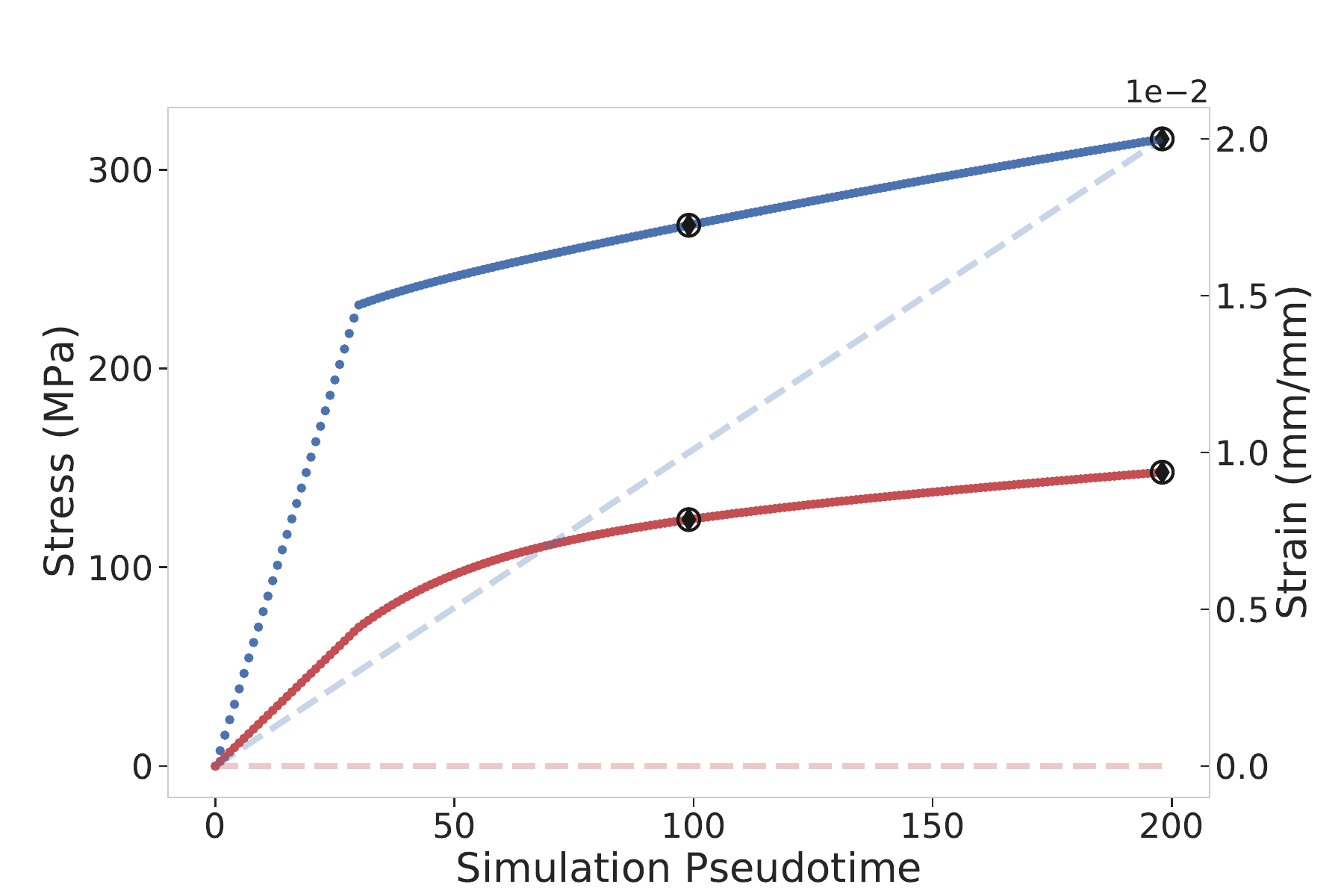}\label{fig: case 1 path A example} }}%
    \sidesubfloat[]{{\includegraphics[width=0.45\textwidth]{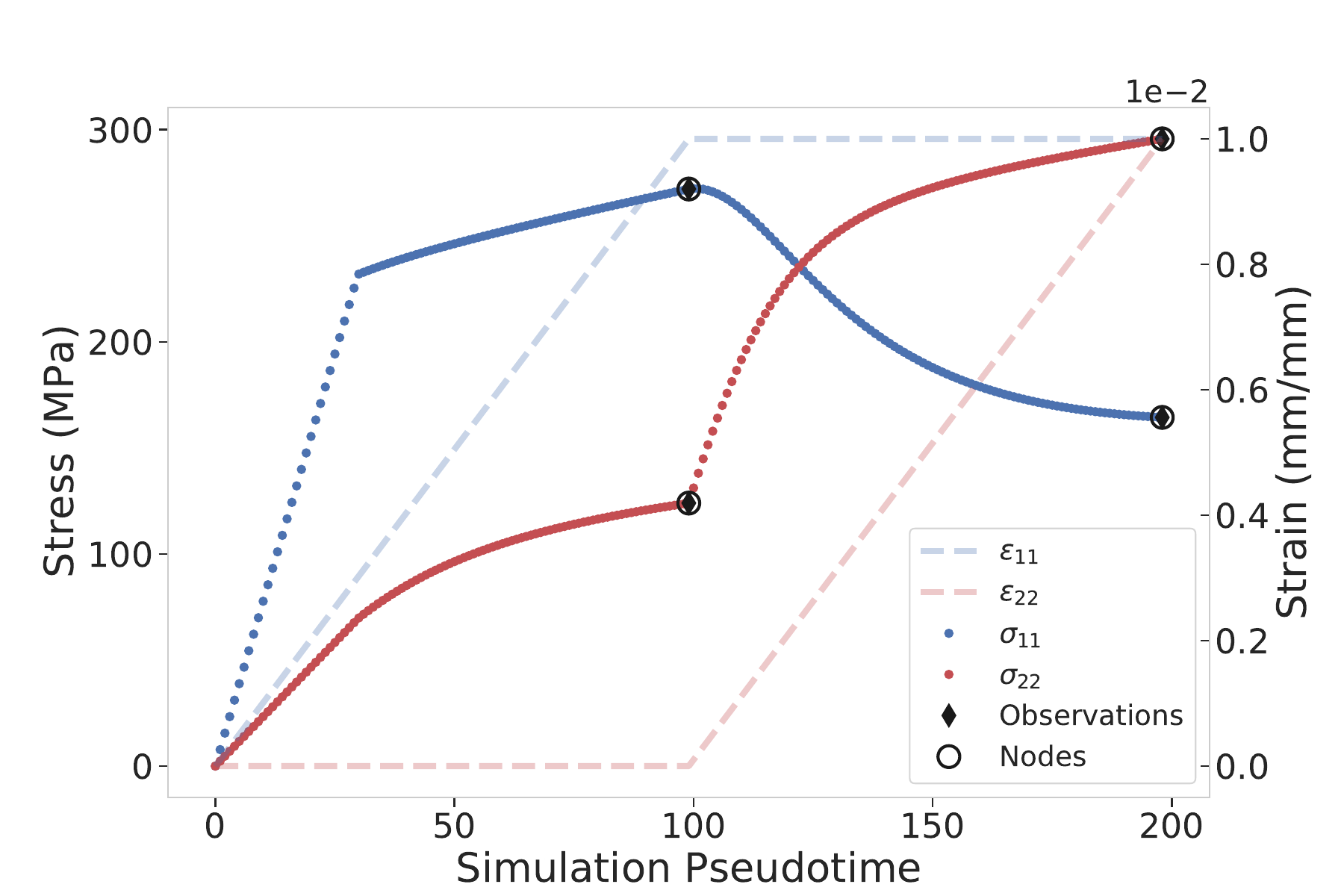}\label{fig: case 1 path B example} }}\\
    \sidesubfloat[]{{\includegraphics[width=0.45\textwidth]{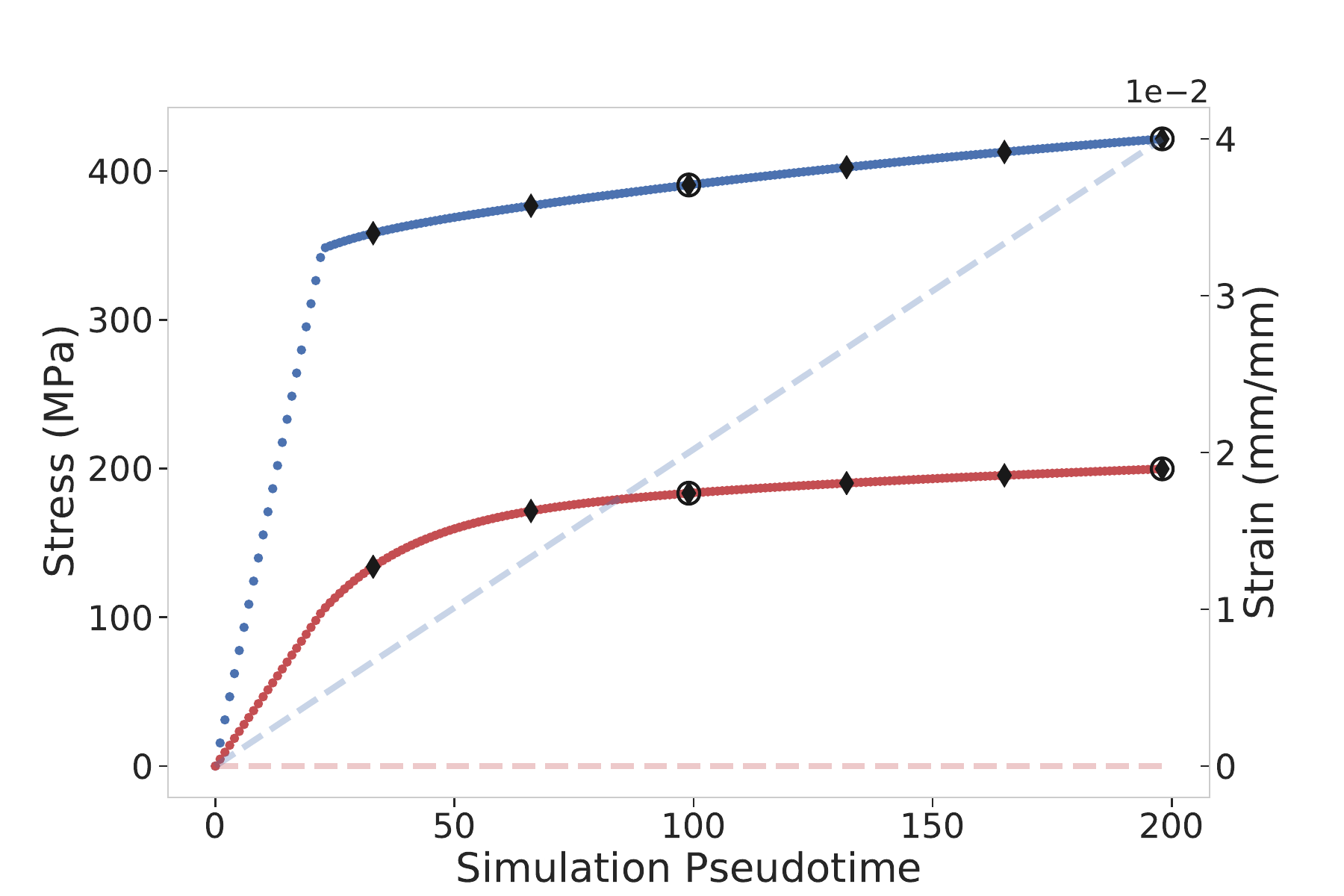}\label{fig: case 5 path A example} }}%
    \sidesubfloat[]{{\includegraphics[width=0.45\textwidth]{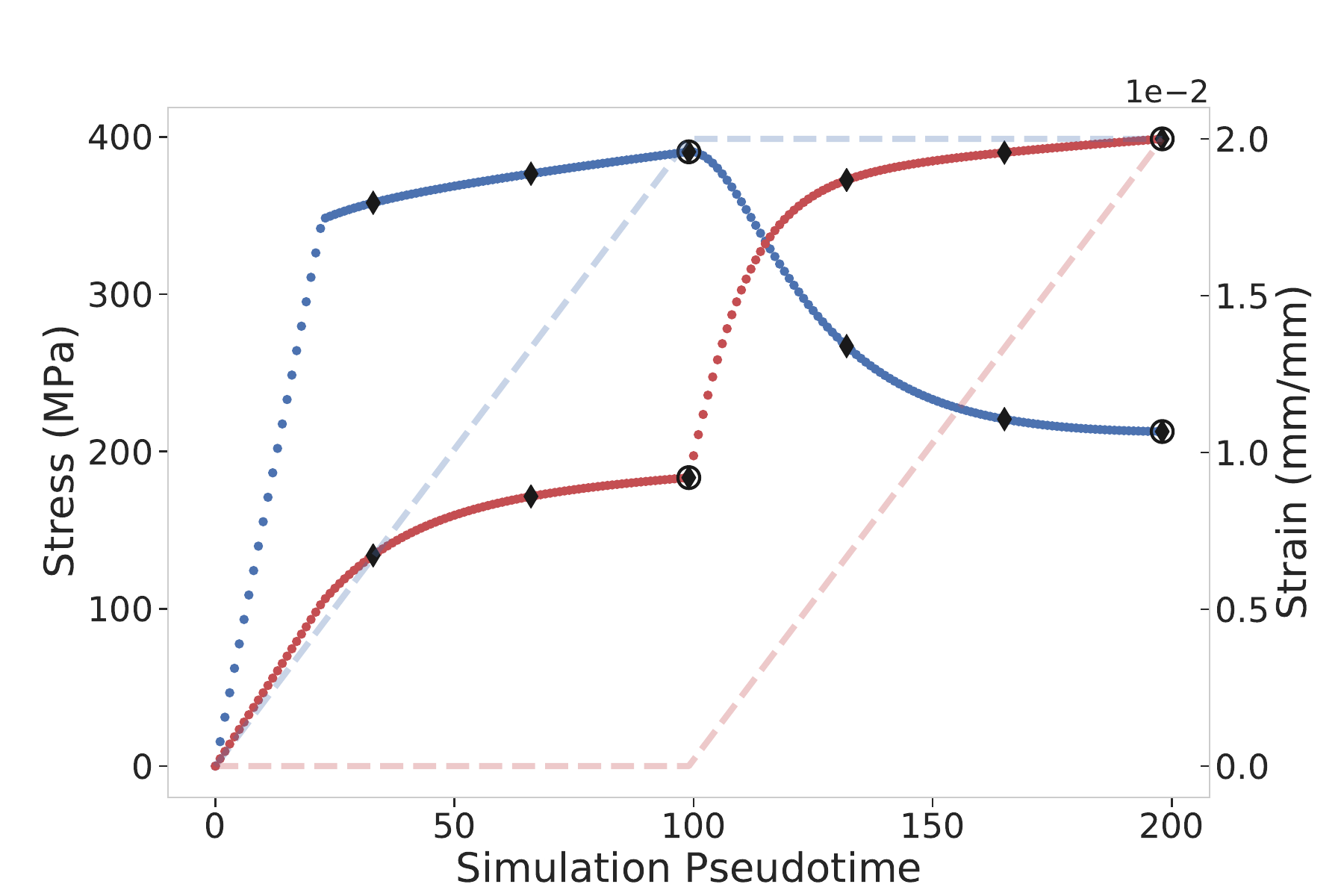}\label{fig: case 5 path B example} }}
    \caption{The left axes of the plots show the resulting in-plane stresses from following load path $A$ ($\des = [\varepsilon_{11}, \varepsilon_{11}]$) (left column) and $B$ ($\des = [\varepsilon_{11}, \varepsilon_{22}]$) (right column) in Fig.~\ref{fig: loadpath_tree} using parameter values from Case~1 in Table~\ref{Tab: exemplar 1 parameter cases} (top row) and Case~5 in Table \ref{Tab: exemplar 2 parameter cases} (bottom row). Locations of the measured data points are shown with black diamonds and tree nodes with open circles. The right axes of the plots show the associated strain paths for each case. The stresses and strain paths are a) load path A, Case~1 in Exemplar~1, b) load path~B, Case~1 in Exemplar~1, c) load path~A, Case~5 in Exemplar~2, and d) load path~B, Case~5 in Exemplar~2.}%
    \label{fig:loadpaths A and B}%
\end{figure}

To aid in the description of the load path tree, the strains and in-plane stress curves from the MPS for load paths $A$ and $B$ from Fig.~\ref{fig: loadpath_tree} are plotted in Fig.~\ref{fig:loadpaths A and B} vs. simulation pseudotime using parameter values from Case~1 in Table~\ref{Tab: exemplar 1 parameter cases} (Figs.~\ref{fig: case 1 path A example}~and~\ref{fig: case 1 path B example}) and Case~5 in Table~\ref{Tab: exemplar 2 parameter cases} (Figs.~\ref{fig: case 5 path A example}~and~\ref{fig: case 5 path B example}). If a single measurement is taken at the end of each load step as in Exemplar~1, then there are four measured data points, which are represented by black diamonds. The measured data are the two in-plane stresses that are measured at each node---represented by open circles---in the load path tree. If multiple measurements are taken during each load step as in Exemplar~2, then there are multiple measured data points between each node with the final measurement coinciding with the node. The stress curves for the experiments represented by load path $A$ ($\des = [\xi_{1}, \xi_{2}] = [\varepsilon_{11}, \varepsilon_{11}$]) resemble traditional uniaxial stress-strain curves, except that here $\varepsilon_{22}$ is fixed at zero, which causes the development of the $\sigma_{22}$ stress. The reversal in the direction of applied strain for load path $B$ ($\des = [\xi_{1}, \xi_{2}] = [\varepsilon_{11}, \varepsilon_{22}]$) produces striking changes in the in-plane stresses, and the exponential nature of the hardening behavior is also easily observed. For clarity, the applied strains are also shown in Fig.~\ref{fig:loadpaths A and B}.

\floatsetup[figure]{style=plain,subcapbesideposition=top,font=small}

\subsection{Surrogate Model Construction}\label{sec: surrogate model construction}

Even with the fast MPS, the computational cost of obtaining the Monte Carlo estimate of the EIG in \eqref{eq: double_nested MC estimator} is very expensive, requiring the model to be evaluated for a number of parameter samples on the order of $10^{6}$ or greater. The need to make the ICC algorithm as efficient as possible---of paramount importance for its future application in quasi real-time characterization experiments and calibration---is a strong motivator for the use of surrogate models to replace the forward model. Not only do surrogates allow for rapid parameter inference, they also make it possible to expeditiously evaluate the EIG in order to determine the next load step to take.

The means chosen in this work for dealing with the history dependence inherent in elastoplastic models is to remove it from consideration by building individual surrogates $\tilde{u}$ for each measured QoI, which are the in-plane stresses.\footnote{With the future target experimental configuration of a cruciform specimen, the QoIs will be resultant forces and DIC data, which can be experimentally measured (or derived from experimental measurements).} For Exemplar~1, the surrogates were built for every node in the load path tree in Fig.~\ref{fig: loadpath_tree}, other than the root node (which represents the undeformed material point). For Exemplar~2, the surrogates were built for the measured QoIs at the specified intermediate pseudotime increments between nodes (as well as at the nodes). The surrogate models approximately map constitutive model parameters to each QoI that can, in principle, be computed from experimental measurements.

Training data for the surrogates was generated by first obtaining samples of the unknown model parameters. For this task, Halton samples \cite{halton1960efficiency}, which consist of a deterministic sequence of prime numbers to produce a space-filling design, were used. For Exemplar~1, 200 training samples were generated within the bounds specified for $F$ and $G$ in Table~\ref{Tab: parameters in exemplars} while keeping all other parameters fixed at the values recorded in the table for Exemplar~1. Likewise, in Exemplar~2, 500 training samples were generated within the bounds specified for $F$, $G$, $\sigma_{y}$, $A$ and $n$, keeping all other parameters fixed at the recorded values. Each parameter sample was used as input to the MPS to produce corresponding output, which was the in-plane stresses. The resulting collection of input-output pairs constitute the training and test data for the surrogates.

The surrogates built for both exemplar problems generally exhibited low error compared to the MPS. The mean absolute percentage error (MAPE) of the surrogates was calculated from 1,000 test samples generated with a Halton sequence. The surrogates in Exemplar~1 had a MAPE ranging from $2\times10^{-4}$\% to $3\times10^{-4}$\% at each node and for each QoI and those from Exemplar~2 had a MAPE ranging from $0.01$\% to $.2$\%.

The surrogate model used in this work is a Gaussian process (GP), which can be used to describe probability distributions over functions. A GP is fully characterized by a mean function $\mu(x)$ and covariance function $k(x, x')$ ($x$ being a random variable), which describes the strength of correlation between inputs as a function of the distance between the points. In this work, an anisotropic squared exponential kernel was used. The reader is guided to \cite{williams2006gaussian} for a thorough introduction to GPs.

A notable benefit of using GPs is that they are a stochastic process, which means they inherently provide a measure of uncertainty of the surrogate representation of the model. The uncertainty surrounding the surrogate model may then be incorporated into the inference infrastructure. However, in this work, for the sake of simplicity, the mean of the GP is used in the inference without accounting for the added uncertainty of the surrogate.

%% file: Exemplar_1.tex
\section{Exemplar 1}\label{sec: exemplar 1}

In this section, Exemplar~1 is presented and discussed. The statistical modeling choices, data generation process and algorithmic settings within the ICC framework are discussed (Sec.~\ref{sec: exemplar 1 set up}). Results are presented (Sec.~\ref{sec: exemplar 1 results}) with an analysis of the adaptive load path selection (Sec.~\ref{sec: exemplar 1 optimal load paths}) and a comparison of the parameter inference from the adaptive setting via the ICC framework vs. two naive static design settings. Comparisons are made between the posterior summaries (Sec.~\ref{sec: exemplar 1 posterior summaries}) and propagated uncertainty (Sec.~\ref{sec: case 1 propagated uncertainty}).

\subsection{Problem Setup}\label{sec: exemplar 1 set up}

The first exemplar problem, introduced in Sec.~\ref{sec: exemplar problems}, considers the anisotropic yield parameters $F$ and $G$ in \eqref{eq: hill effective stress} as unknown, with other parameters fixed as noted in Table~\ref{Tab: parameters in exemplars}, and the parameter vector is defined as $\pars = [F, G]^T$. This section is used to describe the statistical modeling decisions and other algorithm settings specific to this exemplar problem.

The parameters were constrained to take on values within the bounds specified for the surrogate training data (Table~\ref{Tab: parameters in exemplars}) by choosing a truncated normal (TN) prior probability distribution that enforces this constraint, 

\begin{equation}\label{eq: exemplar 1 prior settings}
    \pi(\pars_{d}) = TN\left(\pars_{d}; \mu_{(\pars_{d})}, \delta^{2}_{(\pars_{d})}, a_{(\pars_{d})}, b_{(\pars_{d})}\right), \quad d = 1, \ldots, D.
\end{equation}

The distribution means $\mu$ were chosen to yield an expected value within the bounds, and the variances $\delta^{2}$ were chosen to be moderately diffuse. The support of the distribution, defined by $a$ (the lower bound) and $b$ (the upper bound), was chosen to be the same as that used for the surrogate training. All distribution parameters are detailed in Table~\ref{Tab: exemplar 1 priors}. These prior choices not only restrict the range of values the parameters can take, but also reflect the belief that the parameters are less likely to take on values exactly at the bounds.

\begin{table}[ht!]
\centering
\begin{tabular}{lllll}
\hline
\textbf{Parameter}  & $\boldsymbol{\mu_{\theta}}$ & $\boldsymbol{\delta^{2}_{\theta}}$ & $\boldsymbol{a}$ & $\boldsymbol{b}$ \\ 
\hlineB{4}
F    & 0.5    & 1   & 0.3 & 0.7  \\
G    & 0.5    & 1   & 0.3 & 0.7 \\
\hline
\end{tabular}
\caption{Means $\mu_{\theta}$, variances $\delta^{2}_{\theta}$ and bounds $a$ and $b$ for the truncated normal prior model \eqref{eq: exemplar 1 prior settings} used in Exemplar~1.}\label{Tab: exemplar 1 priors}
\end{table}

Next, synthetic experimental results were modeled to reflect the assumption that the observed data is equal to the model output with added Gaussian error $e$,

\begin{equation}\label{eq: data model}
    \data = \tilde{u}(\pars) + e, \qquad e \sim \mathcal{N}_{N_{obs}}\left(0,\Psi\right).
\end{equation} 

The expression in \eqref{eq: data model} may be written as a multivariate normal model centered at the surrogate replacement of the forward model evaluated at $\pars$, $\tilde{u}(\pars)$, with a covariance matrix $\Psi$ of size $N_{obs} \times N_{obs}$,

\begin{equation}\label{eq: likelihood}
    f\left(\data \given \pars, \des \right) = \mathcal{N}_{N_{obs}}\left(\data; \tilde{u}(\pars), \Psi\right).
\end{equation}

The observation noise variance is assumed to be independent and identically distributed (i.i.d.) such that $\Psi = \psi^{2}\mathbb{I}_{N_{obs}}$, where $\mathbb{I}_{N_{obs}}$ is an $N_{obs} \times N_{obs}$ identity matrix and $\psi^{2}$ is the observation noise variance---assumed to be known and fixed at $\psi^{2} = 10\mbox{MPa}^{2}$. The noise level was chosen to be on the same order as that seen in experimental data.

In \eqref{eq: data model} and \eqref{eq: likelihood}, $\tilde{u}(\pars)$ represents the output of the surrogate model, which is the in-plane stresses $[\sigma_{11}, \sigma_{22}]$, with $\pars$ as input. Since a single measurement is collected at the end of each load step in Examplar 1 (Fig.~\ref{fig:loadpaths A and B}), the observations occur at strains corresponding to the nodes in the load path tree. The number of experimental observations $N_{obs}$ increases with each step in the algorithm as more load steps are added to the load path. Experimental data was simulated by running the MPS at assumed true parameter values $\pars^{true}$ (detailed in Table~\ref{Tab: exemplar 1 parameter cases}) with added Gaussian error according to \eqref{eq: data model} for a specified load path $\des$.

Moving on to algorithmic settings for the ICC framework and calculation of the EIG \eqref{eq: double_nested MC estimator}, in this first exemplar problem, $M$ and $N$ were chosen to be $1\times10^2$ and $1\times10^4$, respectively, and the initial load step before any data was collected was set to be $\xi_{1} = \varepsilon_{11}$. The algorithm was carried out to a total of $T=5$ load steps, with each load step applying a 0.01 mm/mm strain increment, so that the total strain was 0.05 mm/mm at the end of step 5. After the initial load step, the EIG was used within the BOED setup described in Section \ref{subsec: bayesian optimal experimental design} to determine all subsequent load steps in the load path $[\xi_{2}, \xi_{3}, \xi_{4}, \xi_{5}]$.

\subsection{Results}\label{sec: exemplar 1 results}

The ICC algorithm was run for the four different parameter cases detailed in Table~\ref{Tab: exemplar 1 parameter cases}. For all four cases, calibration was performed in an adaptive design setting in which the optimal load path was chosen within the ICC framework as well as in two static design settings chosen based on human intuition \emph{a priori}.\footnote{The static designs were chosen \emph{a priori} in a traditional manner based on human intuition ahead of data collection---not to be confused with any of the guided static designs discussed in Sec.~\ref{sec: introduction}.} The two static cases include one in which the load path was $\varepsilon_{11}$ for each step in the algorithm and, alternatively, one in which the path was $\varepsilon_{22}$ for each step. In total, 12 different combinations (4 parameter cases run under 3 design settings) were studied in Exemplar~1. Since there is randomness present in the data generation process as well as in the EIG estimation, 100 repeat trials were performed for each of the 12 combinations with different instances of noisy data in each trial. Repetition served to make meaningful comparisons in the parameter inference between the adaptive and static designs in an average sense. The repeat trials also serve to establish further confidence in the chosen $M$ and $N$ values for the EIG estimate in \eqref{eq: double_nested MC estimator}. Load paths that result in similar parameter inference over many trials indicate the bias and variance of the estimator are sufficiently small and that the optimal load path is in fact being chosen given the data. Since the data varies from trial to trial due to different instances of the random noise, there may be more than one optimal load path chosen among the different trials. Therefore, when assessing the success of the algorithm, the focus is on the resulting parameter inference and not necessarily on the load path itself. On average, each step in the algorithm took 26 seconds in Exemplar~1, which included the time needed to calculate the EIG as well as to perform the inference.

\subsubsection{Optimal Load Path Selection}\label{sec: exemplar 1 optimal load paths}

\begin{figure}[!htb]
\begin{center}
  \includegraphics[width=0.5\textwidth]{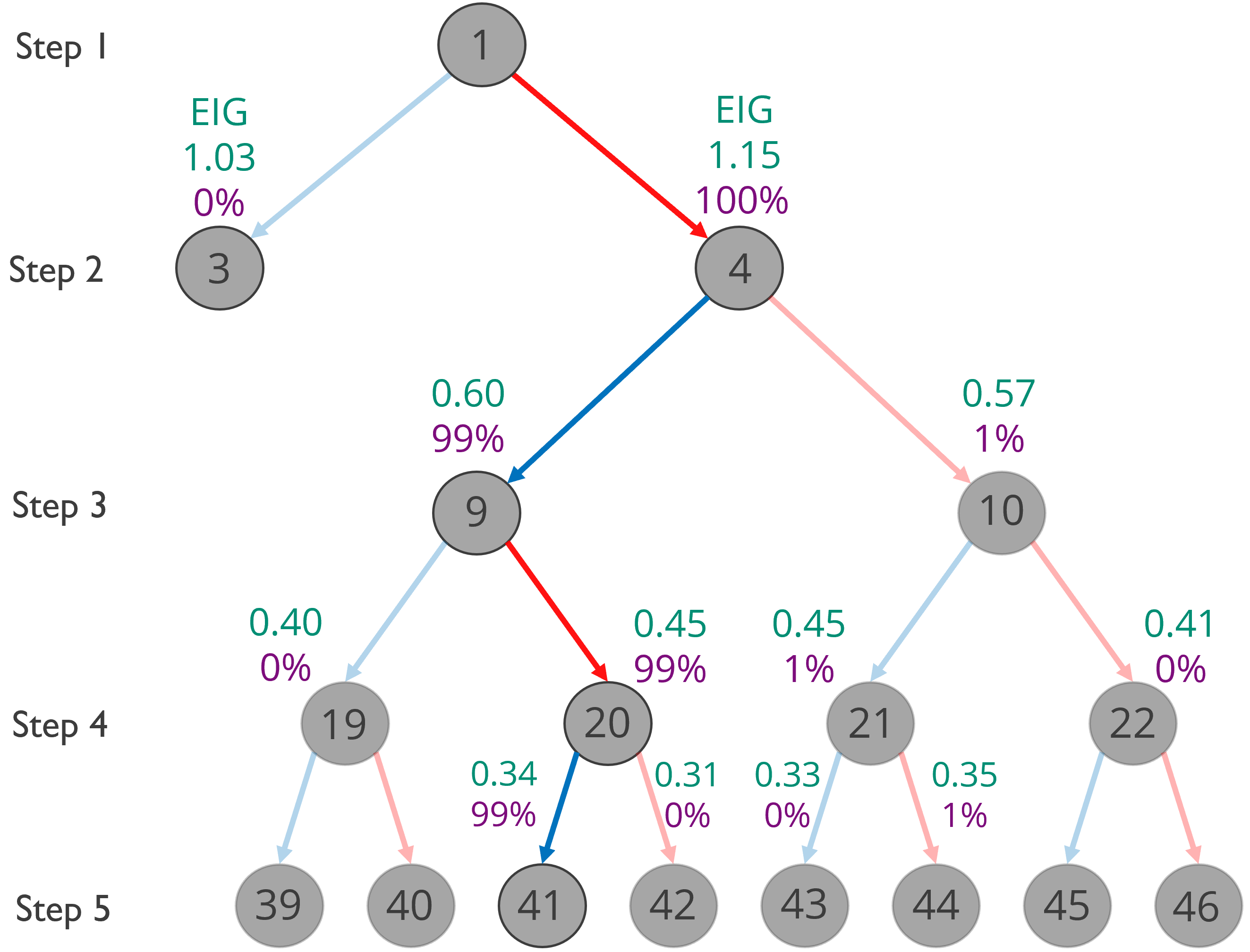}
  \caption{Summary of the load path selection and EIG calculations for Case~1: $F$=0.55 and $G$=0.45. Node numbers correspond to those in Fig.~\ref{fig: loadpath_tree} extended to 5 load steps. The mean EIG over the 100 trials at each node is shown in green, and the percent of trials that went to each node is shown in purple. The most popular path, chosen for 99\% of the trials, is emphasized with darkened arrows. Blue arrows indicate a $\varepsilon_{11}$ load step and red arrows indicate a $\varepsilon_{22}$ load step.}
\label{fig: case 1 Eq_opp_tree_diagram}
\end{center}
\end{figure}

\begin{figure}[!htb]
\begin{center}
  \includegraphics[width=0.35\textwidth]{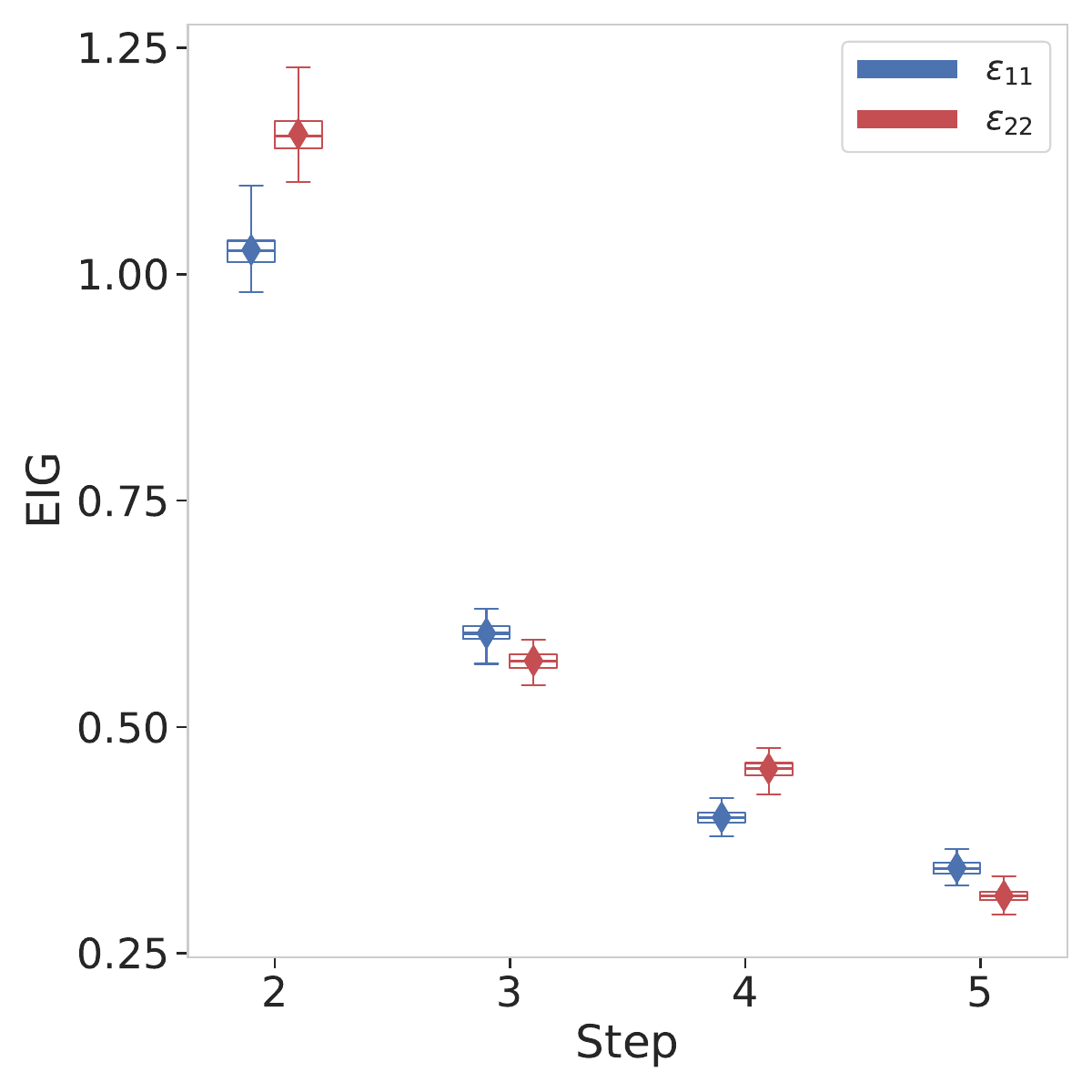}
  \caption{Box plots show the distributions of EIG estimates for each design option over the 99 trials that chose the predominant alternating load path for steps 2-5 ($\xi_{1} = \varepsilon_{11}$ was pre-selected ahead of the ICC algorithm initiation). The box limits are defined by Q1 and Q3, and whiskers extend to the minimum and maximum values. The medians are denoted by a horizontal line inside each box and the means by a diamond marker. The plots are shown are for Case~1: $F$=0.55 and $G$=0.45. The EIG box plots for $\xi = \varepsilon_{11}$ are shown in blue and in red for $\xi = \varepsilon_{22}$. Box plots for each design option are offset for visual clarity.}
\label{fig: case 1 stepwise EIG}
\end{center}
\end{figure}

Details of the load step selection for Case~1---the equal and opposite case in which $F=0.55$ and $G=0.45$---are presented in Figs.~\ref{fig: case 1 Eq_opp_tree_diagram} and \ref{fig: case 1 stepwise EIG}. Fig.~\ref{fig: case 1 Eq_opp_tree_diagram} shows the decision tree for steps 1--5 starting at load step 1 after initially applying an $\varepsilon_{11}$ strain increment (Fig.~\ref{fig: loadpath_tree}). Node numbers are shown that correspond to the same nodes as indicated in Fig.~\ref{fig: loadpath_tree}. Not shown in the tree is the top node (node 0) at the undeformed state nor the right portion of the tree that follows an initial load step of applying an $\varepsilon_{22}$ strain increment---node 2 and all that follow. In the tree diagram, the mean EIG over the 100 trials at each node is shown in green, and the percent of the 100 trials that went to each node is shown in purple.

At step 1, after the initial $\varepsilon_{11}$ load step was specified, the corresponding data was generated, and inference was performed on the unknown parameters to update the prior to the posterior, $\pi(\pars \given \data_{1}, \xi_{1}) \propto f(\data \given \pars, \xi_{1})\pi(\theta)$, where $\xi_{1} = \varepsilon_{11}$. The EIG was then calculated for each of the two load path options, $\varepsilon_{11}$ to go to node 3 or $\varepsilon_{22}$ to go to node 4. In all 100 trials, the optimal load step was determined to be $\varepsilon_{22}$ to go to node 4 at the end of step 2, as this load step had the greater EIG in all trials (mean value of 1.15 over the 100 trials) compared to a $\varepsilon_{11}$ load step (mean EIG value of 1.03 over the 100 trials). At load step 3, in 99\% of the trials, load path $\varepsilon_{11}$ to go to node 9 was determined to be the optimal load step with an average EIG over all trials of 0.60, while 1 trial chose load step $\varepsilon_{22}$ to go to node 10. This process continued for 5 steps.  In the end, the most popular path, chosen for 99\% of the trials, was the alternating load path $\des = [\varepsilon_{11}, \varepsilon_{22}, \varepsilon_{11}, \varepsilon_{22}, \varepsilon_{11}]$, which is emphasized with the darkened arrows.

Figure~\ref{fig: case 1 stepwise EIG} contains box plots summarizing the distributions of the estimated EIG of each design for steps 2-5 of the predominant alternating path shown in Fig.~\ref{fig: case 1 Eq_opp_tree_diagram}. The box bounds are defined by the lower (Q1) and upper (Q3) quartiles of the distributions with whiskers extending to the minimum and maximum EIG estimates of the 99 trials that chose this path. The distribution medians are denoted by a horizontal line within each box and the means by a diamond marker. The EIG noticeably decreases with each step as the possible gain in information from the collection of additional data decreases. There is clear division in the EIG distributions for each load step option at steps 2 and 4, resulting in unanimous step selection at these steps, which was load step $\varepsilon_{22}$. At step 3, overlap in the EIG distributions resulted in 1 path choosing $\varepsilon_{22}$ and the other 99 choosing $\varepsilon_{11}$. There was again overlap in the distributions at load step 5, although all trials chose $\varepsilon_{11}$. Variation in the optimal path selection among different trials may be due in part to the data, as previously mentioned, which varied randomly from trial to trial, in addition to the presence of the EIG estimator bias and variance, which can be reduced with greater computational resources.

Table~\ref{Tab: exemplar 1 design choice} shows the results of the path selection in the adaptive load path design for the 100 different repeat trials for Case~1 as well as for the other 3 cases in Exemplar~1. For Cases 2 and 3, the algorithm chose an alternating design for all 100 trials. In Cases 1 and 4, the alternating design was selected as the optimal one for 99 of the 100 trials. In the singular trial that did not strictly alternate, the $\varepsilon_{22}$ step was repeated once; otherwise, the design choice alternated. Thus, even in the presence of noisy data and randomness in the EIG estimation, an alternating design was clearly preferable for all cases.

\begin{table}[ht!]
\centering
\begin{tabular}{lllllll}
\hline
\textbf{Case No.} & \textbf{Optimal Load Path} & \textbf{Percent} \\ 
\hlineB{4}
\multirow{2}{*}{1}   & $[\varepsilon_{11},\quad \varepsilon_{22}, \quad \varepsilon_{11}, \quad \varepsilon_{22}, \quad \varepsilon_{11}]$   & 99\%  \\
    & $[\varepsilon_{11}, \quad  \varepsilon_{22}, \quad  \varepsilon_{22}, \quad   \varepsilon_{11}, \quad \varepsilon_{22}]$   & 1\%  \\ \hline
2   & $[\varepsilon_{11}, \quad  \varepsilon_{22}, \quad \varepsilon_{11}, \quad \varepsilon_{22}, \quad \varepsilon_{11}]$   & 100\%  \\ \hline
3   & $[\varepsilon_{11}, \quad \varepsilon_{22}, \quad \varepsilon_{11}, \quad \varepsilon_{22}, \quad \varepsilon_{11}]$   & 100\%  \\ \hline
\multirow{2}{*}{4}   & $[\varepsilon_{11}, \quad  \varepsilon_{22}, \quad \varepsilon_{11}, \quad \varepsilon_{22}, \quad \varepsilon_{11}]$   & 99\%  \\
  & $[\varepsilon_{11}, \quad \varepsilon_{22}, \quad \varepsilon_{22}, \quad \varepsilon_{11}, \quad \varepsilon_{22}]$   & 1\%  \\
\hline
\end{tabular}
\caption{Results of the design selection in Exemplar~1 for 4 different parameter cases. Percentages are shown over 100 trials.}\label{Tab: exemplar 1 design choice}
\end{table}

\subsubsection{Posterior Summaries}\label{sec: exemplar 1 posterior summaries}

Functionals of the posterior, such as expected values $\mathbb{E}_{\pars \given \data, \des}$, variances $\mathbb{V}_{\pars \given \data, \des}$ and credible intervals (CI) can be used for parameter estimation and to summarize uncertainty. These values are discussed here for Case~1, while complete results for all four cases are detailed in \ref{ex1.app} for completeness.

\begin{figure}%
    \centering
    \sidesubfloat[\centering]{\includegraphics[width=0.45\textwidth]{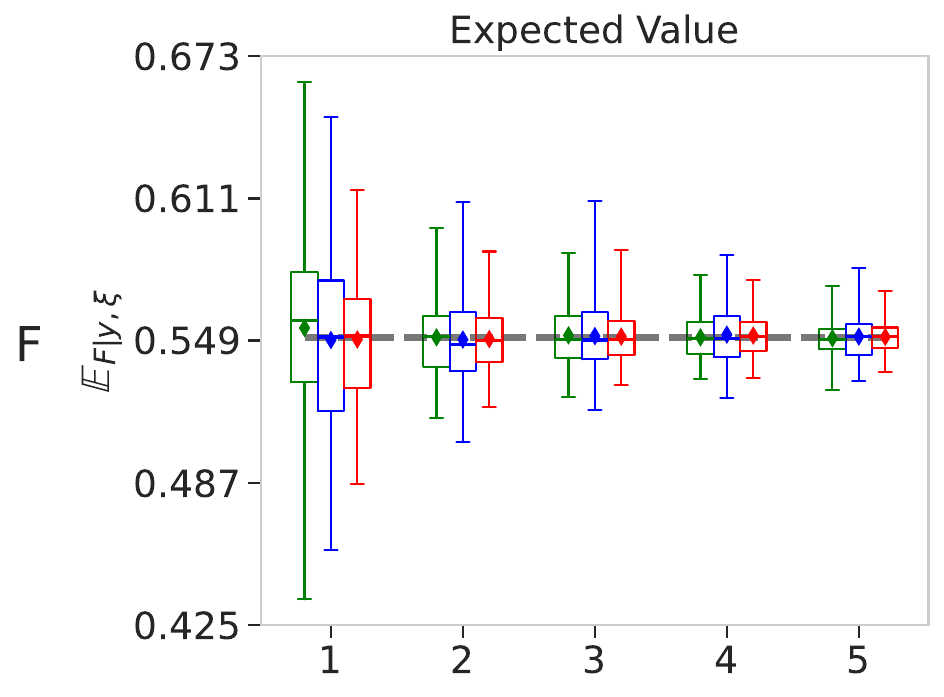}\label{fig: case 1 expected value F} }%
    \sidesubfloat[\centering]{\includegraphics[width=0.45\textwidth]{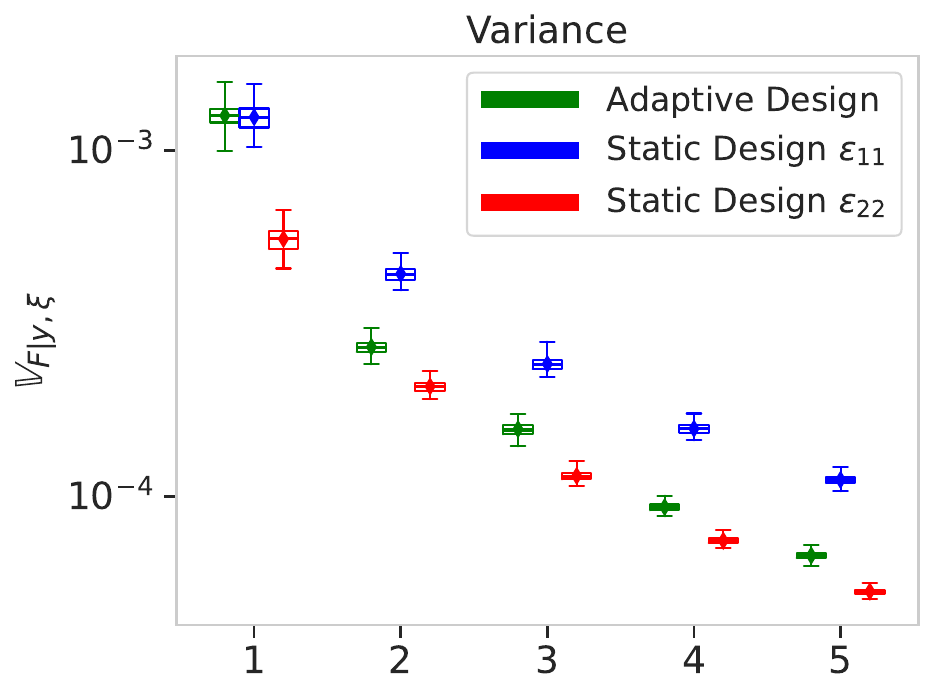}\label{fig: case 1 stdev F} }\\%
    \sidesubfloat[\centering]{\includegraphics[width=0.45\textwidth]{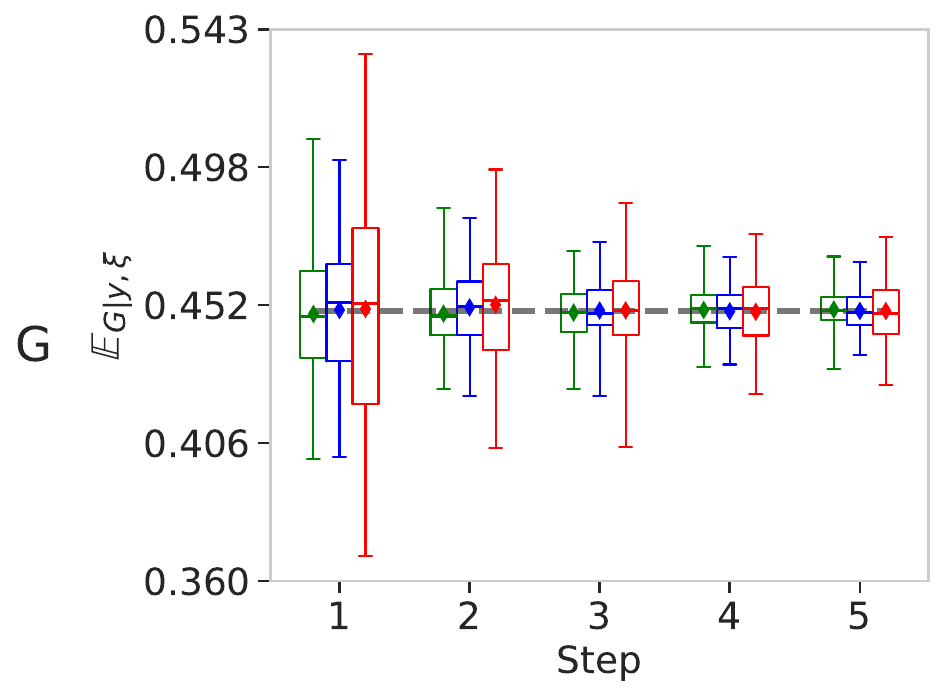}\label{fig: case 1 expected value G} }%
    \sidesubfloat[\centering]{\includegraphics[width=0.45\textwidth]{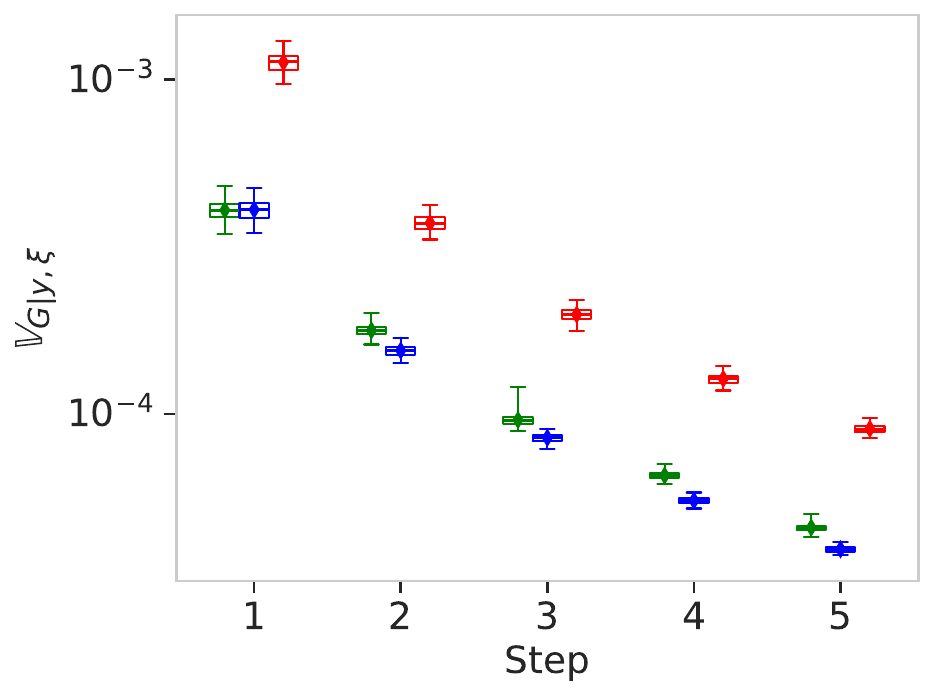}\label{fig: case 1 stdev G} }\\%
    \caption{Box plots are shown which describe the distribution of marginal posterior summaries over the 100 trials for Case~1 at each load step for the adaptive design and two static designs. True parameter values are shown with black dashed lines (Table~\ref{Tab: exemplar 1 parameter cases}). Box plots for the expected value $\mathbb{E}_{\pars \given \data, \xi}$ (left column) and variance $\mathbb{V}_{\pars \given \data, \xi}$ (right column) are shown for parameter $F$ (top row) and parameter $G$ (bottom row). The boxes are bounded by Q1 and Q3 and whiskers extend to the minimum and maximum values in the distribution. The median and mean values are shown with a horizontal line and diamond marker, respectively, for each box plot. The adaptive design results are shown in green, the $\varepsilon_{11}$ design results are shown in blue and the $\varepsilon_{22}$ design results in red. Box plots for each design are offset for visual clarity. } %
    \label{fig: case 1 posterior summary}%
\end{figure}

Box plots showing the distribution of marginal posterior summaries at each step ($t=1,\cdots, 5$) over the 100 repeat trials are plotted in Fig.~\ref{fig: case 1 posterior summary} for Case~1 for the adaptive design as well as the two static designs. The static $\varepsilon_{22}$ load path produced an expected value for parameter $F$, on average, that was closest to the true value (as shown by the mean) with the lowest variability from trial to trial (as shown by the box and whiskers) (Fig.~\ref{fig: case 1 expected value F}), as well as the lowest uncertainty about $F$ at the final load step (Fig.~\ref{fig: case 1 stdev F}). Likewise, the static $\varepsilon_{11}$ load path yielded data that provided an expected value for parameter $G$, on average, that was closest to the true value at the final step with the lowest variability among the trials (Fig.~\ref{fig: case 1 expected value G}), as well as the least amount of uncertainty (Fig.~\ref{fig: case 1 stdev G}). Similar results were obtained for the other three cases as well (see Table~\ref{Tab: exemplar 1 posterior summaries} in \ref{ex1.app}). 

While marginal summaries are useful for understanding the uncertainty and estimating the quality of each individual parameter, the interest of this work is to obtain the optimal joint calibration of the parameters that has the lowest combined uncertainty over all parameters. Two different scalar measures of multi-dimensional dispersion are the generalized variance and the total variance. The generalized variance is the determinant of the covariance matrix, $\det{(\Psi)}$, and is proportional to the volume of space occupied by the distribution through its square root. This metric reflects parameter correlations and takes on smaller values in the presence of high parameter correlation and higher values when there is little to no parameter correlation. High diffusivity in the distribution, and likewise uncertainty, is indicated by large values. The total variance is obtained through the trace of the covariance matrix, $\mbox{trace}(\Psi)$. As its name suggests, by summing all the diagonal components of variance, the total variance provides a measure of the total amount of variation in the distribution.

\begin{figure}%
    \centering
    \sidesubfloat[]{\includegraphics[width=0.45\textwidth]{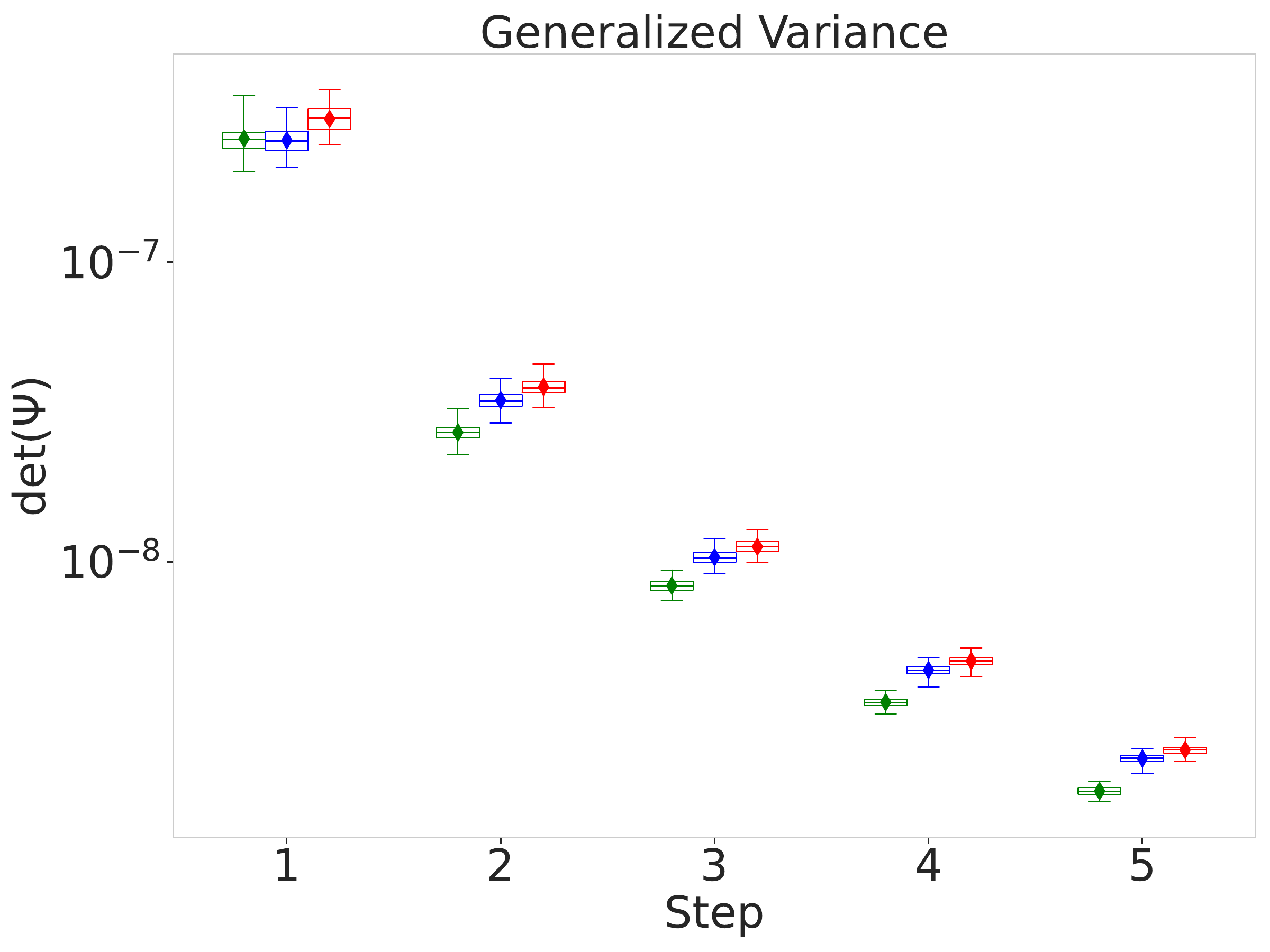}\label{fig: case 1 generalized_variance} }%
    \sidesubfloat[]{\includegraphics[width=0.45\textwidth]{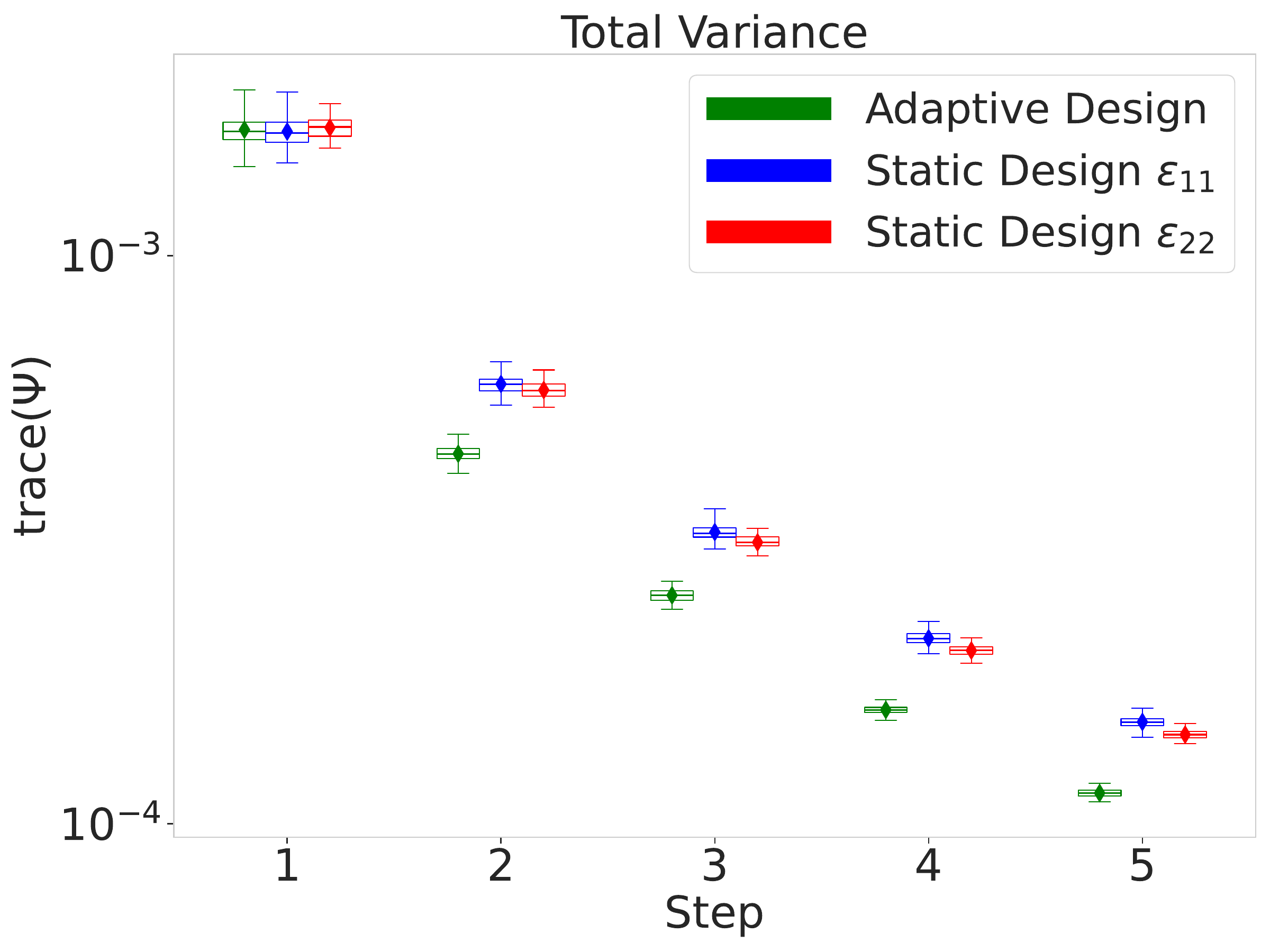}\label{fig: case 1 total_variance} }%
    \caption{Box plots of the generalized (a) and total (b) variance for Case~1: $F$ = 0.55, $G$ = 0.45. The plots summarize the distributions over the 100 repeat trials. The boxes are bounded by Q1 and Q3 and have whiskers that extend to the minimum and maximum values. The median values are indicated with a horizontal line in the box, and the mean values are shown with diamond markers. Box plots for each design are offset for visual clarity.} %
    \label{fig: case 1 1D_cov_metrics}%
\end{figure}

Fig.~\ref{fig: case 1 1D_cov_metrics} shows box plots of the generalized (\ref{fig: case 1 generalized_variance}) and total (\ref{fig: case 1 total_variance}) variance of the posterior for the adaptive and static designs at each step in the algorithm for Case~1. The plots summarize the distribution of results that were obtained over the 100 repeat trials. Both the generalized variance and total variance decreased with every step in the algorithm---corresponding to reduced parameter uncertainty as more data was collected. The adaptive design had a smaller variance by both metrics on average over the 100 trials than either of the static designs at each step. Again, similar results were obtained for the other three cases as well (see Table~\ref{Tab: exemplar 1 1D metrics} in \ref{ex1.app}).

It has been established that the overall parameter uncertainty was lower in the adaptive design for Exemplar~1 compared to the two static designs. However, if the lower uncertainty is accompanied by a greater bias in parameter expected values, then the inference results in a posterior distribution the places greater probability (confidence) in parameter values that are farther from the true values---an undesirable scenario. A useful metric in comparing how close a point $x$ is to a given distribution in multivariate space, often used for outlier detection, is the Mahalanobis distance (MD) \cite{mahalanobis1936mahalanobis},

\begin{equation}
    \mbox{MD}(x,\mu) = \left((x - \mu)^{T}\Psi^{-1}(x - \mu)\right)^{0.5},
\end{equation}\label{eq: mahalanobis distance}

where $\mu$  and $\Psi$ are the mean and covariance of the distribution and $x$ is the point from which the distance is being calculated. If the variables are uncorrelated, the axes of the distribution are orthogonal to each other, and the Mahalanobis distance is equivalent to the Euclidean distance. However, in the case where two or more parameters are correlated, the axes are no longer orthogonal, and the Mahalanobis distance takes this correlation into account. A lower Mahalanobis distance corresponds to the point residing in a region of higher probability in the distribution. The average Mahalanobis distance of the true parameter values $\mbox{MD}(\pars^{true})$ from the posterior distribution at the final step was $1.25$ in the adaptive design and $1.26$ for both the $\varepsilon_{11}$ and $\varepsilon_{22}$ static designs in Case~1. Table~\ref{Tab: exemplar 1 1D metrics} in \ref{ex1.app} reports $\mbox{MD}(\pars^{true})$ for all cases, and in summary, $\mbox{MD}(\pars^{true})$ was less in the adaptive settings than either of the static settings in three of the four presented cases, meaning $\pars^{true}$ sat in a region of higher probability, on average, when the load path was adaptively chosen. In Case 4, the static $\varepsilon_{11}$ design had the lowest MD, followed by the adaptive design and then finally the $\varepsilon_{22}$ design.

These scalar metrics reveal that when the load path is adaptively chosen within the ICC framework, the calibrated parameters have a lower combined uncertainty, and the posterior distribution places a higher probability on the true values (in most cases) when compared to the static designs. While the advantage of the adaptive design for this exemplar problem is demonstrably small (i.e., a marginally lower uncertainty and MD), these results show the success of the algorithm in actively obtaining data that is more informative for parameter calibration than the selected static designs.

\subsubsection{Propagation of Uncertainty to Stress-Strain Space}\label{sec: case 1 propagated uncertainty}

The end goal of obtaining accurate parameter estimates with known levels of uncertainty is to be able to propagate the parameter uncertainty through the model in the application of interest. Here, the parameter uncertainty is propagated through the MPS for demonstrative purposes and compared for each of the design settings. One hundred samples were drawn from the posterior distribution $\tilde{\pars} \sim \pi(\pars \given \data, \des)$ for each of the three experimental designs (adaptive and both static designs) for Case~1 and propagated to the model output space by running the MPS at the sampled parameters $\bf{u}(\tilde{\pars}) = [\sigma_{11}, \sigma_{22}]$.

\begin{figure}[!htb]%
    \centering
    \sidesubfloat[]{\includegraphics[width=0.45\textwidth]{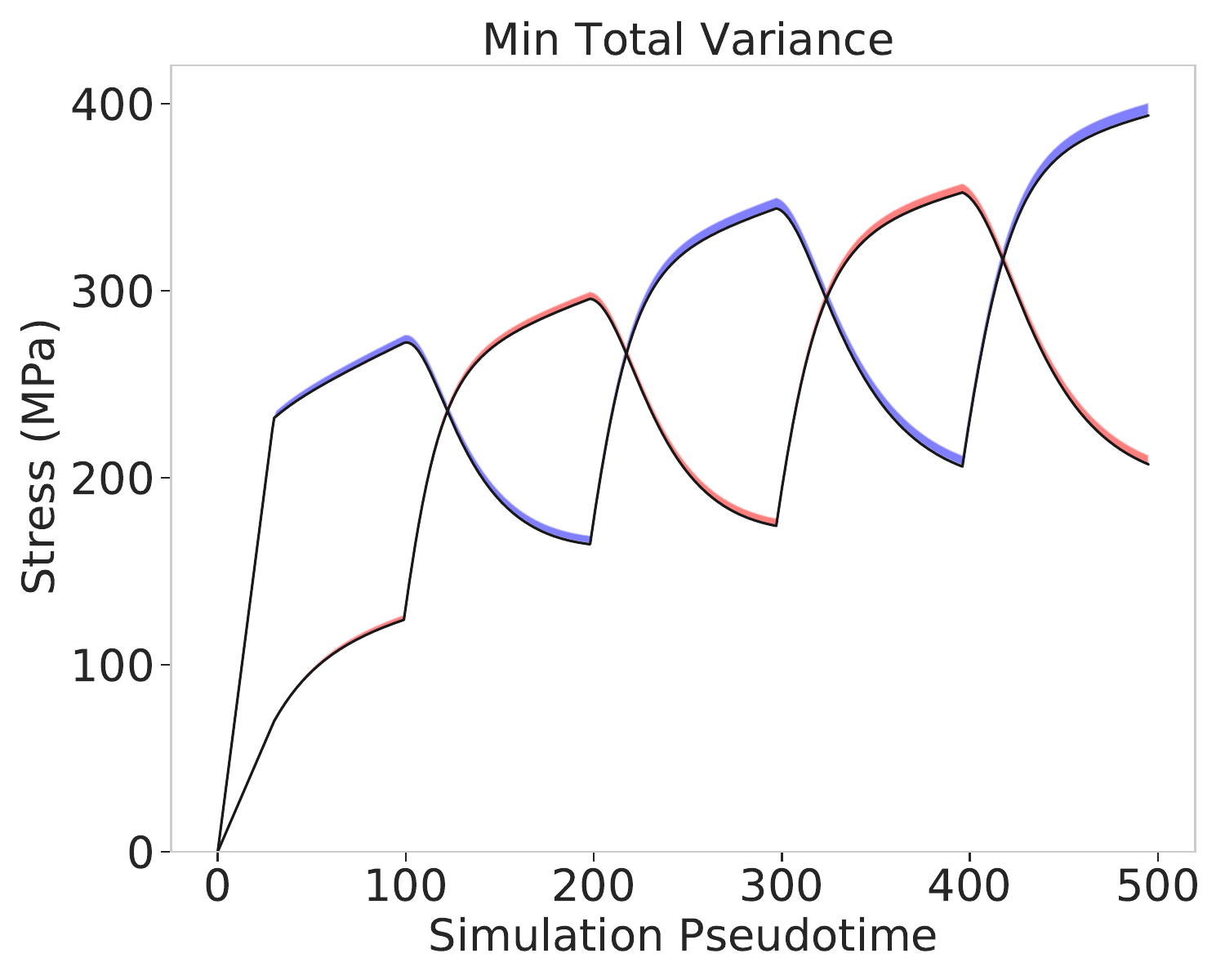}\label{fig: adaptive min total var} }%
    \sidesubfloat[]{\includegraphics[width=0.45\textwidth]{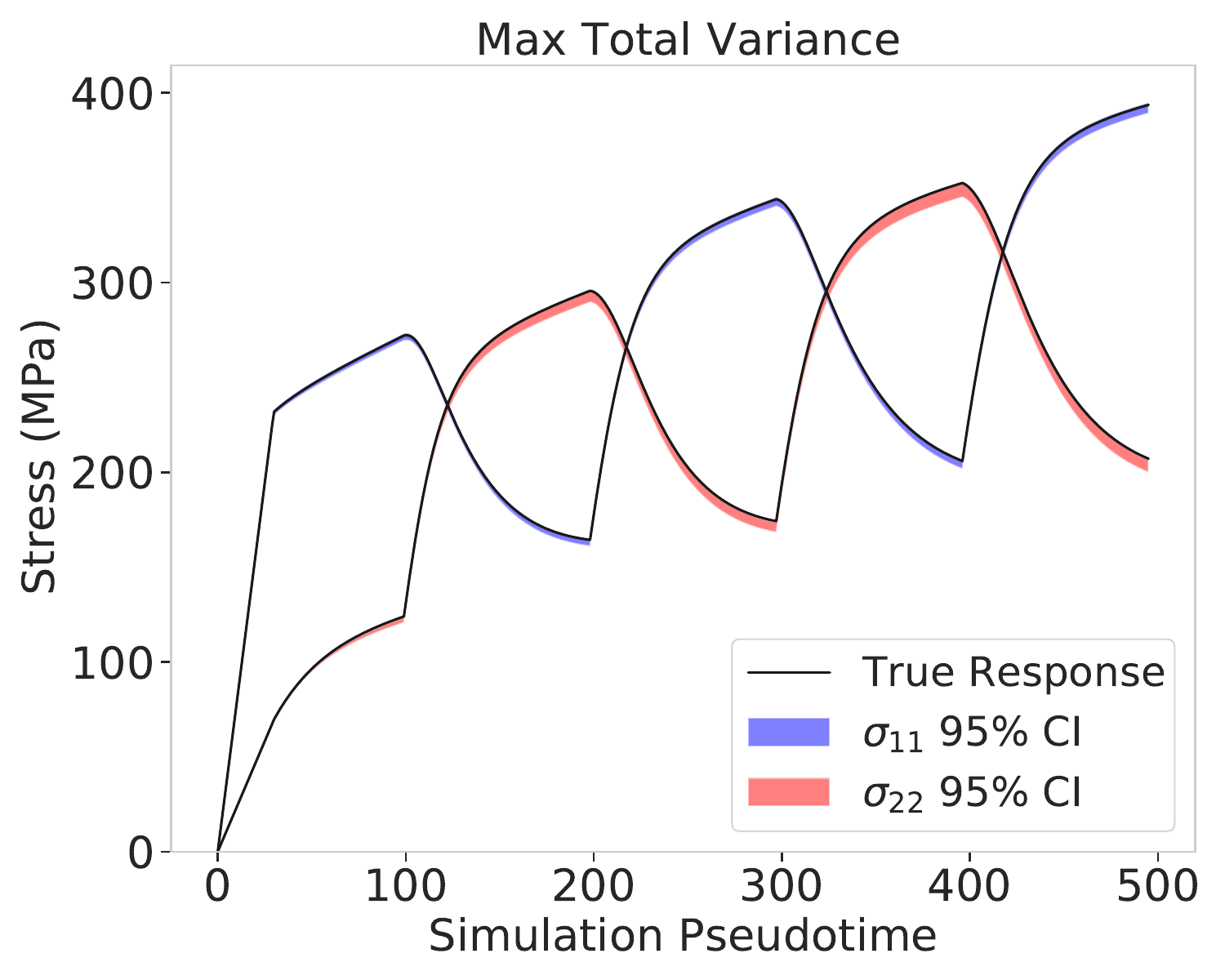}\label{fig: adaptive max total var} }%
    \\
    \sidesubfloat[]{\includegraphics[width=0.45\textwidth]{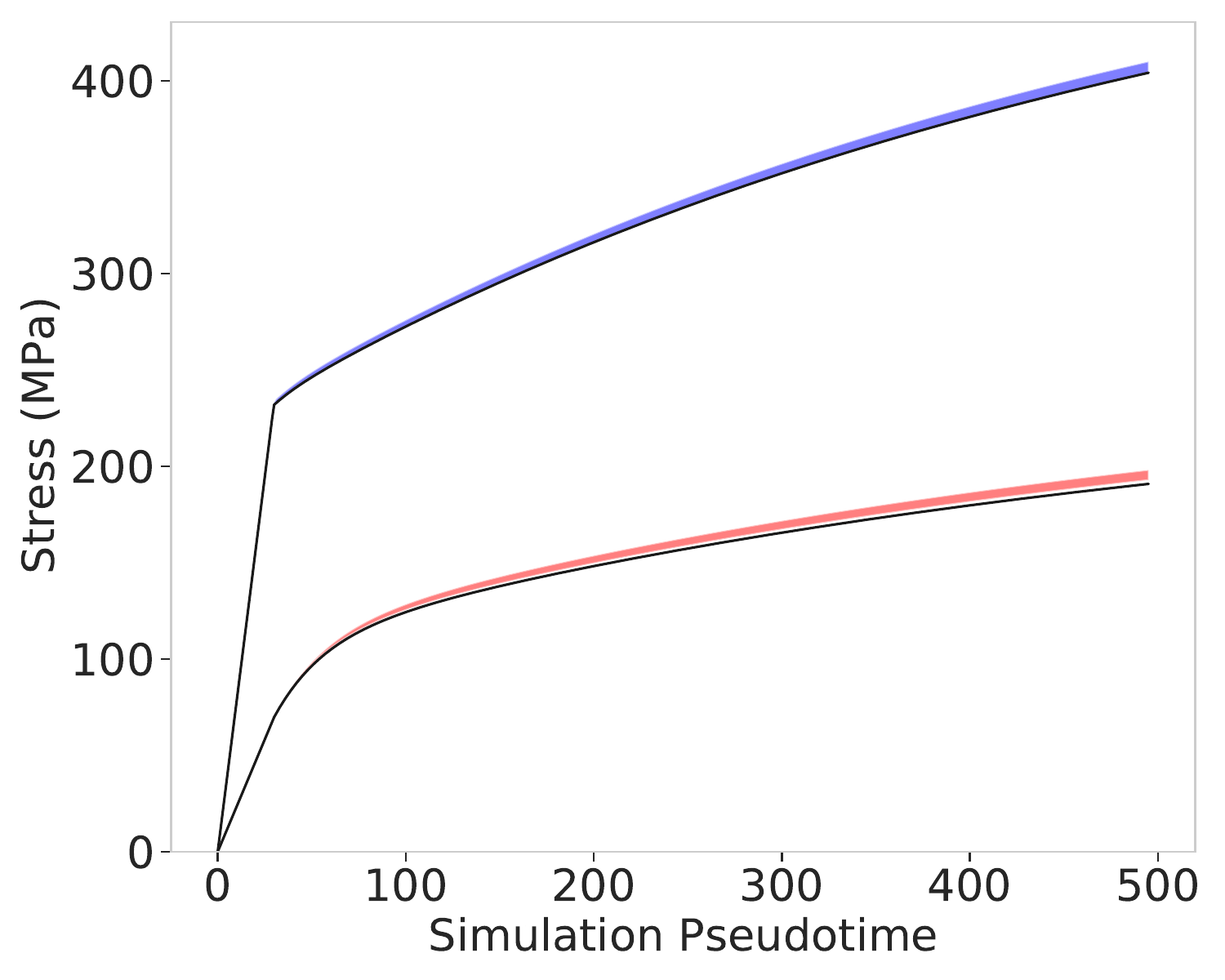}\label{fig: static 11 min total var} }%
    \sidesubfloat[]{\includegraphics[width=0.45\textwidth]{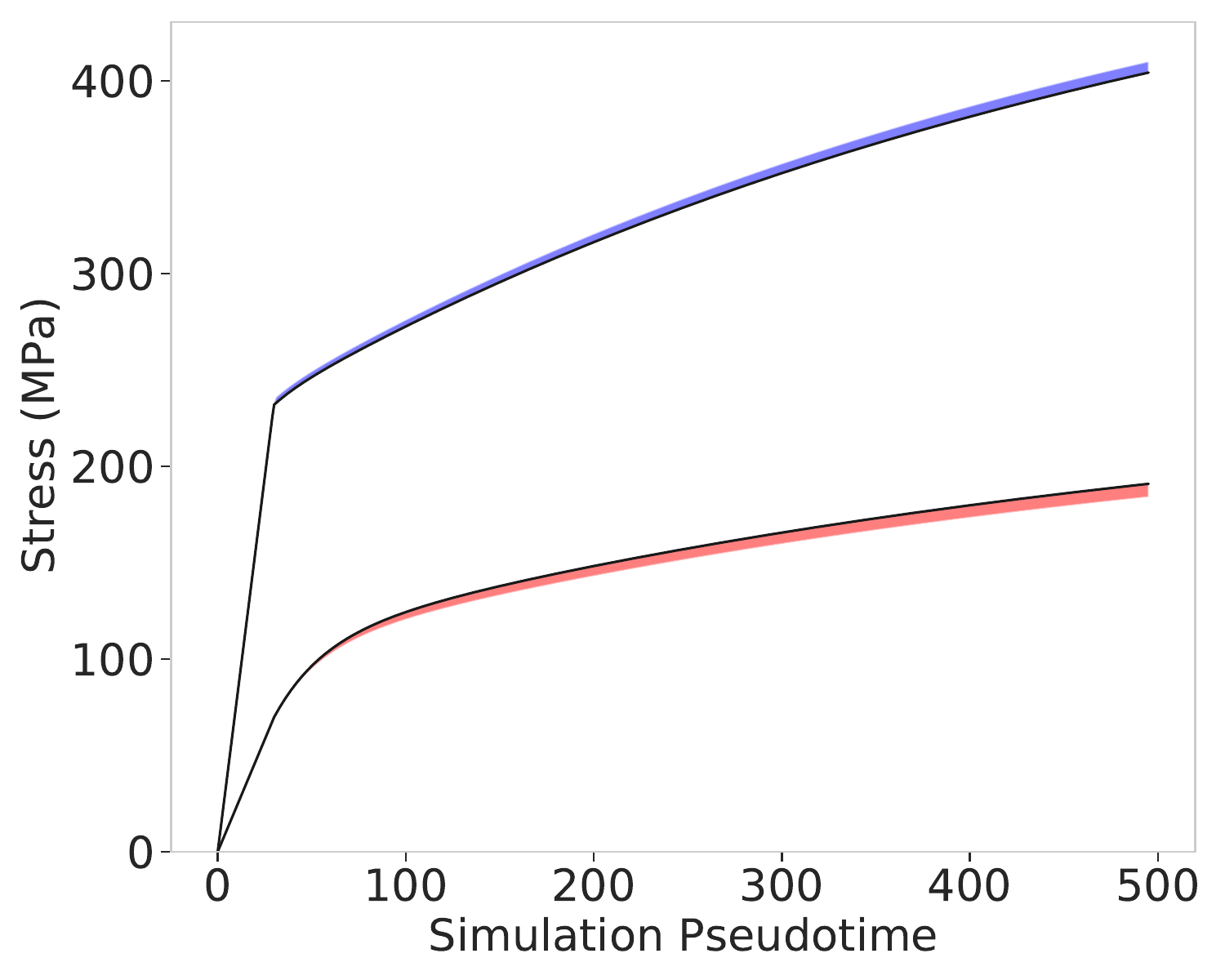}\label{fig: static 11 max total var} }%
    \\
    \sidesubfloat[]{\includegraphics[width=0.45\textwidth]{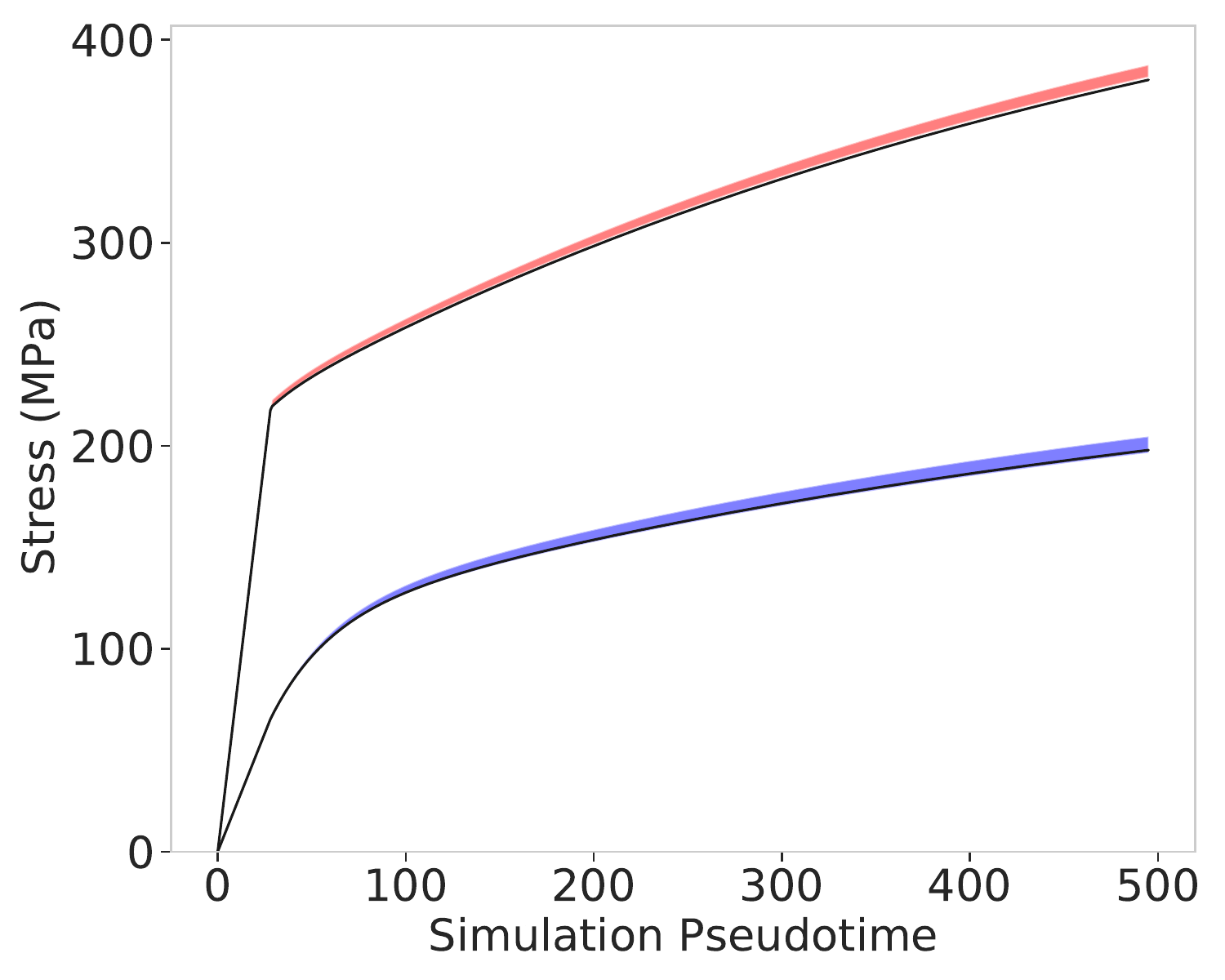}\label{fig: static 22 min total var} }%
    \sidesubfloat[]{\includegraphics[width=0.45\textwidth]{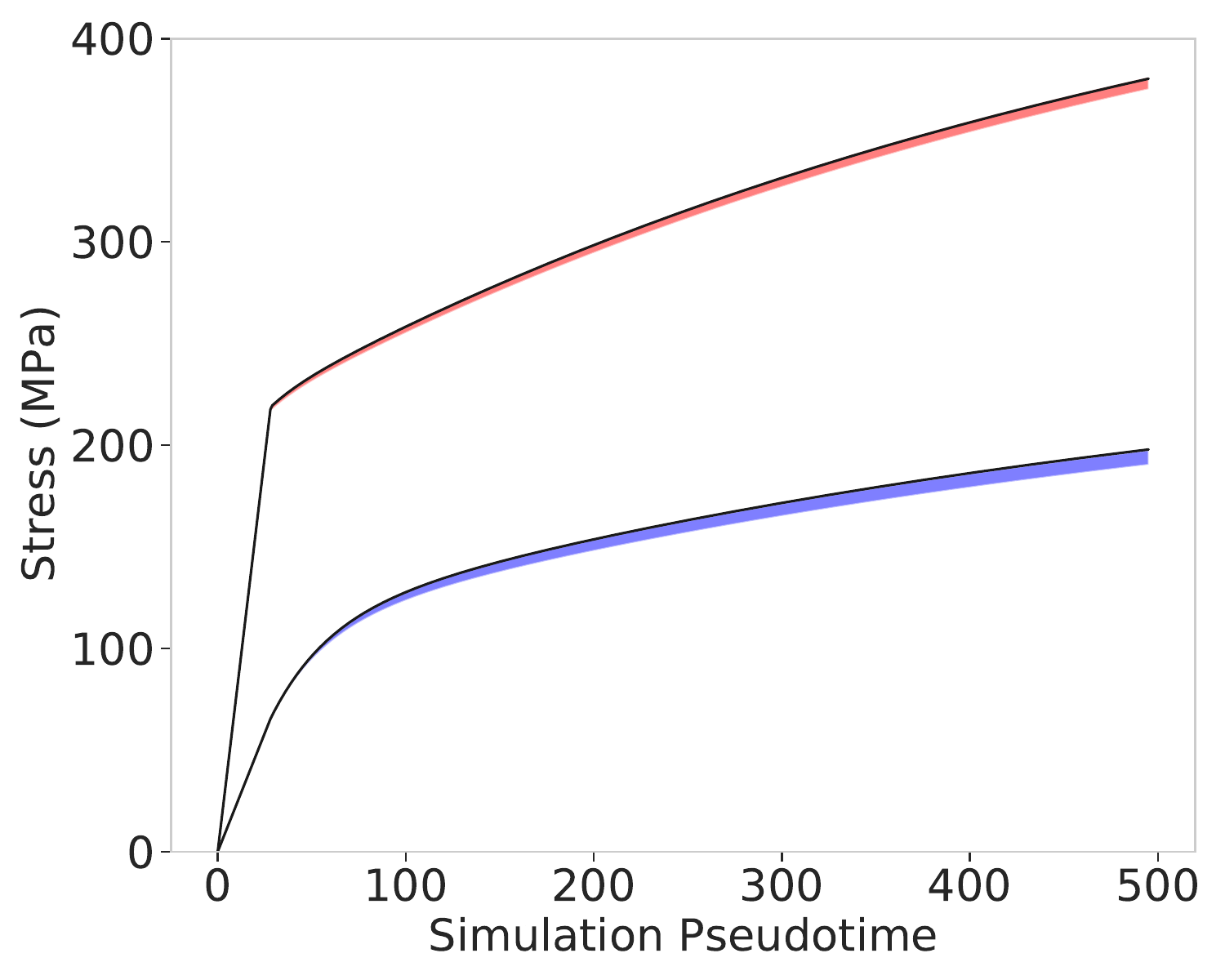}\label{fig: static 22 max total var} }%
    \caption{95\% credible intervals for the posterior draws $\tilde{\pars} \sim \pi(\pars \given \data, \des)$ were calculated point-wise for Exemplar~1, Case~1 and are indicated the the blue ($\sigma_{11}$) and red ($\sigma_{22}$) dashed lines. The model output from the true parameter values is plotted with a black line. The left column of plots show the 95\% CI for the trial in each design setting that had a posterior distribution with the least total variance among all trials. The right column of plots show 95\% CI for the trial in each design setting that had the greatest total variance. Plots (a) and (b) show results for the adaptive design, plots (c) and (d) show results for the static $\varepsilon_{11}$ design and plots (e) and (f) for the static $\varepsilon_{22}$ design.}%
    \label{fig: exemplar 1 posterior draw CI}%
\end{figure}

In Fig.~\ref{fig: exemplar 1 posterior draw CI}, 95\% CIs for computed in-plane stress values from the 100 draws are shown along with the MPS output at the true parameter values $u(\pars^{true})$. The left column of plots shows the 95\% CI for the trial in each design setting that had the least posterior total variance, and the right column of plots shows the 95\% CI for the trial that had the greatest posterior total variance. Plots are shown for both the minimum and maximum total variance in order to illustrate the best and worst case scenario within the 100 trials for each design setting. For all three design settings, there is little difference in the CIs for the two bounding scenarios. There are instances where the true response sits outside the 95\% CIs, most notably in Fig.~\ref{fig: static 11 min total var}~and~\ref{fig: static 22 min total var}. For the adaptive design, both the minimum and maximum total variance came from trials with an optimal design of $\des^{*} = [\varepsilon_{11}, \varepsilon_{22}, \varepsilon_{11}, \varepsilon_{22}, \varepsilon_{11}]$. For all three experimental designs, the low level of posterior uncertainty in the $F$ and $G$ parameters translates to low uncertainty in the stress-strain space, as evidenced by the narrow credible intervals. Thus, for this simplified exemplar, the marginally reduced uncertainty in the parameters obtained with the adaptive design did not translate to a noticeable reduction in the uncertainty of the final QOI (namely, the stress response) when compared to the static designs.

%% file: Exemplar_2.tex
\section{Exemplar~2}\label{sec: exemplar 2}

In this section, details of the statistical modeling choices and algorithmic settings are discussed for Exemplar~2 (Sec.~\ref{sec: exemplar 2 set up}) and results are presented (Sec.~\ref{sec: exemplar 2 results}). The adaptive load paths chosen from within the ICC framework are discussed (Sec.~\ref{sec: exemplar 2 optimal load paths}) and the resulting parameter inference is compared to two naive static design settings via posterior summaries (Sec.~\ref{sec: exemplar 2 posterior summaries}) and propagated uncertainty (Sec.~\ref{sec: exemplar 2 uncertainty propagation}).

\subsection{Problem Setup}\label{sec: exemplar 2 set up}

In the second exemplar, introduced in Sec.~\ref{sec: exemplar problems}, the calibration was made more complex by adding the yield stress $\sigma_{y}$ and the hardening parameters, $A$ and $n$, \eqref{eq: hardening behavior} to the unknown parameter vector $\pars = [F, G, \sigma_y, A, n]^T$.

As in Exemplar~1, the parameters were represented with a truncated normal probability model \eqref{eq: exemplar 1 prior settings} in order to constrain the values they could take to be in agreement with the bounds used for building the surrogates (Table~\ref{Tab: parameters in exemplars}). The means, variances and bounds of the priors on $F$ and $G$ were the same as those used in Exemplar~1. The prior means for the yield stress $\sigma_y$ and hardening parameters $A$ and $n$ were chosen to be within the defined bounds, and variances were chosen that were moderately diffuse. The upper and lower bounds ($a$ and $b$) of the distribution were the same as those used in the surrogate builds (Sec.~\ref{sec: surrogate model construction}). Prior means, variances and bounds are detailed in Table~\ref{Tab: exemplar 2 priors}. The model for the data was chosen to be the same as that used in Exemplar~1 \eqref{eq: likelihood}, and the observation noise variance was assumed to be known at $\psi^{2} = 10 \mbox{MPa}^{2}$.

\begin{table}[ht!]
\centering
\begin{tabular}{lllll}
\hline
\textbf{Parameter} & $\boldsymbol{\mu_{\theta}}$ & $\boldsymbol{\delta^{2}_{\theta}}$ & $\boldsymbol{a}$ & $\boldsymbol{b}$ \\ 
\hlineB{4}
$F$   & 0.5    & 1   & 0.3 & 0.7  \\
$G$   & 0.5    & 1   & 0.3 & 0.7 \\
$A$   & 200    & 1000 & 10  & 400\\
$n$   & 50     & 100 & $1\times10^{-6}$ & 100\\
$\sigma_{y}$   & 250 &  1000 & 50 & 500\\
\hline
\end{tabular}
\caption{Means $\mu_{\theta}$, variances $\delta^{2}_{\theta}$ and bounds $a$ and $b$ for the truncated normal prior model \eqref{eq: exemplar 1 prior settings} used in Exemplar~2. }\label{Tab: exemplar 2 priors}
\end{table}

Exemplar~2, which had 5 unknown parameters and exercised two different phenomenologies (yield and hardening), presented more complexities than Exemplar~1, which only had 2 unknown parameters and exercised only one phenomenology (yield). Specifically, local modes in the posterior made it more difficult to find $\hat{\pars}$ \eqref{eq: MAP}, which was overcome by increasing the number of restarts (initializations from random locations) in the minimization. Parameter uncertainties were greater, leading to a higher susceptibility to arithmetic underflow (\ref{sec: EIG_approx.app}) in the initial steps of the algorithm. To overcome the occurrence of underflow, $M$ in the EIG estimate was increased to $1\times10^4$, which was the same value as $N$. Finally, in order to mitigate problems with parameter identifiability, the number of load steps was increased to $T=7$, with each load step containing a 0.02 mm/mm strain increment, so that the total strain was 0.14 mm/mm at the end of step 7, and data points were collected at three equally-spaced locations between each load step. Thus, three times more data was included compared to Exemplar~1. In total, these modifications allowed the ICC framework to be successful for the more complicated material model of Exemplar~2, as discussed in the following sections. The final deviation from the algorithmic setup in Exemplar~1 is that in Exemplar~2 the initial load step was determined by calculating the EIG, as opposed to Exemplar~1, in which the initial load step was pre-determined to be $\xi_{1} = \varepsilon_{11}$. In Exemplar~2, the time to complete each step in the algorithm (EIG calculation plus inference) was just under 3 minutes on average.

\subsection{Results}\label{sec: exemplar 2 results}

The two parameter cases for Exemplar~2 detailed in Sec.~\ref{sec: exemplar problems} were used in an adaptive setting in which the optimal load path was chosen by calculating the EIG, as well as in two static design settings---one in which the load path was decided \emph{a priori} to be $\varepsilon_{11}$ each step in the algorithm and, alternatively, one in which the path was $\varepsilon_{22}$ for each step (reflecting traditional approaches based on human intuition). In total, 6 different combinations (2 parameter cases run under 3 design settings) were studied in this second exemplar. As in Exemplar~1, 100 repeat trials were performed for each of the 6 combinations.

\subsubsection{Optimal Load Path}\label{sec: exemplar 2 optimal load paths}

Table~\ref{Tab: exemplar 2 design choice} shows the results of the path selection for the adaptive design for the 100 different repeat trials for both parameter cases. Unlike in Exemplar~1, there was not a single dominant load path in either case. Instead, four paths were all relatively likely (chosen 14--35\% of the time), and a fifth path was chosen a few times in each case (2--4\% of the time). Additionally, while the designs were predominately alternating (similar to Exemplar~1), most paths contained at least one repeat load step, typically near the end of the load path.

\begin{figure}
    \centering
    \includegraphics[width=0.45\linewidth]{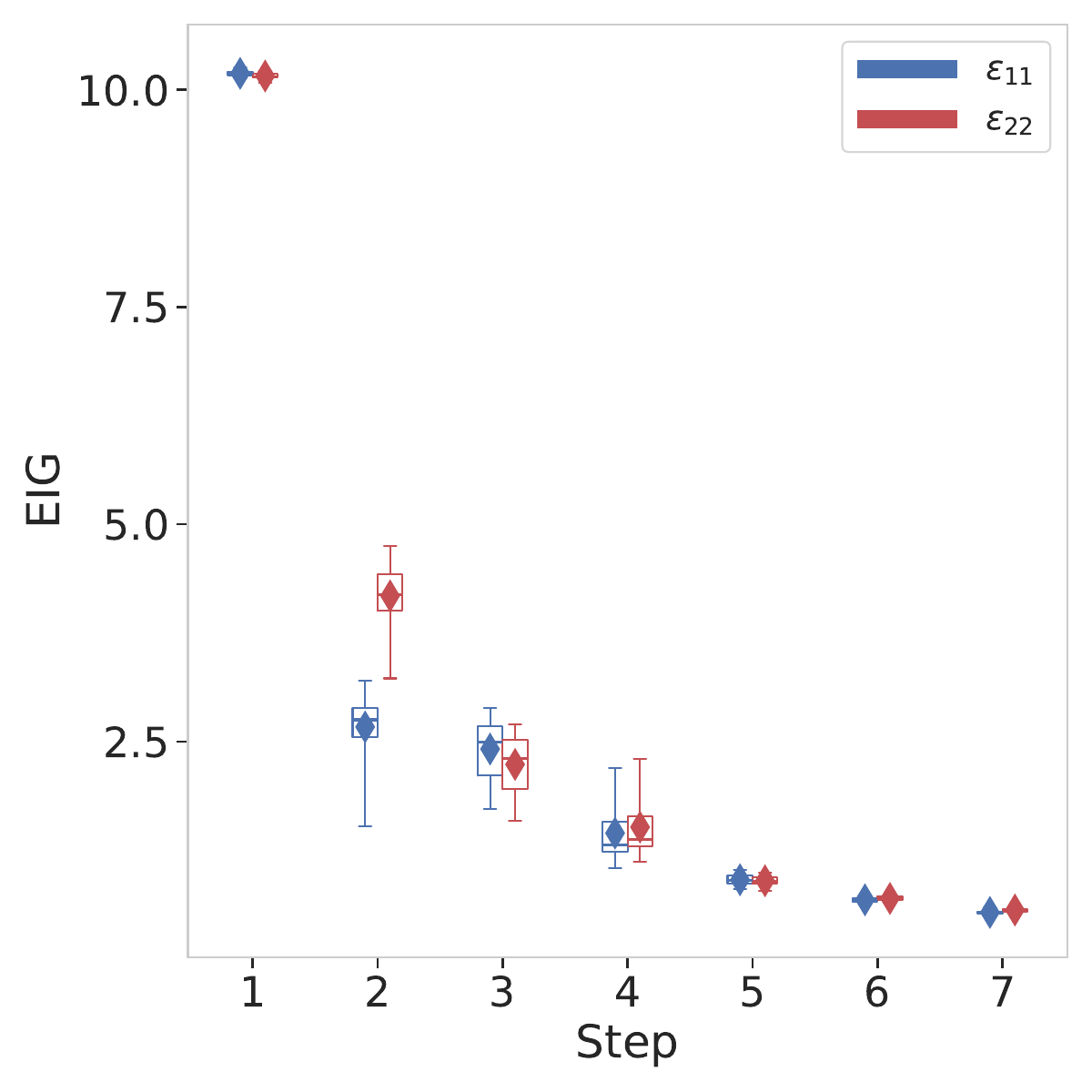}
    \caption{Box plots show the distributions of the EIG estimates at steps 1-7. The represented distributions are over the 31 trials that chose the most popular path in Case~5. Box bounds indicate the upper and lower quartiles and whiskers extend to the minimum and maximum values. The median and mean values are shown with a horizontal line in the box and a diamond marker, respectively. The plots for $\xi = \varepsilon_{11}$ are shown in blue and in red for $\xi = \varepsilon_{22}$. Box plots for each load step option are offset for visual clarity.}
    \label{fig: exemplar 2 EIG}
\end{figure}

Figure.~\ref{fig: exemplar 2 EIG}---shown for the most popular design of Case~5---reveals that at some steps, the distributions of EIG estimates for each of the load step options were very similar, meaning that each load step option was expected to provide nearly the same amount of information. Indeed, as shown in Table~\ref{Tab: exemplar 2 design choice}, the marginal expected values and generalized variance were similar for all five load paths for both cases---indicating that similar information content was obtained from the various optimal load paths.

The higher number of optimal load paths for Exemplar~2 compared to Exemplar~1 may be caused by having more unknown parameters and activating two different phenomenologies. The different instances of noisy data that were used in each trial and the inherent variation in the EIG estimation (see Sec.~\ref{subsec: bayesian optimal experimental design}) may combine with complex parameter interactions to propel the BOED algorithm to select one path over another similar path.

In an actual physical experiment, only a single load path would be selected \textit{in situ}, and this multitude of different optimal paths may not be noticed unless multiple physical experiments were performed. However, as shown in the following sections (as well as in Table~\ref{Tab: exemplar 2 design choice}, where the values from the static designs are recorded below the dashed lines) the paths generated adaptively within the ICC framework produce similar parameter estimates that, notably, are better on the whole than parameter estimates obtained through the static designs.

\begin{table}[!htb]
\centering
\resizebox{\textwidth}{!}{\begin{tabular}{lllllllll}
\hline
\textbf{Case No.} & \textbf{Optimal Load Path} & \textbf{Percent} & $\boldsymbol{\mathbb{E}_{F \given y, \xi}}$ & $\boldsymbol{\mathbb{E}_{G \given y, \xi}}$ & $\boldsymbol{\mathbb{E}_{A \given y, \xi}}$ & $\boldsymbol{\mathbb{E}_{n \given y, \xi}}$ & $\boldsymbol{\mathbb{E}_{\sigma_y \given y, \xi}}$ & \begin{tabular}[c]{@{}l@{}} \textbf{Gen. Var.} \\ $\boldsymbol{(1\times10^{-9})}$ \end{tabular} \\ 
\toprule
\multirow{7}{*}{5} & $[\varepsilon_{11},\quad \varepsilon_{22}, \quad \varepsilon_{11}, \quad \varepsilon_{22}, \quad \varepsilon_{11}, \quad \varepsilon_{22}, \quad \varepsilon_{22}]$ & 32\% & 0.550 & 0.449 & 101 & 20.1 & 299 & 1.19 \\ 
       & $[\varepsilon_{22},\quad \varepsilon_{11}, \quad \varepsilon_{22}, \quad \varepsilon_{11}, \quad \varepsilon_{11}, \quad \varepsilon_{22}, \quad \varepsilon_{22}]$ & 31\% & 0.550 & 0.450 & 101 & 20.4 & 299 & 1.22 \\
       & $[\varepsilon_{11},\quad \varepsilon_{22}, \quad \varepsilon_{11}, \quad \varepsilon_{22}, \quad \varepsilon_{22}, \quad \varepsilon_{11}, \quad \varepsilon_{11}]$  & 19\% & 0.551 & 0.452 & 99.6 & 20.5 & 300 & 1.11 \\
       & $[\varepsilon_{22},\quad \varepsilon_{11}, \quad \varepsilon_{22}, \quad \varepsilon_{11}, \quad \varepsilon_{22}, \quad \varepsilon_{11}, \quad \varepsilon_{11}]$  & 14\% & 0.547 & 0.446 & 98.5 & 19.4 & 301 & 1.30 \\
       & $[\varepsilon_{11},\quad \varepsilon_{22}, \quad \varepsilon_{11}, \quad \varepsilon_{22}, \quad \varepsilon_{11}, \quad \varepsilon_{11}, \quad \varepsilon_{22}]$  & 4\% & 0.549 & 0.454 & 100 & 20.3 & 300 & 1.15 \\
       \cdashlinelr{2-9}
       & $[\varepsilon_{11},\quad \varepsilon_{11}, \quad \varepsilon_{11}, \quad \varepsilon_{11}, \quad \varepsilon_{11}, \quad \varepsilon_{11}, \quad \varepsilon_{11}]$ & NA & 0.550 & 0.460 & 101 & 20.3 & 301 & 160 \\ 
       & $[\varepsilon_{22},\quad \varepsilon_{22}, \quad \varepsilon_{22}, \quad \varepsilon_{22}, \quad \varepsilon_{22}, \quad \varepsilon_{22}, \quad \varepsilon_{22}]$ & NA & 0.545 & 0.450 & 99.6 & 20.0 & 299 & 116 \\
 \midrule
\multirow{7}{*}{6} & $[\varepsilon_{11},\quad \varepsilon_{22}, \quad \varepsilon_{11}, \quad       \varepsilon_{22}, \quad \varepsilon_{11}, \quad \varepsilon_{22}, \quad \varepsilon_{22}]$  &  35\% & 0.550 & 0.450 & 300 & 20.0 & 100 & 0.351 \\ 
       & $[\varepsilon_{22},\quad \varepsilon_{11}, \quad \varepsilon_{22}, \quad \varepsilon_{11}, \quad \varepsilon_{22}, \quad \varepsilon_{11}, \quad \varepsilon_{11}]$  & 27\% &  0.547 & 0.448 & 299 & 20.0 & 99.6 & 0.370 \\
       & $[\varepsilon_{11},\quad \varepsilon_{22}, \quad \varepsilon_{11}, \quad \varepsilon_{22}, \quad \varepsilon_{11}, \quad \varepsilon_{22}, \quad \varepsilon_{11}]$ & 22\% & 0.552 & 0.450 & 300 & 19.9 & 101 & 0.353 \\
       & $[\varepsilon_{22},\quad \varepsilon_{11}, \quad \varepsilon_{22}, \quad \varepsilon_{11}, \quad \varepsilon_{22}, \quad \varepsilon_{11}, \quad \varepsilon_{22}]$  & 14\% & 0.551 & 0.451 & 301 & 19.8 & 100 & 0.390 \\
       & $[\varepsilon_{11},\quad \varepsilon_{22}, \quad \varepsilon_{11}, \quad \varepsilon_{22}, \quad \varepsilon_{11}, \quad \varepsilon_{11}, \quad \varepsilon_{22}]$  & 2\% & 0.558 & 0.459 & 302 & 20.1 & 102 & 0.340 \\
       \cdashlinelr{2-9}
       & $[\varepsilon_{11},\quad \varepsilon_{11}, \quad \varepsilon_{11}, \quad \varepsilon_{11}, \quad \varepsilon_{11}, \quad \varepsilon_{11}, \quad \varepsilon_{11}]$ & NA & 0.550 & 0.503 & 310 & 20.7 & 103 & $6.68 \times 10^4$ \\ 
       & $[\varepsilon_{22},\quad \varepsilon_{22}, \quad \varepsilon_{22}, \quad \varepsilon_{22}, \quad \varepsilon_{22}, \quad \varepsilon_{22}, \quad \varepsilon_{22}]$ & NA & 0.542 & 0.450 & 297 & 19.9 & 98.7 & 66.1 \\
       \midrule
\bottomrule
\end{tabular}}
\caption{Results of the design selection in Exemplar~2 for parameter Cases 5 and 6 are tabulated as well as the percentage of trials (out of 100) that chose each load path. Marginal posterior expected values and generalized variances are recorded after the final load step for each optimal load path chosen. Reported values are averages over the trials that chose each load path. Below the dashed lines are the results from each of the two static designs ($\varepsilon_{11}$ and $\varepsilon_{22}$) for each case.}\label{Tab: exemplar 2 design choice}
\end{table}

\subsubsection{Posterior Summaries}\label{sec: exemplar 2 posterior summaries}

Box plots showing the distribution of marginal posterior summaries at each step ($t=1,\cdots, 7$) over the 100 repeat trials are plotted in Fig.~\ref{fig: case 5 posterior summary} for parameter Case~5 and are tabulated for both Cases 5 and 6 at the final load step in Table~\ref{Tab: exemplar 2 posterior summaries}. The plots reveal that the static $\varepsilon_{22}$ load path was suboptimal for inference on parameter $F$, as seen by an expected value that is demonstrably further from the true value than either of the other two design settings (with much greater variability over the 100 trials) as well as a higher uncertainty (Figs.~\ref{fig: case 5 expected value F} \& \ref{fig: case 5 stdev F}). Likewise, the static $\varepsilon_{11}$ design was sub-optimal for inference on parameter $G$ compared to the other two settings (Figs.~\ref{fig: case 5 expected value G} \& \ref{fig: case 5 stdev G}). The adaptive load path yielded inference on parameter $\sigma_{y}$ with an expected value consistently closer to the true value at the final load step than either of the static designs as well as a significantly lower amount of uncertainty (Figs.~\ref{fig: case 5 expected value sigy} \& \ref{fig: case 5 stdev sigy}). Expected values among all three designs were similar for parameters $A$ and $n$, with $A$ having lower uncertainty in the adaptive setting (Figs.~\ref{fig: case 5 expected value A}-\ref{fig: case 5 stdev n}).  

\begin{figure}%
    \centering
    \sidesubfloat[\centering]{\includegraphics[width=0.4\textwidth]{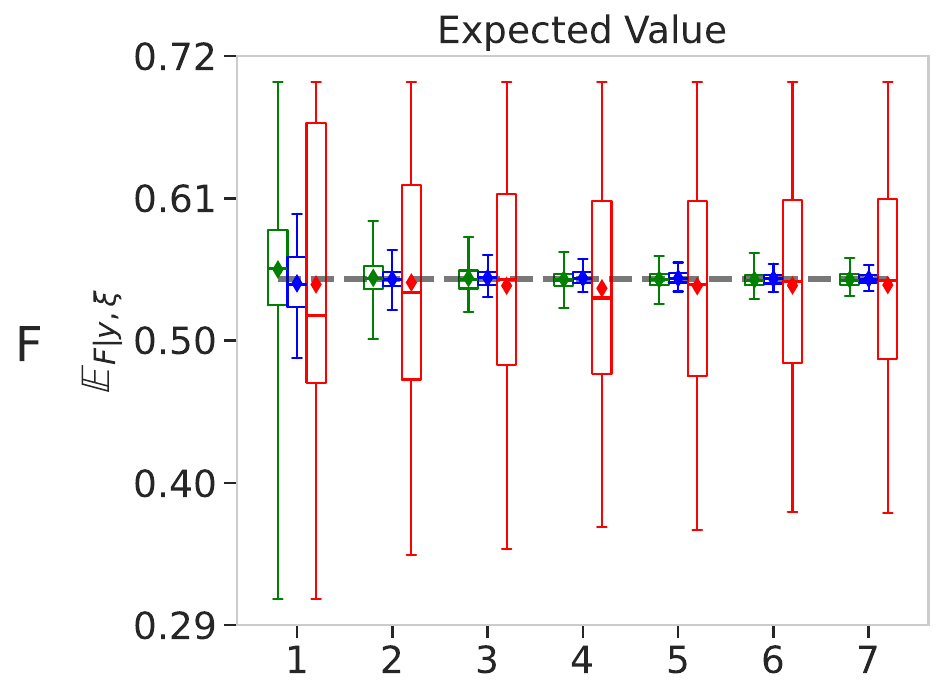}\label{fig: case 5 expected value F} }%
    \sidesubfloat[\centering]{\includegraphics[width=0.4\textwidth]{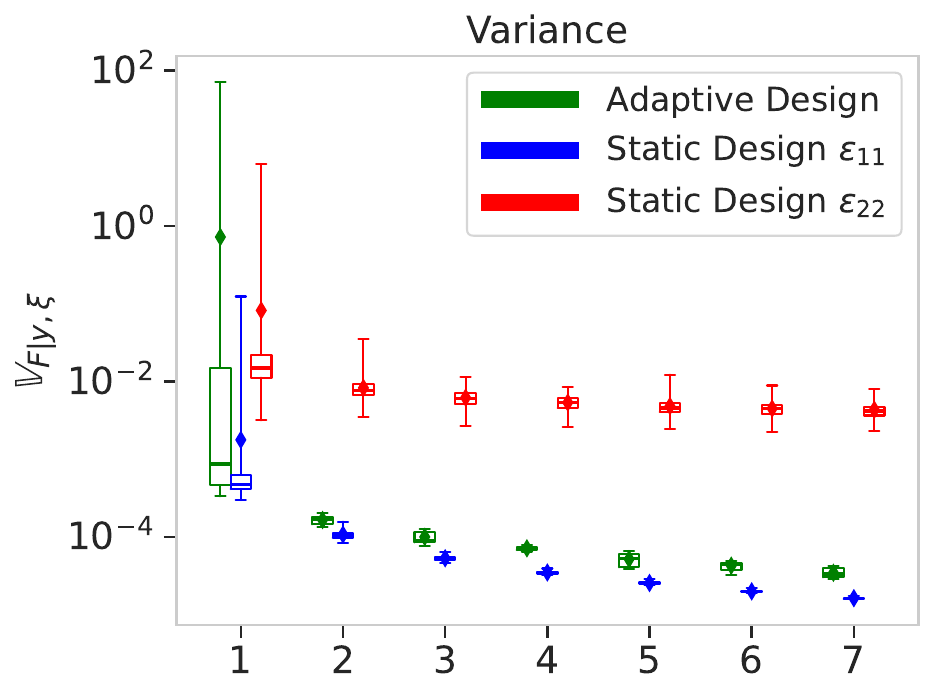}\label{fig: case 5 stdev F} }\\%
    \sidesubfloat[\centering]{\includegraphics[width=0.4\textwidth]{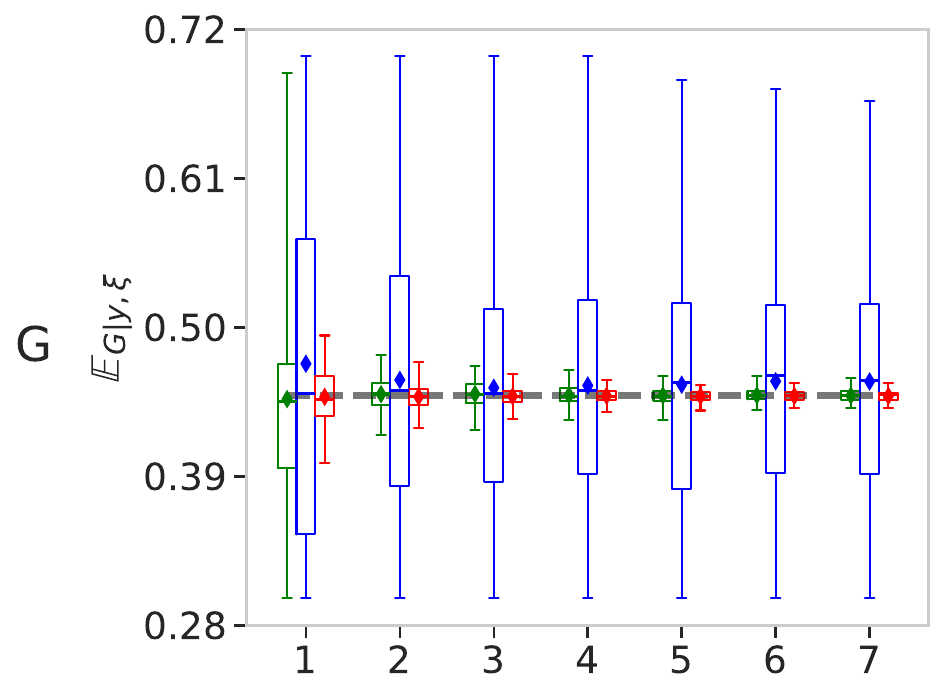}\label{fig: case 5 expected value G} }%
    \sidesubfloat[\centering]{\includegraphics[width=0.4\textwidth]{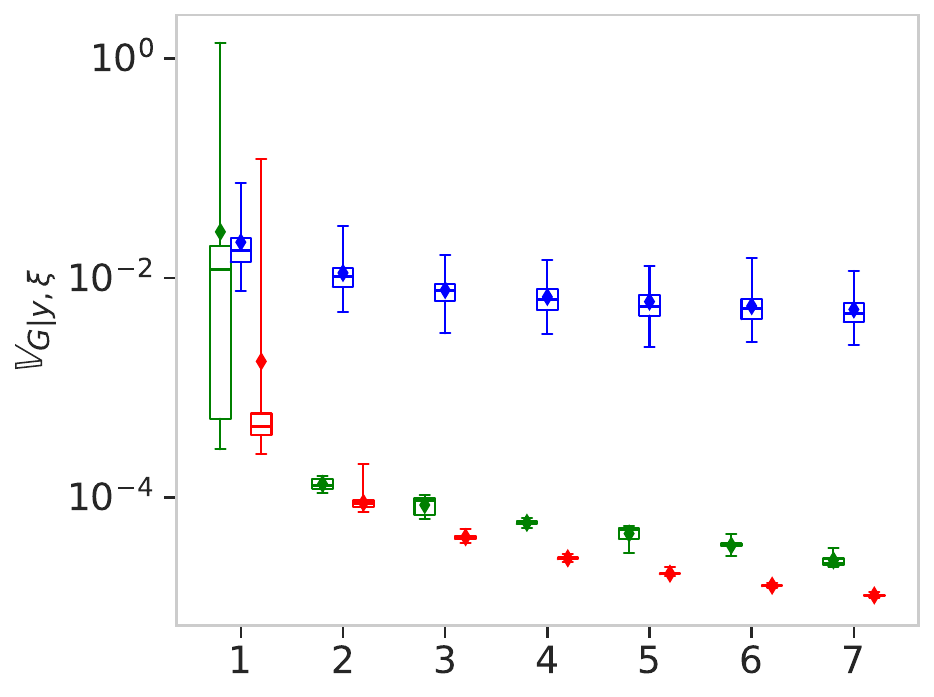}\label{fig: case 5 stdev G} }\\%
    \sidesubfloat[\centering]{\includegraphics[width=0.4\textwidth]{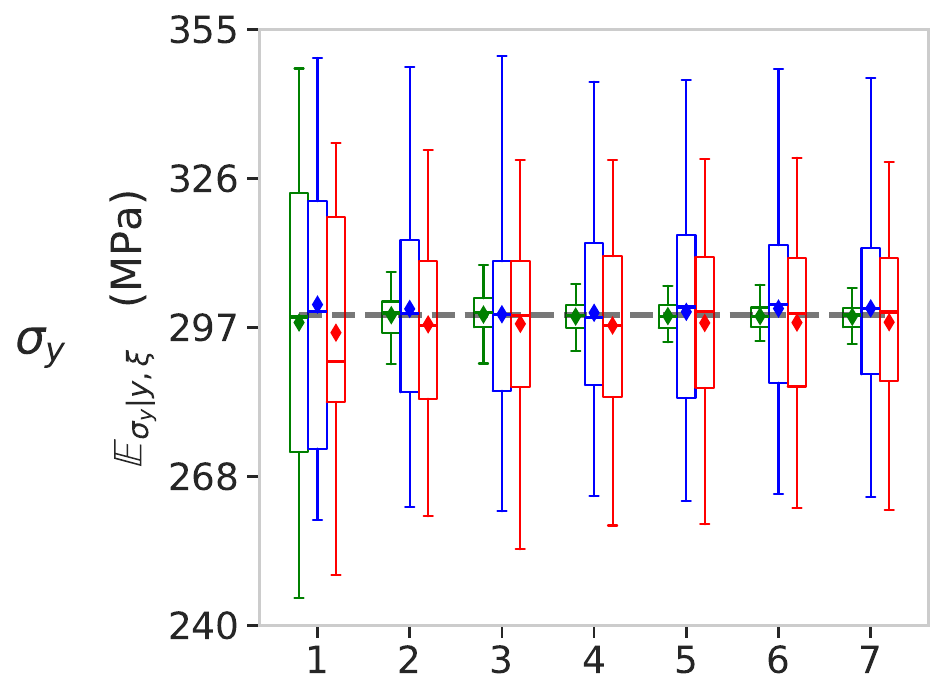}\label{fig: case 5 expected value sigy} }%
    \sidesubfloat[\centering]{\includegraphics[width=0.4\textwidth]{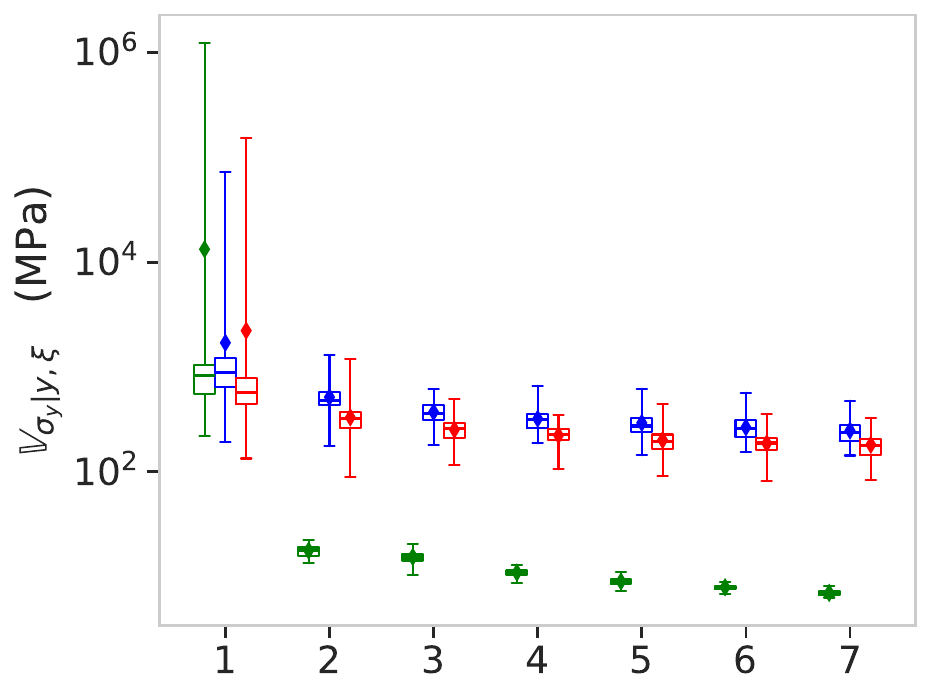}\label{fig: case 5 stdev sigy} }\\%
    \sidesubfloat[\centering]{\includegraphics[width=0.4\textwidth]{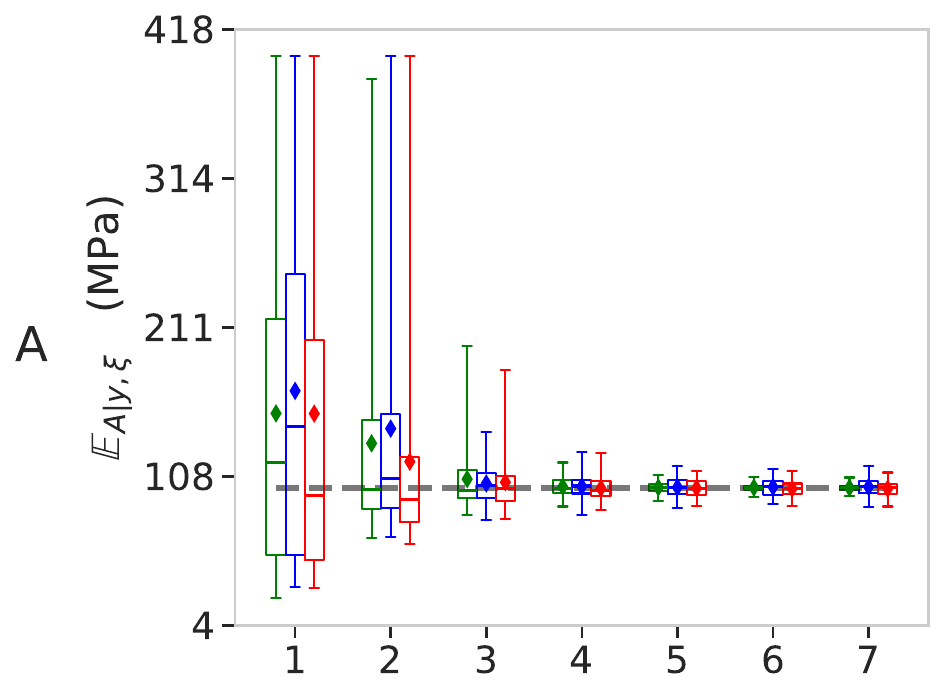}\label{fig: case 5 expected value A} }%
    \sidesubfloat[\centering]{\includegraphics[width=0.4\textwidth]{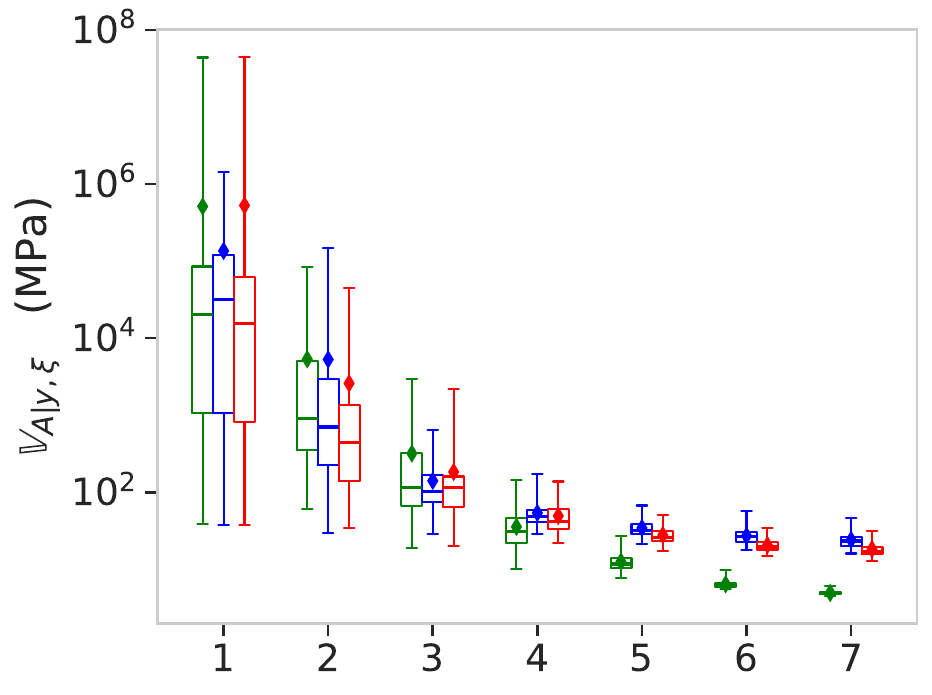}\label{fig: case 5 stdev A} }\\%
    \phantomcaption 
\end{figure}
\clearpage

\begin{figure}%
    \ContinuedFloat
    \centering
    \sidesubfloat[\centering]{\includegraphics[width=0.4\textwidth]{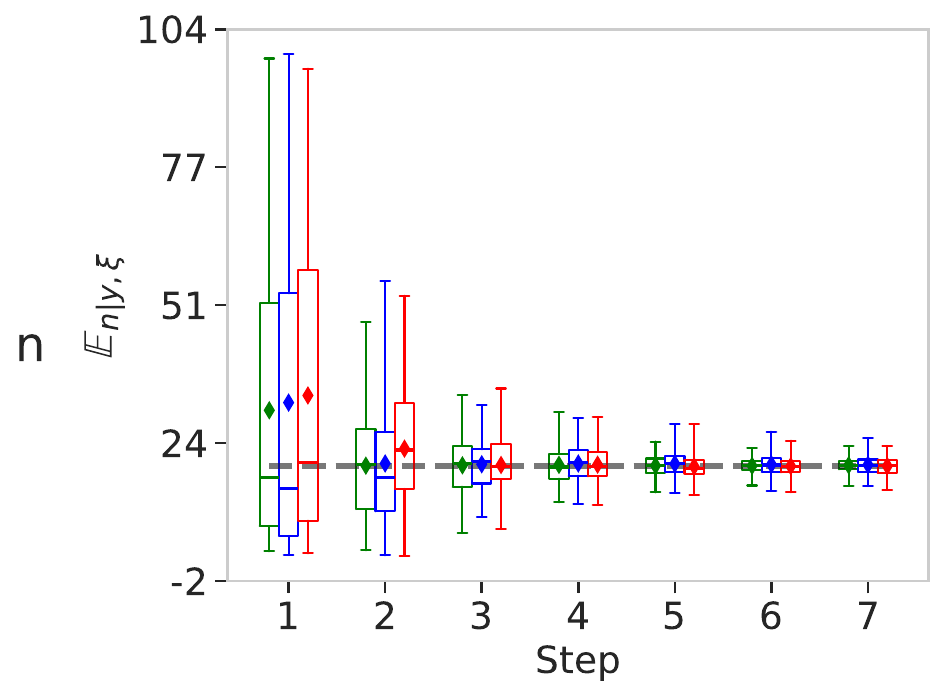}\label{fig: case 5 expected value n} }%
    \sidesubfloat[\centering]{\includegraphics[width=0.4\textwidth]{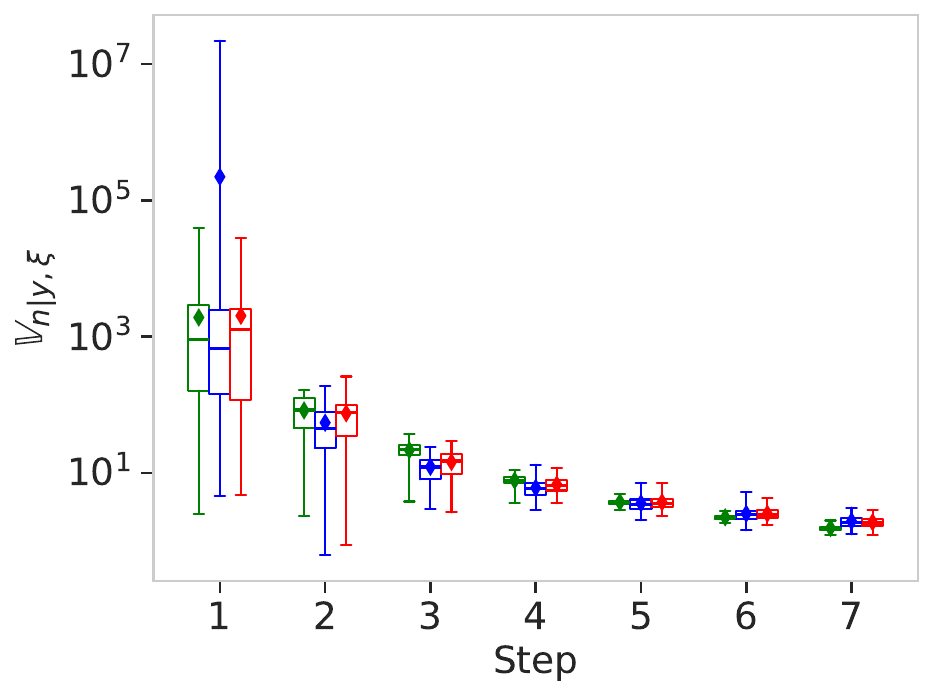}\label{fig: case 5 stdev n} }\\%
    \caption{Box plots are shown which describe the distribution of marginal posterior summaries over the 100 trials for Case~5 at each load step for the adaptive design and two static designs. True parameter values (Table~\ref{Tab: exemplar 2 parameter cases}) are shown with black dashed lines. Box plots for the expected value $\mathbb{E}_{\pars \given \data, \xi}$ (left column) and variance $\mathbb{V}_{\pars \given \data, \xi}$ (right column) are shown for each parameter. The boxes are bounded by Q1 and Q3 and whiskers extend to the minimum and maximum values in the distribution. The median and mean values are shown with a horizontal line and diamond marker, respectively, for each box plot. The adaptive design results are shown in green, the $\varepsilon_{11}$ design results are shown in blue and the $\varepsilon_{22}$ design results in red. Box plots for each design are offset for visual clarity.} %
    \label{fig: case 5 posterior summary}%
\end{figure}

\begin{figure}%
    \centering
    \sidesubfloat[]{\includegraphics[width=0.45\textwidth]{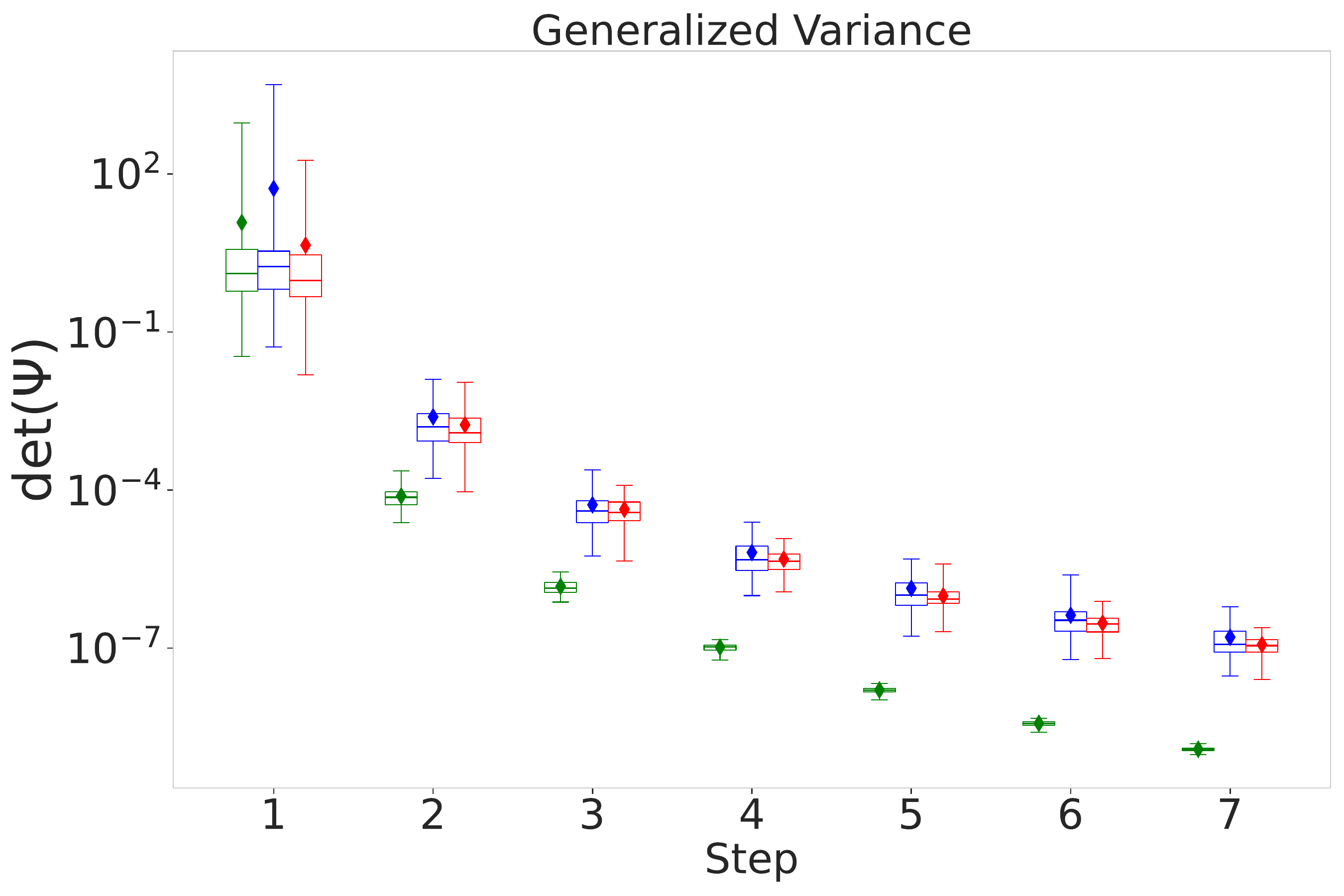}\label{fig: case 5 generalized variance} }%
    \sidesubfloat[]{\includegraphics[width=0.45\textwidth]{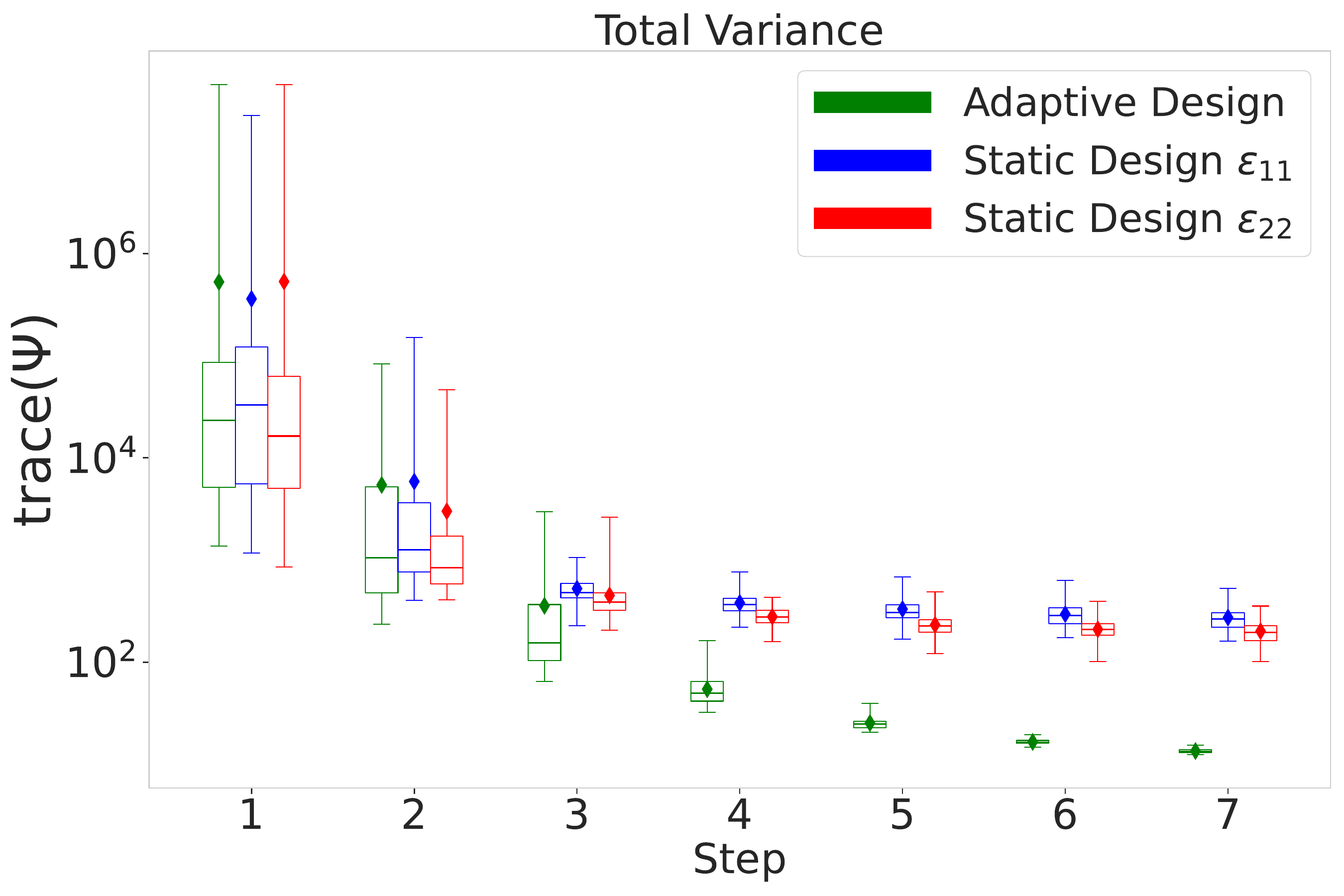}\label{fig: case 5 total variance} }%
    \caption{Box plots of the generalized (a) and total (b) variance for Case~5: $F$ = 0.55, $G$ = 0.45, $A$ = 100, $n$ = 20, $\sigma_y$ = 300. The plots summarize the distributions over the 100 repeat trials. The boxes are bounded by Q1 and Q3 and have whiskers that extend to the minimum and maximum values. The median values are indicated with a horizontal line in the box, and the mean values are shown with diamond markers. Box plots for each design are offset for visual clarity.}%
    \label{fig: case 5 1D uncertainty metrics}%
\end{figure}

Fig.~\ref{fig: case 5 1D uncertainty metrics} shows box plots of the generalized (\ref{fig: case 5 generalized variance}) and total (\ref{fig: case 5 total variance}) variance of the posterior for the adaptive and static designs at each step in the algorithm for Case~5. The plots summarize the distribution of results that were obtained over the 100 repeat trials. The adaptive design had a significantly smaller variance by both metrics than either of the static designs. Similar results were obtained for Case~6 (see Table~\ref{Tab: exemplar 2 1D metrics} in \ref{ex2.app}).

The most significant difference in Case~6 compared to Case~5 in the adaptive setting was seen in the inference for parameters $\sigma_{y}$ and $A$. Since the true values of $A$ and $\sigma_{y}$ changed significantly, the coefficient of variation (Cv) (a standardized and unitless measure of variation) is referred to for a reliable comparison of parameter variability. Recall that Case~5 had a lower value of $A$ and a higher value of $\sigma_{y}$ ($A = 100$ MPa and $\sigma_{y} = 300$ MPa) compared to Case~6 ($A = 300$ MPa and $\sigma_{y} = 100$ MPa).  The coefficients of variation followed the opposite trends, with the Cv for $A$ being higher than the Cv for $\sigma_{y}$ for Case~5 ($\mbox{Cv}_{A} = 0.022$ and $\mbox{Cv}_{\sigma_{y}} = 0.010$) and the Cv for $A$ being lower than the Cv for $\sigma_{y}$ for Case~6 ($\mbox{Cv}_{A} =  0.009$ and $\mbox{Cv}_{\sigma_{y}} = 0.026$).  These results suggest that the dominant phenomenology (e.g., yield in Case~5 and hardening in Case~6) is more readily identified with a lower amount of normalized variability compared to the non-dominant contributor.

A second key difference is that parameter inference was especially difficult in Case~6 for the static design settings, and not all parameters were successfully inferred. This is reflected in the considerable variances reported in Table~\ref{Tab: exemplar 2 design choice} and Tables \ref{Tab: exemplar 2 posterior summaries} \& \ref{Tab: exemplar 2 1D metrics} in \ref{ex2.app}. In particular, some of the credible intervals for $F$ and $G$ go outside the bounds that were set and yield values that are not physically possible.\footnote{A drawback of using the Laplace method for approximating the posterior is that while the parameter bounds can be enforced when finding $\hat{\pars}$ (the posterior mean), the bounds are not considered when calculating of the covariance $\Sigma^{L} = \bs{H}(\hat{\pars})^{-1}$, which may result in parameter uncertainties that go beyond the defined bounds if parameter uncertainties are high.} Even if the trials that were unsuccessful are removed from consideration in the average values (Tables~\ref{Tab: exemplar 2 posterior summaries 4 outliers removed} \& \ref{Tab: exemplar 2 1D metrics 4 outliers removed} in \ref{ex2.app}), the 95\% CIs of $G$ still fall outside the given bounds and variances are still large. Overall, Case~6 proved to be a much more difficult setting to perform inference with a static design. In contrast, the adaptive design reliably provided parameter estimates in close agreement with the true values with low uncertainty (Table~\ref{Tab: exemplar 2 posterior summaries} in \ref{ex2.app}).

The average MD of the true parameter values from the posterior distribution at the final step is also shown in Table~\ref{Tab: exemplar 2 1D metrics}. The adaptive design yielded a smaller MD than both static designs in both cases. Not only did the adaptive design provide parameter inference with lower uncertainty in Exemplar~2, it also yielded a posterior distribution that represented $\pars^{true}$ better than either of the static designs.  

Taken together, these results reinforce the advantages of the \textit{in situ} feedback loop and BOED algorithm over traditional static designs chosen based on human intuition \textit{a priori}. By ensuring that the necessary data is collected, the adaptive designs are able to calibrate material models in situations when static designs fail completely, and they provide greater accuracy and precision in the parameter values overall.

\subsubsection{Propagation of Uncertainty to Stress-Strain Space}\label{sec: exemplar 2 uncertainty propagation}

\begin{figure}[!htb]%
    \centering
    \sidesubfloat[]{{\includegraphics[width=0.45\textwidth]{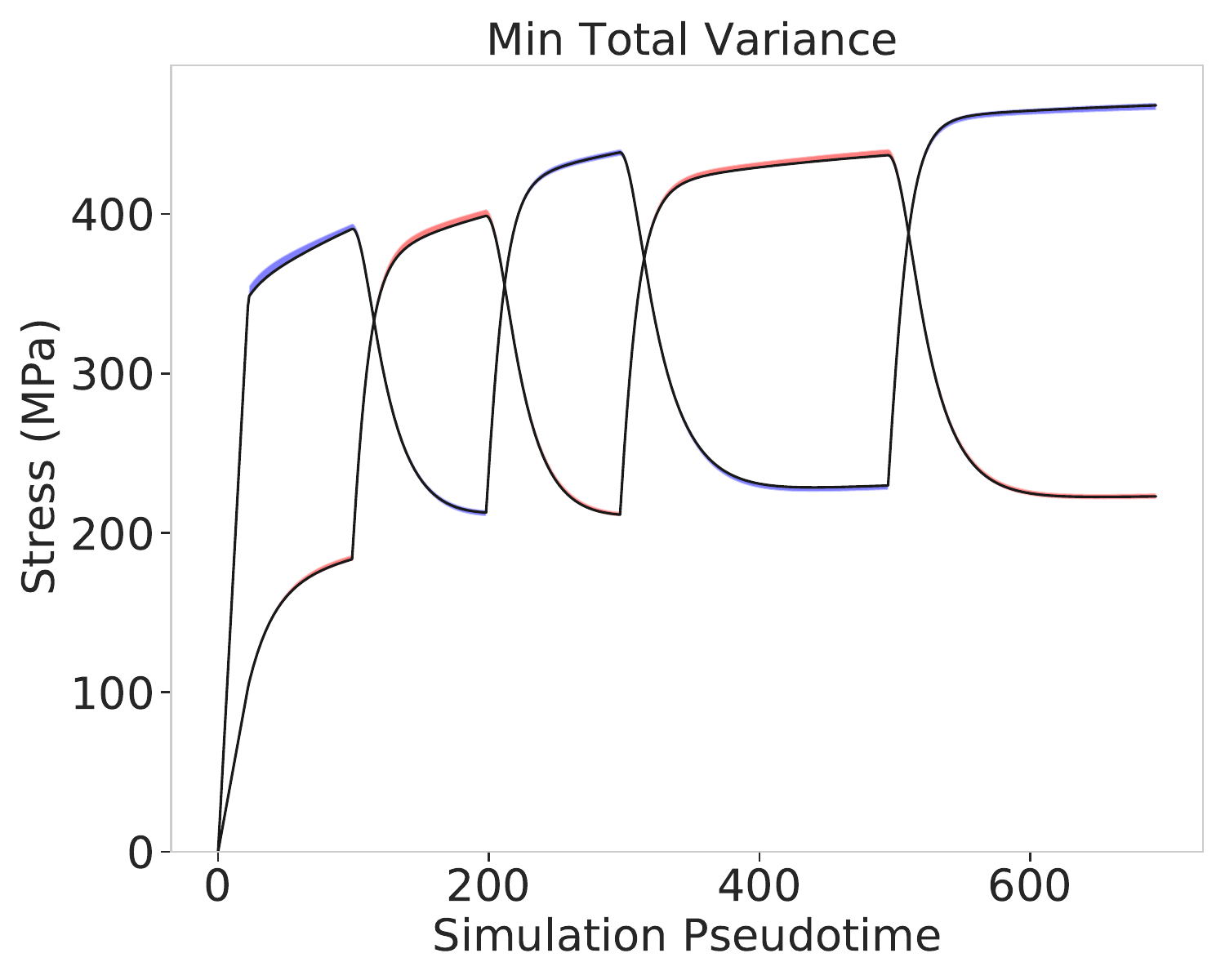} }}%
    \sidesubfloat[]{{\includegraphics[width=0.45\textwidth]{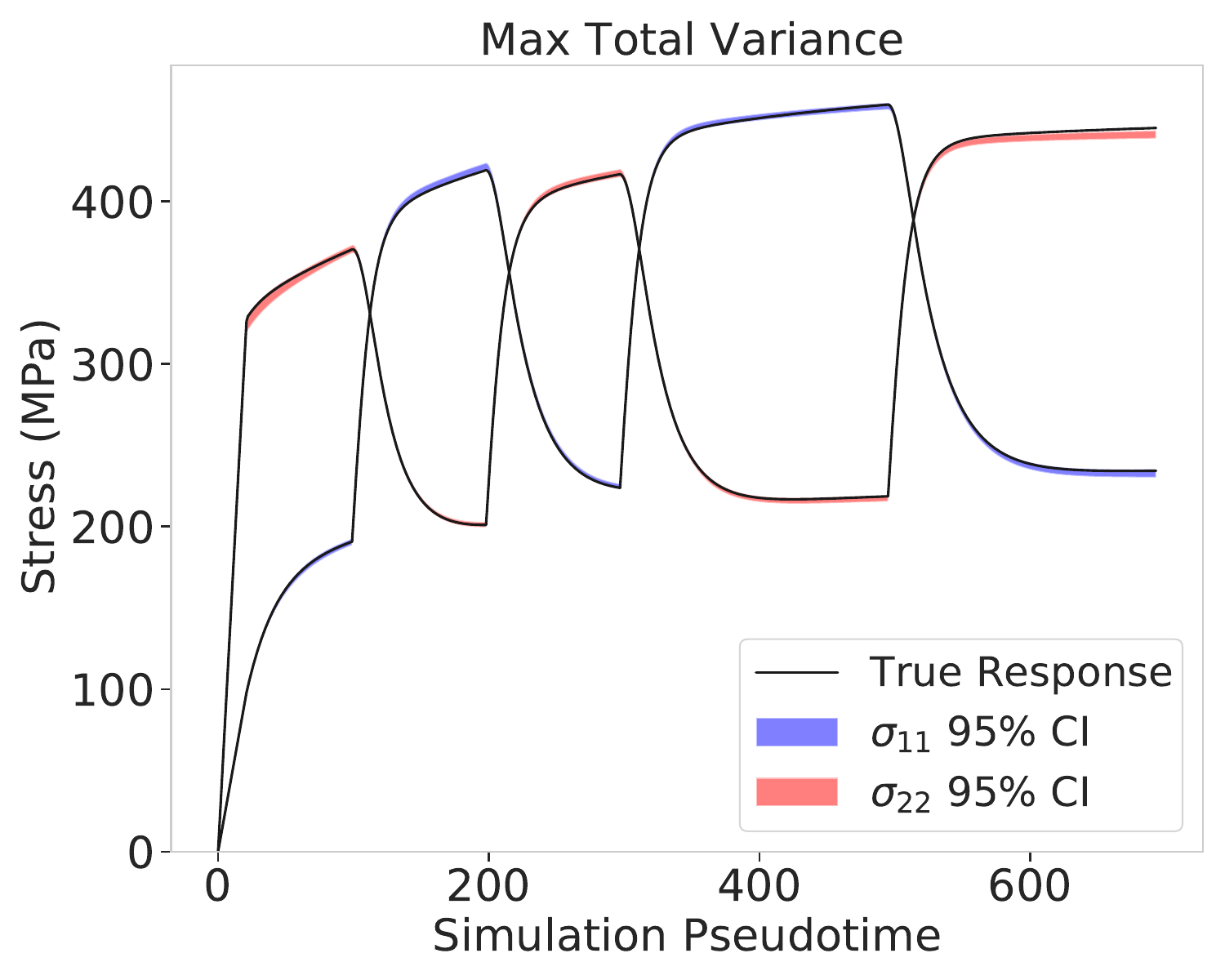} }}%
    \\
    \sidesubfloat[]{{\includegraphics[width=0.45\textwidth]{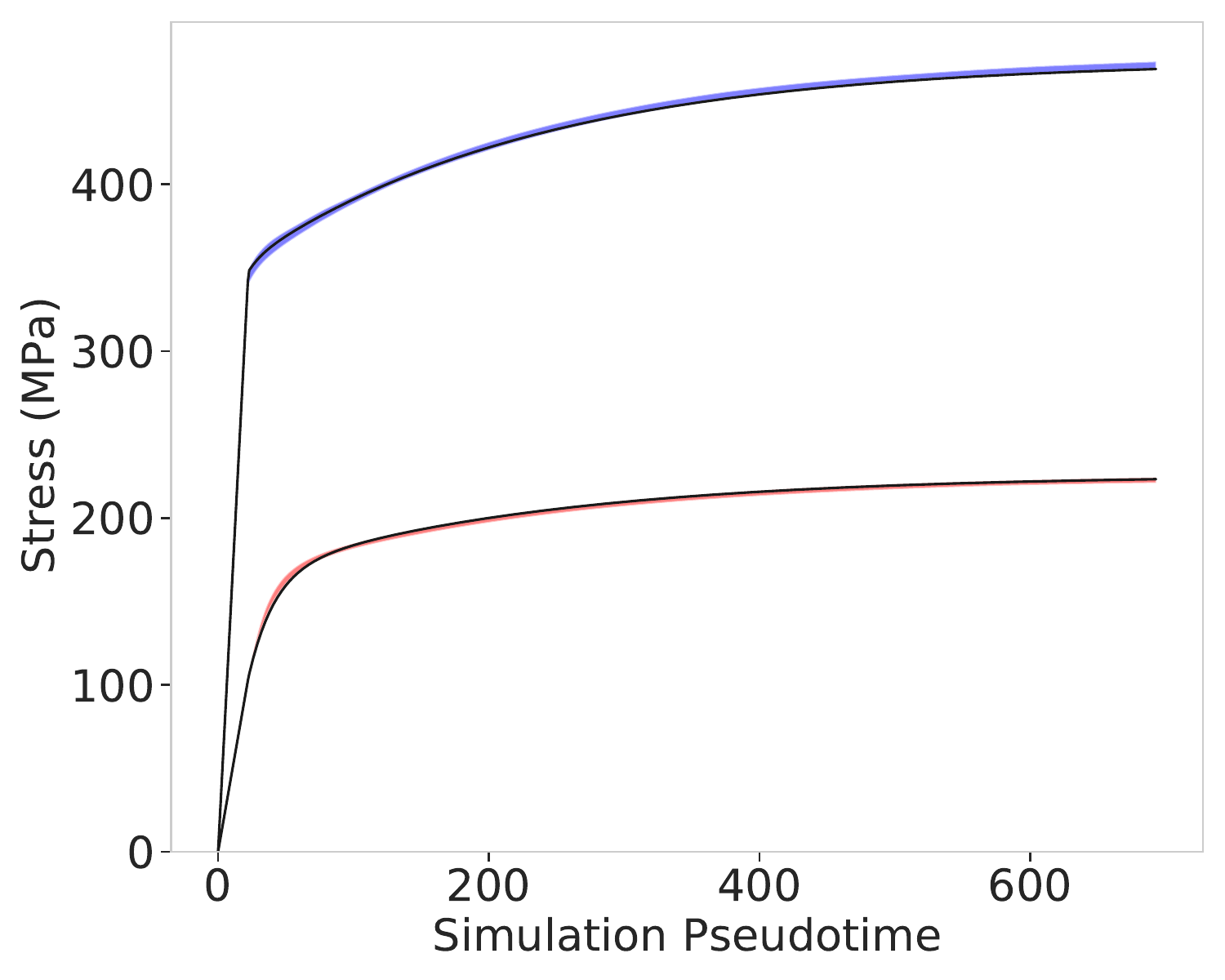} }}%
    \sidesubfloat[]{{\includegraphics[width=0.45\textwidth]{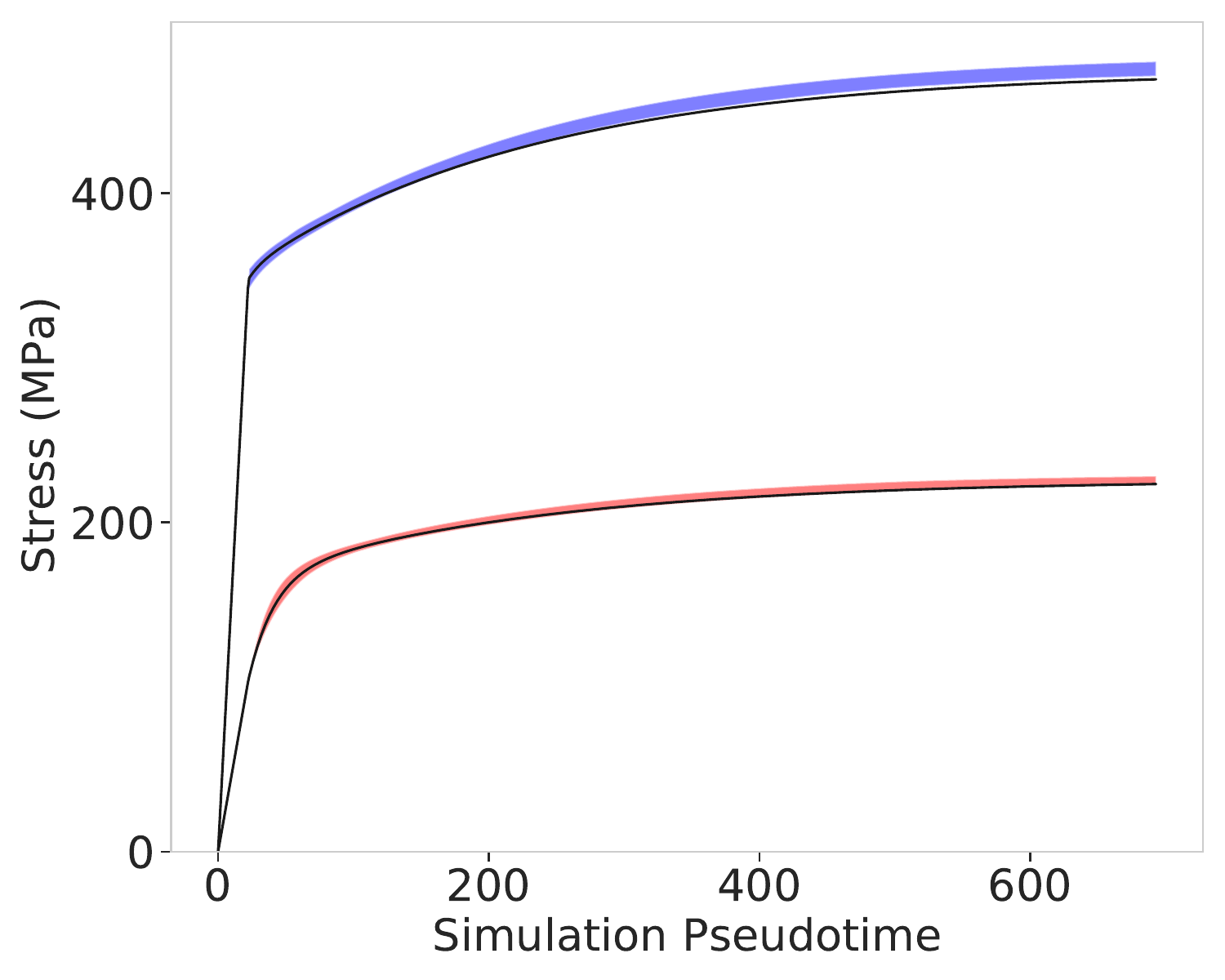} }}%
    \\
    \sidesubfloat[]{{\includegraphics[width=0.45\textwidth]{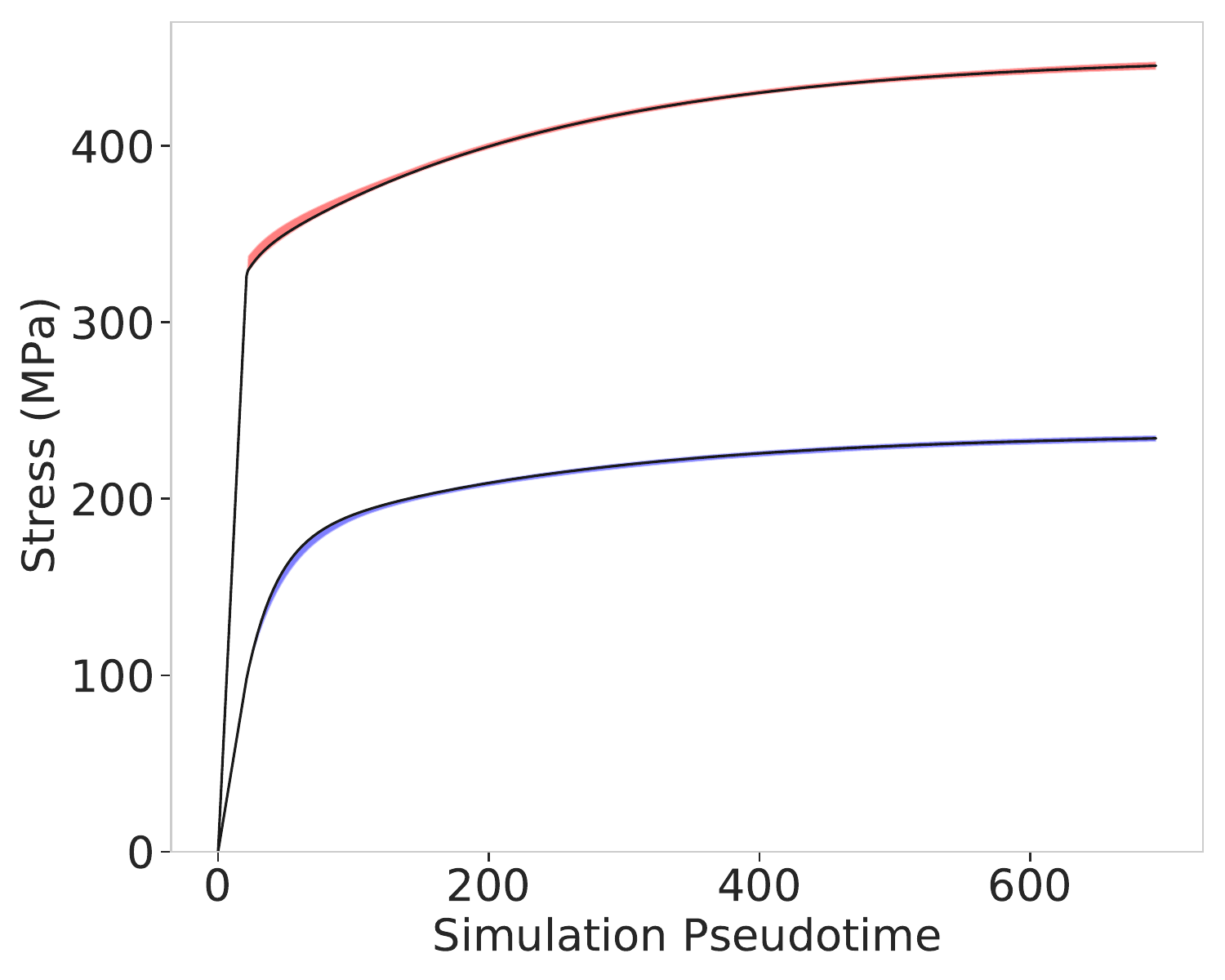} }}%
    \sidesubfloat[]{{\includegraphics[width=0.45\textwidth]{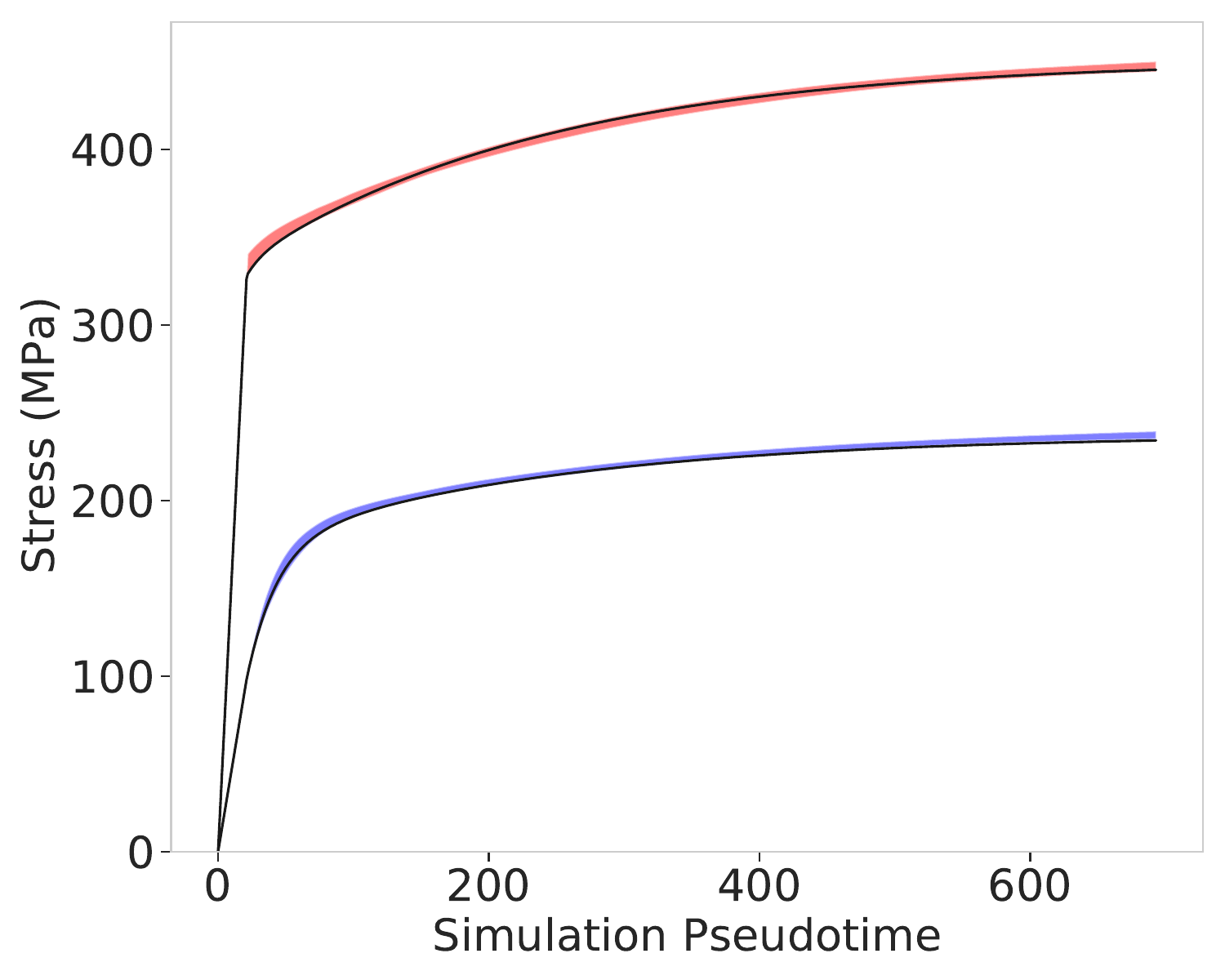} }}%
    \caption{95\% credible intervals for the posterior draws $\tilde{\pars} \sim \pi(\pars \given \data, \des)$ were calculated point-wise for Exemplar~2, Case~5 and are indicated the the blue ($\sigma_{11}$) and red ($\sigma_{22}$) dashed lines. The model output from the true parameter values is plotted with a black line. The left column of plots show the 95\% CI for the trial that had a posterior distribution with the least total variance among all trials for each design setting. The right column of plots show 95\% CI for the trial that had the greatest total variance. Plots (a) and (b) show results for the adaptive design, plots (c) and (d) show results for the static $\varepsilon_{11}$ design and plots (e) and (f) for the static $\varepsilon_{22}$ design. Box plots for each design are offset for visual clarity.}%
    \label{fig: exemplar 2 posterior draw CI}%
\end{figure}

One hundred samples from the posterior distribution were drawn and propagated to the model output space as was done in Exemplar~1 (Sec.~\ref{sec: case 1 propagated uncertainty}). In Fig.~\ref{fig: exemplar 2 posterior draw CI}, 95\% credible intervals for computed in-plane stress values from the 100 draws are shown. The left column of plots shows the 95\% CI for the trial in each design setting that had the least posterior total variance, and the right column of plots shows the 95\% CI for the trial that had the greatest posterior total variance. For the adaptive design, the minimum total variance came from a trial with an optimal design of $\des^{*} = [\varepsilon_{11}, \varepsilon_{22}, \varepsilon_{11}, \varepsilon_{22}, \varepsilon_{22}, \varepsilon_{11}, \varepsilon_{11}]$, and the maximum total variance came from a trial with an optimal design of $\des^{*} = [\varepsilon_{22}, \varepsilon_{11}, \varepsilon_{22}, \varepsilon_{11}, \varepsilon_{11}, \varepsilon_{22}, \varepsilon_{22}]$. At a high level, all three designs provided credible intervals that generally encompassed the true response.

The 95\% CIs for Exemplar~2 Case~6 are shown in Fig.~\ref{fig: exemplar 2 case 6 posterior draw CI} in \ref{ex2.app}. The difficulty of the static designs to perform reliable inference for this case is reflected in the wide CIs for the trials that had the maximum total variance. In summary, for Case~5, the reduced uncertainty in the material model parameters obtained with the adaptive design did not translate to a noticeable reduction in the uncertainty of the final QOI; however, in Case~6, the reduced parameter uncertainty did so translate. Thus, the adaptive design still has a great potential to reduce the uncertainty of the final QOI for more realistic applications that move beyond the material point simulator and utilize real experimental data. Such investigations are the subject for future work.

%% file: Conclusions.tex
\section{Conclusions}\label{sec: conclusion}

This paper presented a first demonstration of the ICC framework for a constitutive model calibration problem in which the load path of a material point---representing the center of a cruciform specimen---was optimized with respect to parameter uncertainty. The adaptive design of the load path was segmented into the selection of individual load steps of a specified increment of strain along a specified axis of the material point, and the optimal path was determined by calculating the EIG of each candidate load step. The benefits of choosing the load path in an adaptive manner within the framework as opposed to choosing a static load path were demonstrated with two exemplar problems of varying complexity. 

In both exemplars, the adaptive load path designs yielded posterior distributions with lower uncertainty than in static settings on average when the uncertainty was considered holistically over all parameters. This result was supported by the total and generalized variances reported for the designs at each step in the algorithm. There was also evidence that the adaptive designs on average lead to posterior distributions that were more reliable in capturing the true parameter values. This greater reliability was supported by a lower $MD(\pars^{true})$ on average for the adaptive designs, which indicates that the true parameter values resided in regions of higher probability for the adaptive designs compared to the static designs. 

In both exemplar problems, an alternating load path was clearly preferred in the adaptive designs. This alternating tendency was likely due to the yield anisotropy, which benefits from probing different directions in order to obtain estimates of the parameters. However, an in depth analysis of the contributing factors for the predominantly chosen load paths was not performed. The optimal load path is specific to the load path tree structure, which in this work was a binary tree where each parent node had two children nodes corresponding to applying a positive increment of strain along one of two directions ($\varepsilon_{11}$ or $\varepsilon_{22}$). Any alternative tree structure would likely result in optimal load paths which deviate from what was observed in this work. Although the binary tree structure provided a simple starting point, the ICC framework is not restricted to this tree structure and is easily adaptable to alternatives. Additional children could be introduced at each node in the tree, which could include compression or simultaneous strain increments along the two axes, for example. In such a case, the ICC framework would operate in the same manner, with the exception that the EIG would need to be calculated for any additional design options.

This initial effort was the significant first step of a greater objective to demonstrate the ICC framework on a cruciform specimen being actively controlled in a bi-axial load frame in real-time with the calibration of FEA models using full-field DIC data. With a run-time of less than 3 minutes per step in Exemplar~2 (the more complicated problem), which includes the time required for the EIG calculation plus inference, the efficiency of the developed algorithm makes it feasible to be used in a quasi real time scenario. The importance of time efficiency is emphasized for future live demonstrations of the ICC framework which introduce lab time considerations as well as anticipated challenges with holding a sample completely still in the load frame while the algorithm is running. Unwanted sample movement during the experiment may result in boundary conditions and/or material responses (e.g. creep) that are unaccounted for in the material model, thus increasing any model form error that may be present and negatively impacting the results of the parameter inference. The 100 repeat trials that were performed in this work were utilized for demonstrative purposes only and will not be a part of future applications. With live experiments, the ICC algorithm will be run once, and a single specimen will be tested to collect data for calibration.
 
 \newpage

%% file: Acknowledgements.tex
\section{Acknowledgements}\label{sec: acknowledgements}
This work was supported by the Laboratory Directed Research and Development program at Sandia
National Laboratories, a multimission laboratory managed and operated by National Technology
\& Engineering Solutions of Sandia, LLC, a wholly owned subsidiary of Honeywell International
Inc., for the U.S. Department of Energy’s National Nuclear Security Administration under
contract DE-NA0003525.

This paper describes objective technical results and analysis. Any subjective views or opinions
that might be expressed in the paper do not necessarily represent the views of the U.S. Department
of Energy or the United States Government.

%% file: app_A.tex
\section{Algorithmic Supplementary Material}\label{alg.app}
\setcounter{figure}{0}  
\setcounter{table}{0}  

\subsection{Posterior Approximation with MCMC}\label{sec: MCMC_posterior_approx.app}
Markov Chain Monte Carlo (MCMC) is a commonly used method for approximating the posterior distribution and obtaining posterior summaries such as parameter expected values, variances and credible intervals. In this approach, a Markov chain is constructed through sampling such that its stationary distribution matches the posterior distribution of interest. The trade-off for the high-accuracy of this method is computational time, as it can require millions of forward model evaluations. In the setting of material model calibration, MCMC simulations may be applied for computationally cheap models, but they become prohibitively expensive for high-fidelity constitutive models and simulations (such as FEA). There exist a number of ways to improve simulation efficiency, which include the use of conjugate priors \cite{robert2007bayesian, bernardo2009bayesian}, advanced sampling techniques \cite{altekar2004parallel, andrieu2008tutorial, tierney1999some, green2001delayed}, or using a surrogate model $\surr$ in place of the expensive forward model $\pred$. However, even with these adjustments, an MCMC simulation would be too expensive for the quasi real-time characterization and calibration setting in which the ICC framework will eventually operate.

\subsection{Approximation of the Expected Information Gain (EIG)}\label{sec: EIG_approx.app}

A nested Monte Carlo estimator is used in this work to approximate the EIG, 

\begin{equation}\label{eq: double_nested MC estimator II}
    \begin{aligned}
    \widehat{EIG}_{MC}(\des) = \frac{1}{N} \sum_{n=1}^{N} \log \frac{f\left(\data_n \given \pars_{n,0},\des\right)}{\frac{1}{M} \sum_{m=1}^{M} f\left(\data_n \given \pars_{n,m},\des\right)}, \quad \pars_{n,m} \sim \pi(\pars), \quad \data_{n} \sim f(\data \given \pars_{n,0}, \des),
    \end{aligned}
\end{equation}

where samples from the data distribution and the prior are used to evaluate the inner and outer sum of the Monte Carlo estimator. $N$ and $M$ determine the number of samples that are used to evaluate the outer and inner sums, and the estimator variance and bias scale inversely with the magnitude of $N$ and $M$, respectively \cite{ryan2003estimating}. The estimator requires a total of $N(1 + M)$ model evaluations to calculate the EIG, so $N$ and $M$ must be chosen judiciously to balance the estimator quality and computational cost. Using the same $\pars_{n,:}$ parameter samples for every iteration of the outer loop reduces the number of model evaluations down to $(N + M)$. While this simplification may introduce some additional bias, some studies have shown that minimal levels of error are incurred \cite{huan2013simulation, huan2010accelerated}; thus, using the same inner samples for each iteration of the outer loop is the approach taken in this work. Some alternative approaches include calculating the evidence with Laplace's approximation \cite{long2013fast, ryan2003estimating} or using a linear approximation of the forward model to improve the efficiency of EIG estimation \cite{eckels2023optimal}.

In some cases, a numerical issue known as arithmetic underflow may be encountered when computing the inner loop in \eqref{eq: double_nested MC estimator}. In particular, arithmetic underflow may occur if the distribution of the evidence is narrow (i.e., low variability), and the prior distribution is not sufficiently centered on the posterior distribution. This situation may lead to a scenario where the calculated likelihood of all data samples $\data_n$ are below machine precision, resulting in an approximation for the evidence term equal to 0. Arithmetic underflow is especially likely for small values of $M$, so a simple way to avoid it is to increase the value of $M$ or to use the same parameter samples for the inner and outer loops \cite{huan2010accelerated}, possibly at the cost of a greater bias in the estimate. Underflow can also be handled by using importance sampling with a biasing distribution other than the prior. One approach is to define the biasing distribution as the Laplace approximation to the posterior for each outer sample  \cite{beck2018fast, goda2020multilevel}. Similarly, in \cite{feng2015optimal} the biasing distribution was calculated with a self-normalized approximation to the posterior. In this work, underflow was avoided by choosing $M$ to be sufficiently large.

\subsection{Adaptive BOED}\label{sec: adaptive_boed.app}

In cases, such as in this work, in which multiple experiments are to be performed $\des = [\xi_{1}, \dots, \xi_{T}]$, where $T$ is the number of experiments, an adaptive BOED determines the optimal experimental design by incorporating previously collected data into the design selection process.

Following closely the theory laid out in \cite{foster2019variational}, assume the $T$ experiments yield data $\data_1, \cdots, \data_T$. Under the assumption of conditional independence of the data given the unknown parameters and the designs, 
the following relationship for the joint distribution over the data and parameters holds,

\begin{equation}\label{eq: joint marginal multiple experiments}
    \pi(\data_{1:T}, \pars \given \xi_{1:T}) = \pi(\pars) \prod_{t=1}^T f(\data_t \given \pars, \xi_t).
\end{equation}

This relationship is leveraged when calculating the EIG for the $t^{\text{th}}$ experiment, $\xi_t$, which takes into account data obtained from experiments $1:t-1$.  For all $t > 1$, $\pi(\pars)$ in \eqref{eq: double_nested MC estimator} is replaced by the posterior conditional on all previous experiments and collected data, $\pi(\pars \given \data_{1:t-1}, \des_{1:t-1})$, and samples from the data distribution are drawn for the $t^{th}$ experiment, $\data_{n} \sim f(\data_t \given \pars, \xi_t)$. Thus, the joint distribution of the data and parameters given the designs at the $t^{\text{th}}$ step is written as

\begin{equation}\label{eq: adaptive boed}
    \begin{aligned}
    \pi(\data_{1:t}, \pars \given \des_{1:t}) &= \pi(\pars) f(\data_1 \given \pars, \xi_1)f(\data_2 \given \pars, \xi_2)\cdots  f(\data_{t-1} \given \pars, \xi_{t-1}) f(\data_t \given \pars, \xi_t) \\
    &= \pi(\pars \given \data_{1:t-1}, \des_{1:t-1})f(\data_{t} \given \pars, \xi_{t}).
    \end{aligned}
\end{equation}

\subsection{ICC Implementation}\label{sec: ICC_algorithm.app}

The pseudocode for the proposed algorithm is presented in this section to outline the steps of the ICC framework. To initiate the procedure, surrogates are pre-built offline ahead of entering the feedback loop (Fig.~\ref{fig: framework} and Sec.~\ref{sec: surrogate model construction}). The characterization and calibration cycle is then initiated by determining the initial experimental design (load step). Algorithm~\ref{alg: ICC} assumes the initial load step is determined by calculating the EIG, as in Examplar~2 (Sec.~\ref{sec: exemplar 2}), although the first load step may alternatively be pre-determined, as was done in Exemplar~1 (Sec.~\ref{sec: exemplar 1}). At each decision point $t=1,\dots,T$, the next load step in the load path is determined by calculating the EIG {\eqref{eq: double_nested MC estimator}} for each proposed load step $\xi_{t} \in [\varepsilon_{11}, \varepsilon_{22}]$. To do this, parameter samples $\pars_{n,0}$ are drawn from the prior for the outer loop of the EIG approximation and data $\data_{n}$ is simulated from the data distribution. In the ICC implementation, the same outer parameter samples were used for each proposed load step ($\varepsilon_{11}$ or $\varepsilon_{22}$). Parameter samples $\pars_{n,m}$ are then drawn for the inner loop of the EIG approximation. The same inner loop samples were used for each iteration of the outer loop $n=1,\dots,N$ (Sec.~\ref{sec: EIG_approx.app}) and for each proposed load step. The load step that has the maximum EIG is chosen, and data $y_{t}$ is generated. The posterior is approximated and becomes the prior for the EIG approximation in the next step {\eqref{eq: joint marginal multiple experiments}} and {\eqref{eq: adaptive boed}}. After a pre-determined number of load steps, inference on the parameters is complete and summaries of the resulting posterior distribution are used to describe parameter uncertainty.

\begin{algorithm}
\caption{ICC Framework.}\label{alg: ICC}
\begin{algorithmic}[1]
\State Input: pre-build surrogates offline, $\mathbf{\tilde{u}}(\cdot)$.
\State Set $q(\pars) = \pi(\pars)$. 
\For{t=1:T}
\For{$i = 1:\mbox{length}(\Xi)$}
\For{$n = 1:N$}
\If{i == 1}
\State Draw parameter samples $\pars_{n,0} \sim q(\pars)$ for the outer loop of the EIG approximation.
\If{n == 1}
\For{$m = 1:M$}
\State Draw parameter samples $\pars_{n,m} \sim q(\pars)$ for the inner loop of the EIG approximation.
\EndFor
\Else 
\State Set $\pars_{n,:} = \pars_{1, :}$
\EndIf
\EndIf
\State Simulate data sample $y_{n}$ from the data distribution $y_{n} \sim f(y \given \pars_{n,0}, \xi_{i})$.
\EndFor
\State Estimate $\widehat{EIG}_{MC}(\xi_{i})$ via \eqref{eq: double_nested MC estimator}.
\EndFor
\State $\xi_{t} = \argmax_{\xi \in \Xi} \widehat{EIG}_{MC}(\xi)$.
\State Produce data $y_{t}$.
\State Calculate $\pi(\pars \given \data_{1:t}, \des_{1:t}) \propto f(\data_{1:t}\given \pars, \des_{1:t})\pi(\pars)$ via \eqref{eq: bayes_rule} and \eqref{eq: laplace approx}.
\State Set $q(\pars) = \pi(\pars \given \data_{1:t}, \des_{1:t})$.
\EndFor
\State Output: Approximation of the posterior $\pi(\pars \given \data_{1:T}, \des_{1:T})$.
\end{algorithmic}
\end{algorithm}

%% file: app_B.tex
\section{Exemplar~1 Supplementary Material}\label{ex1.app}
\setcounter{figure}{0}  
\setcounter{table}{0}  

\noindent The complete results for all 12 parameter and design settings after the final load step are recorded in Table~\ref{Tab: exemplar 1 posterior summaries}. Along with the true parameter values, the posterior summaries averaged over all 100 repeat trials are presented. Expected values $\mathbb{E}_{\pars \given \data, \xi}$, posterior variances $\mathbb{V}_{\pars \given \data, \xi}$, equal-tailed 95\% credible intervals (CI)---calculated from quantiles of the posterior---the mean absolute percentage error of the MAP estimate  $\mbox{MAPE}(\hat{\pars})$ and the coefficient of variation of each parameter (Cv) are recorded. The $\mbox{MAPE}(\hat{\pars})$ was obtained by calculating the normalized difference between the true and estimated parameter values, $\mbox{MAPE}(\hat{\pars}) = (|\pars^{true}_{d} - \hat{\pars}_{d}|/\pars^{true}_{d})\times 100, \quad d=1,\dots,D$. The coefficient of variation is a standardized and unitless measure of variation, defined by the ratio of the standard deviation and the mean. It is useful in comparing the degree of variation among different parameters, especially when they have magnitudes on different scales as in Exemplar~2. In all 4 parameter cases as well as for the adaptive and both static designs, the $\mbox{MAPE}(\hat{\pars})$ was less than a quarter of a percent, variances were on the order of $1 \times 10^{-4}$ to $1 \times 10^{-5}$, and the credible intervals contained the true values. The coefficient of variation was nearly equal for $F$ and $G$ in all cases in the adaptive setting, notably greater for $F$ than $G$ in the static $\varepsilon_{11}$ setting and vice versa in the $\varepsilon_{22}$ setting. There was no notable difference in the marginal summaries among the different parameter cases.

\begin{table}[!htb]
\centering
\resizebox{\linewidth}{!}{%
\begin{tabular}{lllllllll}
\hline
\textbf{Case No.} & \textbf{Design} &  \textbf{Parameter} & $\boldsymbol{\theta^{true}}$ & $\boldsymbol{\mathbb{E}_{\theta \given y, \xi}}$ & $\boldsymbol{\mathbb{V}_{\theta \given y, \xi}}$ & \textbf{95\% CI} & \textbf{MAPE($\boldsymbol{\hat{\pars}}$)} & \textbf{Cv}\\ 
\toprule
\multirow{6}{*}{1} & Adaptive & $F$ & 0.55 & 0.5496 & 6.75 $\times 10^{-5}$ & $(0.534, 0.566)$ & 0.064 & 0.015 \\ 
                   & & $G$ & 0.45 & 0.4504 & 4.57 $\times 10^{-5}$ & $(0.437, 0.464)$ & 0.081 & 0.015 \\
                   & Static $\varepsilon_{11}$ & $F$ & 0.55 & 0.5504 & 1.11 $\times 10^{-4}$ & $(0.530, 0.571)$ & 0.076 & 0.019 \\ 
                   & & $G$ & 0.45 & 0.4499 & 3.94 $\times 10^{-5}$ & $(0.438, 0.462)$ & 0.020 & 0.014 \\
                   & Static $\varepsilon_{22}$ & $F$ & 0.55 & 0.5503 & 5.32 $\times 10^{-5}$ & $(0.536, 0.565)$ & 0.046 & 0.013 \\  
                   & & $G$ & 0.45 & 0.4499 & 9.03 $\times 10^{-5}$ & $(0.431, 0.469)$ & 0.026 & 0.021 \\ \midrule
\multirow{6}{*}{2} & Adaptive & $F$ & 0.60 & 0.5996 & 8.38 $\times 10^{-5}$ & $(0.582, 0.618)$ & 0.070 & 0.015 \\ 
                   & & $G$ & 0.50 & 0.5505 & 5.69 $\times 10^{-5}$ &  $(0.486, 0.515)$ & 0.090 & 0.015 \\
                   & Static $\varepsilon_{11}$ & $F$ & 0.60 & 0.6005 & 1.44 $\times 10^{-4}$ & $(0.577, 0.624)$ & 0.081 & 0.020 \\ 
                   & & $G$ & 0.50 & 0.4999 & 4.82 $\times 10^{-5}$ & $(0.486, 0.514)$ & 0.020 & 0.014 \\
                   & Static $\varepsilon_{22}$ & $F$ & 0.60 & 0.6003 & 6.43 $\times 10^{-5}$ & $(0.585, 0.616)$ & 0.047 & 0.013 \\  
                   & & $G$ & 0.50 & 0.4999 & 1.18 $\times 10^{-4}$ & $(0.479, 0.521)$ & 0.022 & 0.022 \\ \midrule
\multirow{6}{*}{3} & Adaptive & $F$ & 0.60 & 0.5996 & 9.04 $\times 10^{-5}$ & $(0.581, 0.618)$  & 0.071 & 0.016 \\ 
                   & & $G$ & 0.60 & 0.6005 & 8.01 $\times 10^{-5}$ & $(0.583, 0.618)$  & 0.088 & 0.015 \\
                   & Static $\varepsilon_{11}$ & $F$ & 0.60 & 0.6005 & 1.68 $\times 10^{-4}$ & $(0.575, 0.626)$ & 0.090 & 0.022 \\ 
                   & & $G$ & 0.60 & 0.5999 & 6.71 $\times 10^{-5}$ & $(0.584, 0.616)$ & 0.017 & 0.014 \\
                   & Static $\varepsilon_{22}$ & $F$ & 0.60 & 0.6003 & 6.72 $\times 10^{-5}$ & $(0.584, 0.616)$ & 0.051 & 0.014 \\  
                   & & $G$ & 0.60 & 0.5999 & 1.67 $\times 10^{-4}$ & $(0.575, 0.625)$  & 0.024 & 0.022 \\ \midrule
\multirow{6}{*}{4} & Adaptive & $F$ & 0.69 & 0.6886 & 1.05 $\times 10^{-4}$ & $(0.668, 0.709)$  & 0.211 & 0.015 \\ 
                   & & $G$ & 0.43 & 0.4308 & 4.65 $\times 10^{-5}$ & $(0.417, 0.444)$ & 0.184 & 0.016 \\
                   & Static $\varepsilon_{11}$ & $F$ & 0.69 & 0.6885 & 1.73 $\times 10^{-4}$ & $(0.663, 0.714)$ & 0.221 & 0.019 \\ 
                   & & $G$ & 0.43 & 0.4305 & 3.91 $\times 10^{-5}$ & $(0.418, 0.443)$  & 0.128 & 0.015 \\
                   & Static $\varepsilon_{22}$ & $F$ & 0.69 & 0.6895 & 8.17 $\times 10^{-5}$ & $(0.672, 0.707)$ & 0.070 & 0.013 \\  
                   & & $G$ & 0.43 & 0.4305 & 1.02 $\times 10^{-4}$ & $(0.411, 0.450)$ & 0.114 & 0.024 \\
\bottomrule
\end{tabular}}
\caption{Posterior summaries at the final step for all 12 parameter and design combinations in Exemplar~1. The posterior expected value ($\mathbb{E}_{\theta \given y, \xi}$), variance ($\mathbb{V}_{\theta \given y, \xi}$), 95\% credible intervals (CI), MAPE of the MAP probability estimate and coefficient of variation (Cv) were averaged over all 100 trials.}\label{Tab: exemplar 1 posterior summaries}
\end{table}

\noindent Table~\ref{Tab: exemplar 1 1D metrics} shows the scalar posterior summaries of uncertainty with the total and generalized variance. The Mahalanobis distance (MD) of $\pars^{true}$, which measures the distance of the true parameter values from the posterior distribution, is also included. Values reported were averaged over all 100 repeat trials and were calculated from the posterior distribution after the final load step. Similar to the results for Case~1 shown in Fig.~\ref{fig: case 1 1D_cov_metrics}, in all cases, both the total and generalized variance are smaller for the adaptive design compared to either of the static designs, showing that a load path determined adaptively provides a greater reduction in uncertainty in a holistic sense when all parameters are considered simultaneously.

\begin{table}[!htb]
\centering
\begin{tabular}{lllll}
\hline
\textbf{Case No.} & \textbf{Design}  & \begin{tabular}[c]{@{}l@{}} \textbf{Tot. Var.} \\ $\boldsymbol{(1\times10^{-4})}$ \end{tabular} & \begin{tabular}[c]{@{}l@{}} \textbf{Gen. Var.}\\ $\boldsymbol{(1\times10^{-9})}$ \end{tabular} & \begin{tabular}[c]{@{}l@{}} $\boldsymbol{\mbox{\textbf{MD}}(\theta^{true})}$ \end{tabular} \\ 
\toprule
\multirow{3}{*}{1} & Adaptive  & 1.13 & 1.72 & 1.25  \\ 
                   & Static $\varepsilon_{11}$ & 1.43 & 2.37 & 1.26    \\ 
                   & Static $\varepsilon_{22}$ & 1.51 & 2.21 & 1.26  \\  \midrule
                   
\multirow{3}{*}{2} & Adaptive & 1.41 & 2.84 & 1.24   \\ 
                   & Static $\varepsilon_{11}$ & 1.93 & 3.76 & 1.26 \\ 
                   & Static $\varepsilon_{22}$ & 1.82 & 4.00 & 1.26  \\  \midrule

\multirow{3}{*}{3} & Adaptive & 1.71 & 4.58 & 1.24   \\ 
                   & Static $\varepsilon_{11}$ & 2.34 & 6.38 & 1.26  \\ 
                   & Static $\varepsilon_{22}$ & 2.34 & 6.38 & 1.26   \\  \midrule

\multirow{3}{*}{4} & Adaptive & 1.52 & 2.92 & 1.17   \\ 
                   & Static $\varepsilon_{11}$ & 2.12 & 3.69 & 1.13  \\ 
                   & Static $\varepsilon_{22}$& 1.84 & 4.36 & 1.19  \\  \midrule
\bottomrule
\end{tabular}
\caption{Scalar posterior summaries for  Exemplar~1. Values reported are average over 100 trials.}\label{Tab: exemplar 1 1D metrics}
\end{table}

%% file: app_C.tex
\section{Exemplar~2 Supplementary Material}\label{ex2.app}
\setcounter{figure}{0}  
\setcounter{table}{0}  

\noindent Table~\ref{Tab: exemplar 2 posterior summaries} contains marginal posterior summaries for Exemplar~2, where $\pars = [F, G, A, n, \sigma_{y}]^{T}$. Posterior expected values ($\boldsymbol{\mathbb{E}_{\theta \given y, \xi}}$), variances ($\boldsymbol{\mathbb{V}_{\theta \given y, \xi}}$), 95\% credible intervals (CIs), the mean absolute percentage error of the MAP estimate (MAPE($\hat{\pars}$)) and the coefficient of variation (Cv) were averaged over the 100 repeat trials for each parameter case and design option and were calculated after the final load step. In Case~5, $\pars^{true}$ is contained within the 95\% CIs for both the adaptive and static designs. The marginal variances reveal that the parameter uncertainty was lower in the adaptive setting than the static ones for all parameters except for $F$ when compared to the $\varepsilon_{11}$ design and $G$ when compared to the $\varepsilon_{22}$ design---where the static designs shown a slightly lower average uncertainty. The most significant difference was in $\sigma_{y}$, where the adaptive design on average had a variance of 6.98 and the static designs had variances of 247.59 and 180.90. In Case~6, $\pars^{true}$ is contained within the 95\% CIs for both the adaptive and static designs, but in the static designs, some of the parameter 95\% CIs went outside the bounds defined in Table~\ref{Tab: parameters in exemplars}. Some trials for the static $\varepsilon_{11}$ design in Case~6 were unsuccessful in identifying the model parameters given the data. Marginal posterior values with the unsuccessful trials removed from the average values are reported in Table~\ref{Tab: exemplar 2 posterior summaries 4 outliers removed}.

\begin{table}[!htb]
\centering
\resizebox{\textwidth}{!}{\begin{tabular}{lllllllll}
\hline
\textbf{Case No.} & \textbf{Design} &  \textbf{Parameter} & $\boldsymbol{\theta^{true}}$ & $\boldsymbol{\mathbb{E}_{\theta \given y, \xi}}$ & $\boldsymbol{\mathbb{V}_{\theta \given y, \xi}}$ & \textbf{95\% CI} & \textbf{MAPE($\boldsymbol{\hat{\pars}}$)} & $\textbf{Cv}$\\ 
\toprule
\multirow{15}{*}{5} & Adaptive & $F$ & 0.55 & 0.5496 & 3.46 $\times 10^{-5}$ & $(0.538, 0.561)$ & 0.075 & 0.011 \\ 
                   & & $G$ & 0.45 & 0.4499 & 2.69 $\times 10^{-5}$ & $(0.440, 0.460)$ & 0.026 & 0.012 \\
                   & & $A$ & 100.0 & 100.223 & 4.96 & $(95.86, 104.59)$ & 0.022 & 0.022 \\
                   & & $n$ & 20.0 & 20.17 & 1.56 & $(17.73, 22.62)$ & 0.866 & 0.062 \\
                   & & $\sigma_{y}$ & 300.0 & 299.65 & 6.98 & $(294.48, 304.83)$ & 0.115 & 0.009 \\
                   & Static $\varepsilon_{11}$ & $F$ & 0.55 & 0.5501 & 1.63 $\times 10^{-5}$ & $(0.542, 0.558)$ & 0.025 & 0.007 \\
                   & & $G$ & 0.45 & 0.4602 & 5.18 $\times 10^{-3}$ & $(0.321, 0.599)$ & 2.277 & 0.157 \\
                   & & $A$ & 100.0 & 100.86 & 24.23 & $(91.28, 110.45)$ & 0.862 & 0.049 \\
                   & & $n$ & 20.0 & 20.25 & 1.97 & $(17.52, 22.99)$ & 1.261 & 0.069 \\
                   & & $\sigma_{y}$ & 300.0 & 301.27 & 247.59 & $(270.68, 331.85)$ & 0.423 & 0.052 \\
                   & Static $\varepsilon_{22}$ & $F$ & 0.55 & 0.5454 & 4.33 $\times 10^{-3}$ & $(0.417, 0.673)$ & 0.831 & 0.122 \\
                   & & $G$ & 0.45 & 0.4496 & 1.29 $\times 10^{-5}$ & $(0.443, 0.457)$ & 0.096 & 0.008 \\ 
                   & & $A$ & 100.0 & 99.63 & 18.36 & $(91.26, 107.99)$ & 0.372 & 0.043 \\
                   & & $n$ & 20.0 & 20.01 & 1.90 & $(17.31, 22.70)$ & 0.032 & 0.069 \\
                   & & $\sigma_{y}$ & 300.0 & 298.60 & 180.90 & $(272.50, 324.70)$ & 0.467 & 0.045 \\ \midrule
\multirow{15}{*}{6} & Adaptive & $F$ & 0.55 & 0.5501 & 4.72 $\times 10^{-5}$ & $(0.537, 0.564)$ & 0.025 & 0.012 \\ 
                   & & $G$ & 0.45 & 0.4497 & 3.87 $\times 10^{-5}$ & $(0.438, 0.462)$ & 0.065 & 0.014 \\
                   & & $A$ & 300.0 & 300.028 & 8.30 & $(294.38, 305.68)$ & 0.009 & 0.010 \\
                   & & $n$ & 20.0 & 19.95 & 0.22 & $(19.04, 20.86)$ & 0.261 & 0.023 \\
                   & & $\sigma_{y}$ & 100.0 & 100.13 & 6.87 & $(95.00, 105.27)$ & 0.134 & 0.026 \\
                   & Static $\varepsilon_{11}$ & $F$ & 0.55 & 0.5496 & 4.05 $\times 10^{-4}$ & $(0.534, 0.565)$ & 0.071 & 0.014 \\
                   & & $G$ & 0.45 & 0.5034 & 4.48 & $(-0.431, 1.438)$ & 11.878 & 0.755 \\
                   & & $A$ & 300.0 & 309.60 & 1.59 $\times 10^{5}$ & $(129.77, 489.44)$ & 3.200 & 0.270 \\
                   & & $n$ & 20.0 & 20.74 & 844.60 & $(7.68, 33.80)$ & 3.719 & 0.292 \\
                   & & $\sigma_{y}$ & 100.0 & 102.57 & 9.71 $\times 10^{3}$ & $(54.68, 150.47)$ & 2.573 & 0.223 \\
                   & Static $\varepsilon_{22}$ & $F$ & 0.55 & 0.5421 & 1.09 $\times 10^{-2}$ & $(0.357, 0.727)$ & 1.44 & 0.193 \\
                   & & $G$ & 0.45 & 0.4496 & 1.79 $\times 10^{-5}$ & $(0.441, 0.458)$ & 0.084 & 0.009 \\
                   & & $A$ & 300.0 & 297.31 & 446.59 & $(260.50, 334.13)$ & 0.896 & 0.065 \\
                   & & $n$ & 20.0 & 19.86 & 2.21 & $(17.21, 22.50)$ & 1.274 & 0.068 \\
                   & & $\sigma_{y}$ & 100.0 & 98.73 & 52.06 & $(86.11, 111.34)$ & 1.274 & 0.068 \\ \midrule
\bottomrule
\end{tabular}}
\caption{Posterior summaries at the final step for all 6 parameter and design combinations in Exemplar~2. The posterior expected value ($\mathbb{E}_{\theta \given y, \xi}$), variance ($\mathbb{V}_{\theta \given y, \xi}$), 95\% credible intervals (CI), MAPE of the MAP probability estimate and coefficient of variation (Cv) were averaged over all 100 trials.}\label{Tab: exemplar 2 posterior summaries}
\end{table}

\begin{table}[!htb]
\centering
\resizebox{\textwidth}{!}{\begin{tabular}{lllllllll}
\hline
\textbf{Case No.} & \textbf{Design} &  \textbf{Parameter} & $\boldsymbol{\theta^{true}}$ & $\boldsymbol{\mathbb{E}_{\theta \given y, \xi}}$ & $\boldsymbol{\mathbb{V}_{\theta \given y, \xi}}$ & \textbf{95\% CI} & \textbf{MAPE($\boldsymbol{\hat{\pars}}$)} & $\textbf{Cv}$\\ 
\toprule

\multirow{5}{*}{6} & Static $\varepsilon_{11}$ & $F$ & 0.55 & 0.5496 & 2.16 $\times 10^{-5}$ & $(0.541, 0.559)$ & 0.069 & 0.008 \\
                   & & $G$ & 0.45 & 0.4953 & 1.52 $\times 10^{-2}$ & $(0.285, 0.706)$ & 10.072 & 0.229 \\
                   & & $A$ & 300.0 & 307.87 & 636.26 & $(264.60, 351.15)$ & 2.624 & 0.072 \\
                   & & $n$ & 20.0 & 20.63 & 3.26 & $(17.50, 23.76)$ & 3.158 & 0.077 \\
                   & & $\sigma_{y}$ & 100.0 & 102.11 & 70.22 & $(87.41, 116.80)$ & 2.105 & 0.074 \\ \midrule
\bottomrule
\end{tabular}}
\caption{Average Case~6 marginal posterior summaries for the $\varepsilon_{11}$ static design with the unsuccessful trials removed.}\label{Tab: exemplar 2 posterior summaries 4 outliers removed}
\end{table}

\noindent Table~\ref{Tab: exemplar 2 1D metrics} shows the scalar posterior summaries of uncertainty with the total and generalized variance and the Mahalanobis distance (MD) of $\pars^{true}$. Values reported were averaged over all 100 repeat trials and were calculated from the posterior distribution after the final load step. Scalar posterior summaries with the unsuccessful trials in the Case~6 static $\varepsilon_{11}$ design removed are reported in Table~\ref{Tab: exemplar 2 1D metrics 4 outliers removed}.

\begin{table}[!htb]
\centering
\begin{tabular}{lllll}
\hline
\textbf{Case No.} & \textbf{Design}  & \textbf{Tot. Var.} & \begin{tabular}[c]{@{}l@{}} \textbf{Gen. Var.} \\ $\boldsymbol{(1\times10^{-9})}$ \end{tabular} & $\boldsymbol{\mbox{\textbf{MD}}(\theta^{true})}$ \\ 
\toprule
\multirow{3}{*}{5} & Adaptive  & 13.50 & 1.20 & 2.14  \\ 
                   & Static $\varepsilon_{11}$ & 273.80 & 160.00 & 3.00  \\ 
                   & Static $\varepsilon_{22}$ & 201.17 & 116.24 & 2.75  \\   \midrule
                   
\multirow{3}{*}{6} & Adaptive & 15.39 & 0.36 & 2.21  \\ 
                   & Static $\varepsilon_{11}$ & 170,013 & 6.68 $\times 10^{4}$ & 5.13  \\ 
                   & Static $\varepsilon_{22}$ & 500.87 & 66.12 & 5.73 \\  \midrule
\bottomrule
\end{tabular}
\caption{1D posterior summaries for  Exemplar~2. Values reported are average over 100 trials.}\label{Tab: exemplar 2 1D metrics}
\end{table}

\begin{table}[!htb]
\centering
\begin{tabular}{lllll}
\hline
\textbf{Case No.} & \textbf{Design}  & \textbf{Tot. Var.} & \begin{tabular}[c]{@{}l@{}} \textbf{Gen. Var.} \\ $\boldsymbol{(1\times10^{-9})}$ \end{tabular} & $\boldsymbol{\mbox{\textbf{MD}}(\theta^{true})}$ \\ 
\toprule
\multirow{1}{*}{6} & Static $\varepsilon_{11}$ & 709.76 & 150.57 & 5.00 \\ \midrule
\bottomrule
\end{tabular}
\caption{Average Case~6 1D posterior summaries for the $\varepsilon_{11}$ static design with the unsuccessful trials removed.}\label{Tab: exemplar 2 1D metrics 4 outliers removed}
\end{table}

\begin{figure}[!htb]%
    \centering
    \sidesubfloat[]{{\includegraphics[width=0.45\textwidth]{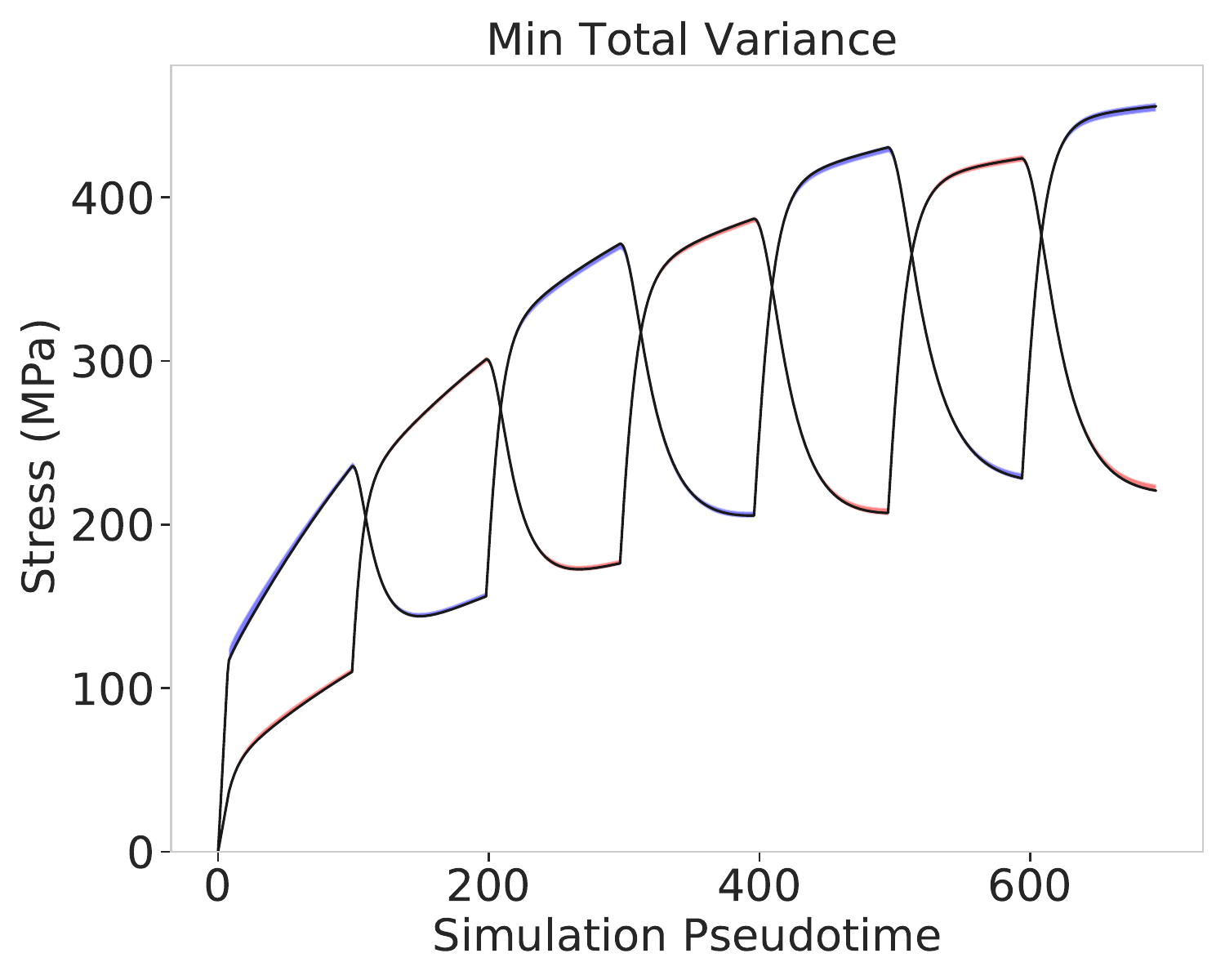} }}%
    \sidesubfloat[]{{\includegraphics[width=0.45\textwidth]{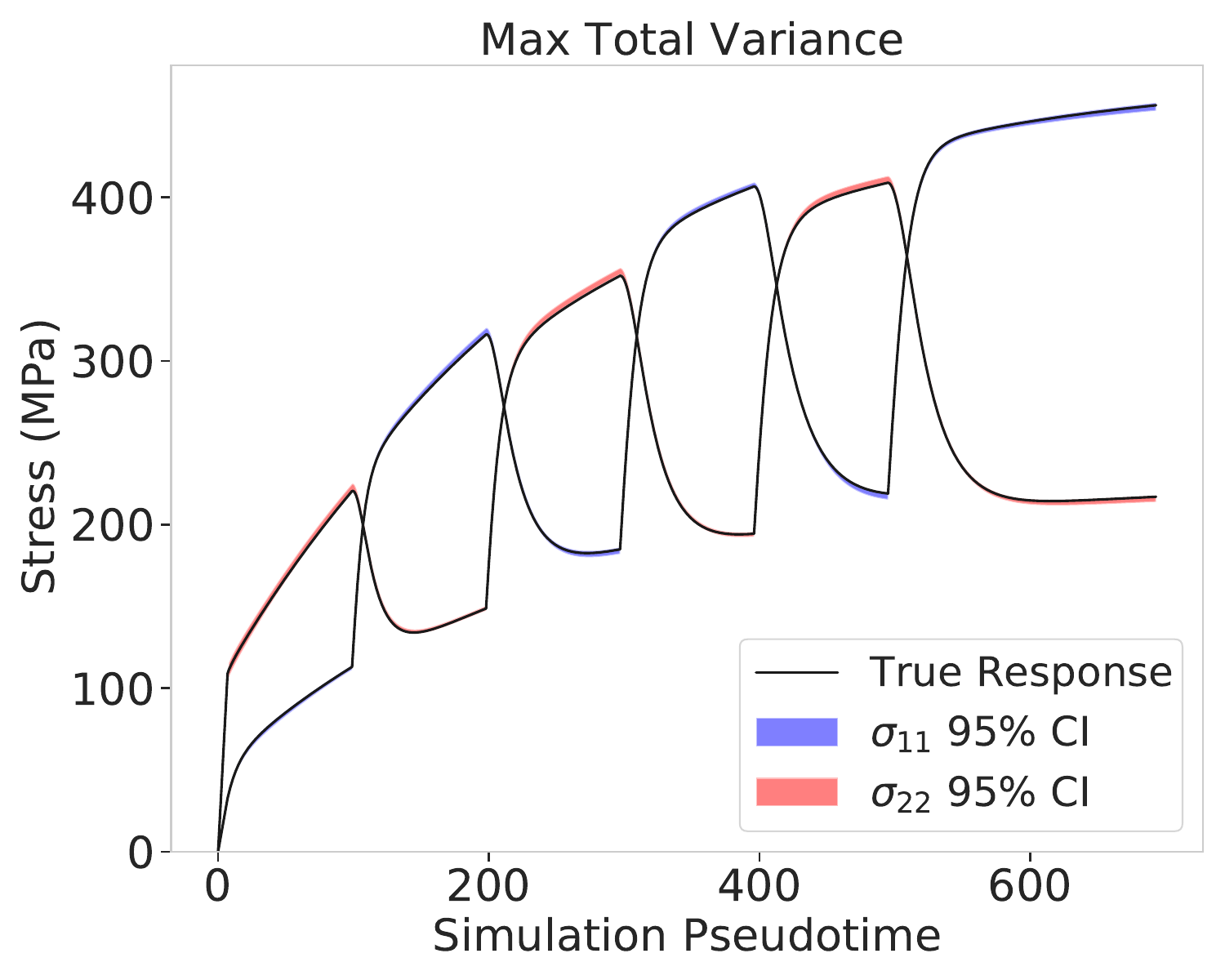} }}%
    \\
    \sidesubfloat[]{{\includegraphics[width=0.45\textwidth]{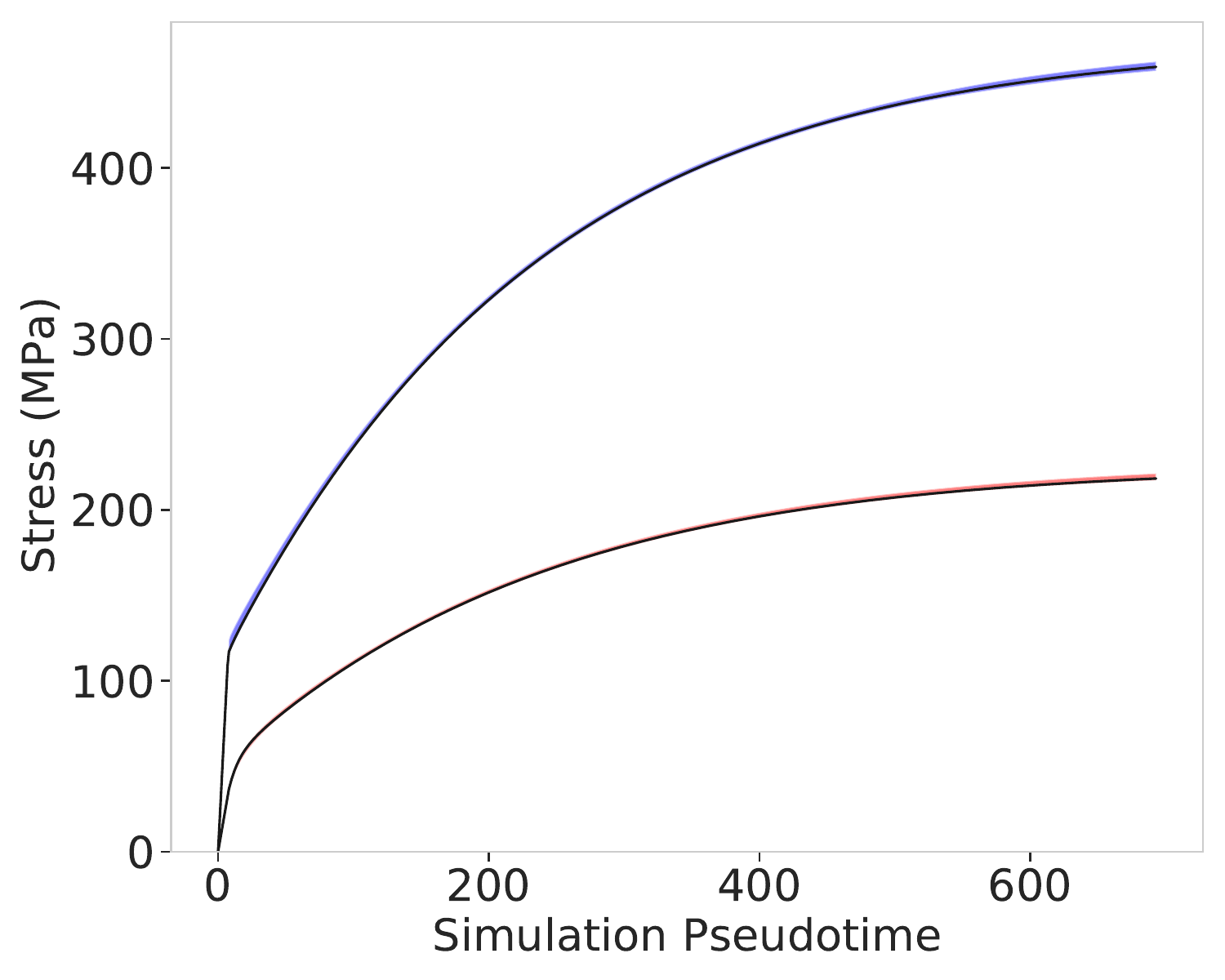} }}%
    \sidesubfloat[]{{\includegraphics[width=0.45\textwidth]{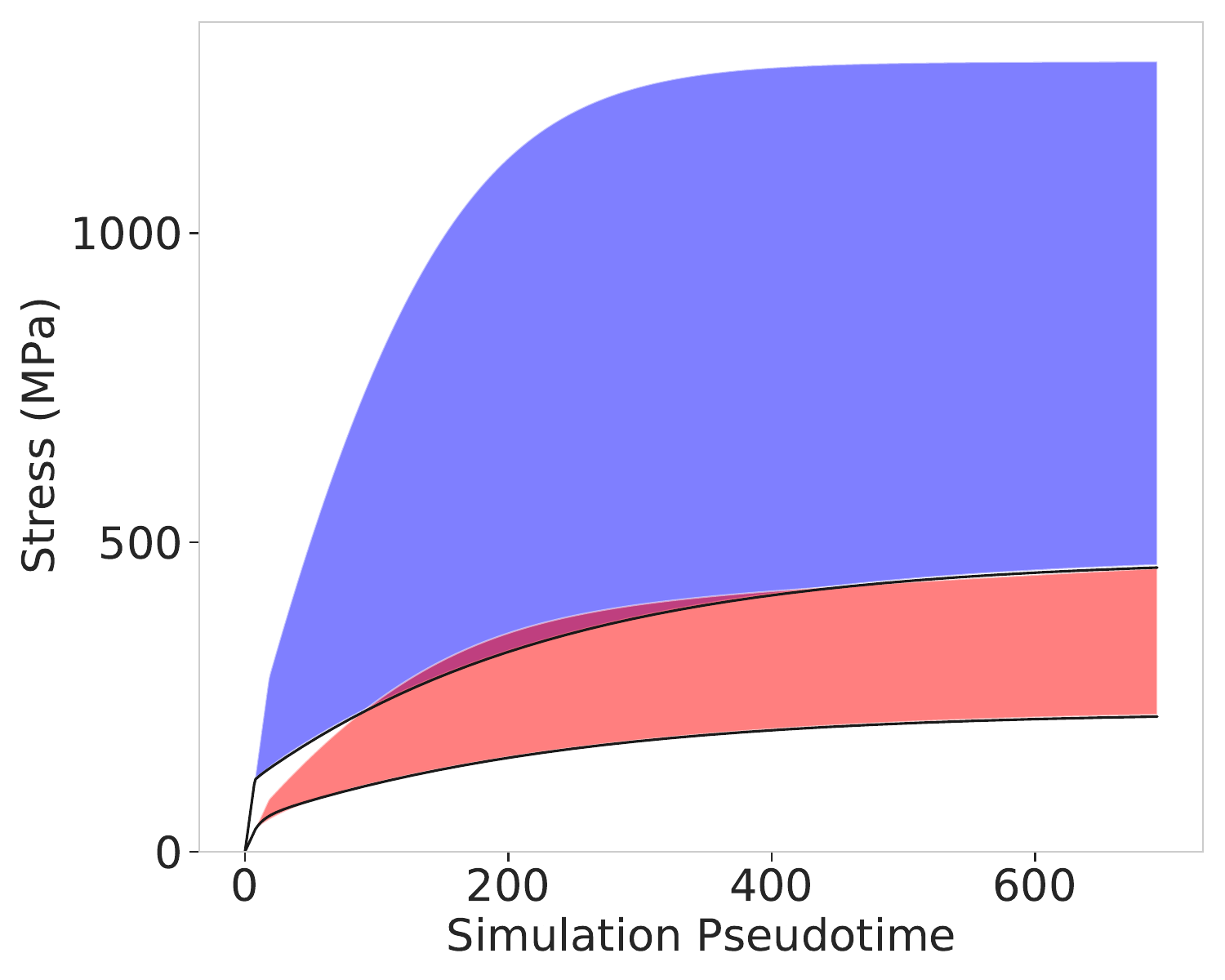} }}%
    \\
    \sidesubfloat[]{{\includegraphics[width=0.45\textwidth]{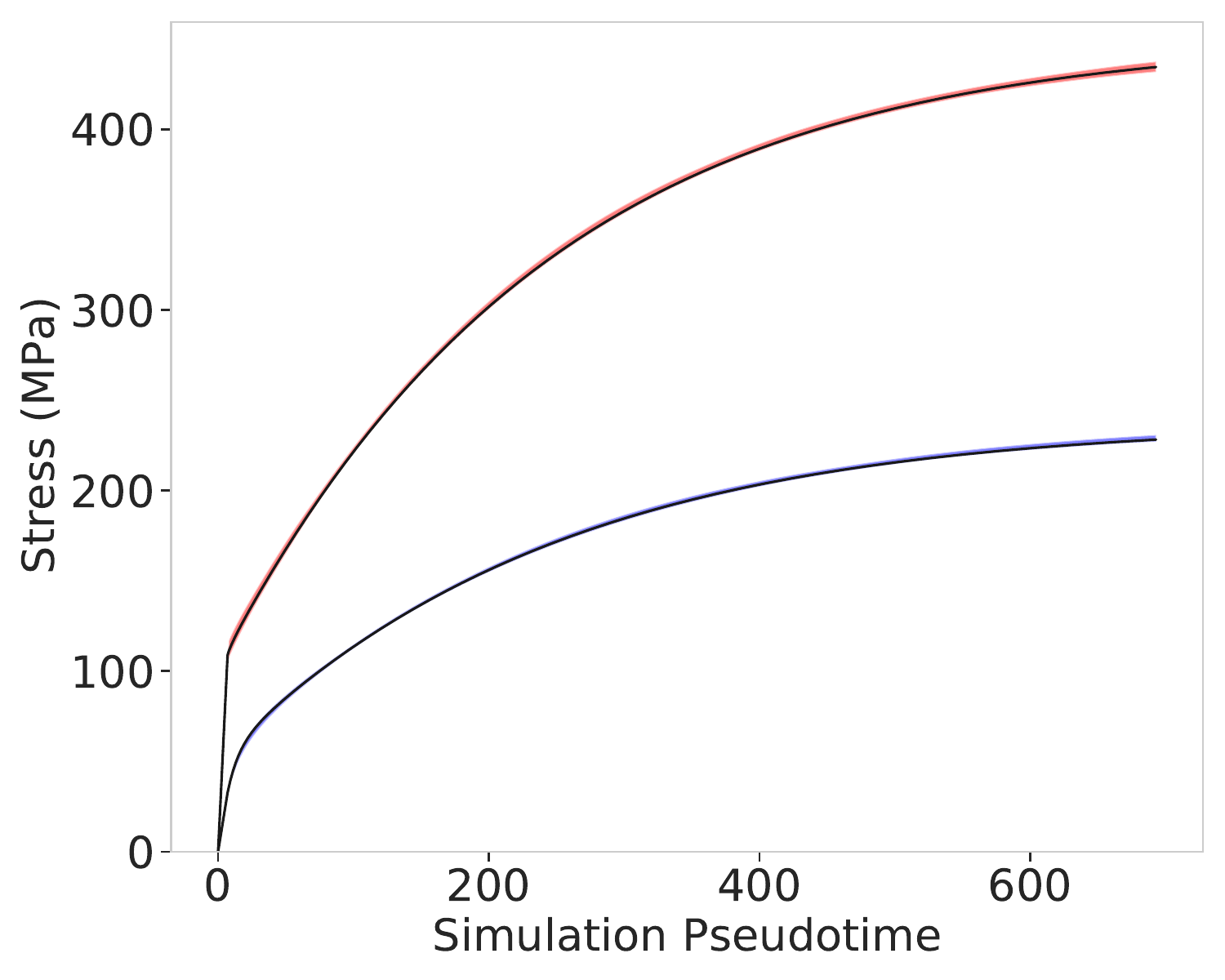} }}%
    \sidesubfloat[]{{\includegraphics[width=0.45\textwidth]{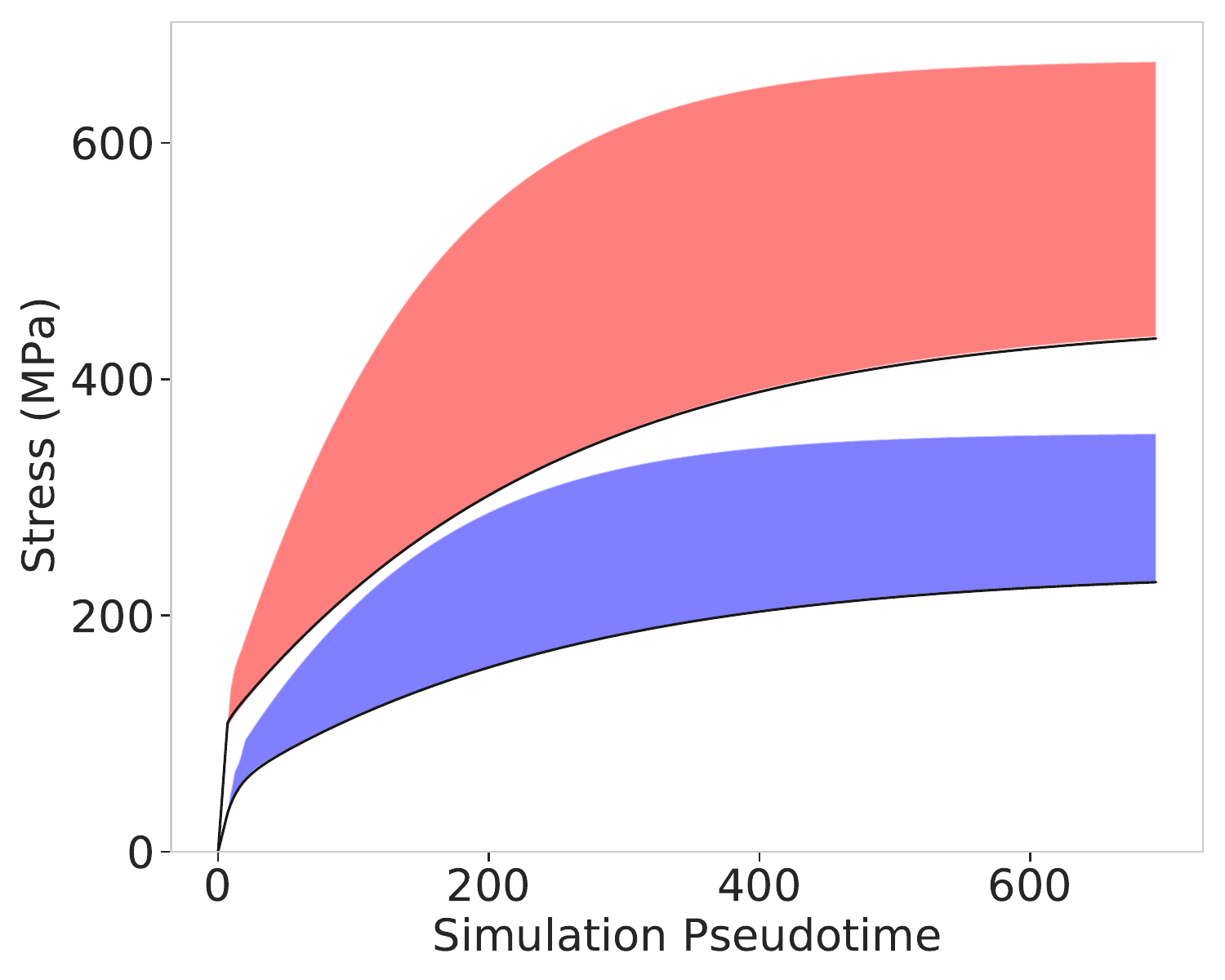} }}%
    \caption{95\% credible intervals for the posterior draws $\tilde{\pars} \sim \pi(\pars \given \data, \des)$ were calculated point-wise for Exemplar~2, Case~6 and are indicated the the blue ($\sigma_{11}$) and red ($\sigma_{22}$) dashed lines. The model output from the true parameter values is plotted with a black line. The left column of plots show the 95\% CI for the trial that had a posterior distribution with the least total variance among all trials for each design setting. The right column of plots show 95\% CI for the trial that had the greatest total variance. Plots (a) and (b) show results for the adaptive design, plots (c) and (d) show results for the static $\varepsilon_{11}$ design and plots (e) and (f) for the static $\varepsilon_{22}$ design.}%
    \label{fig: exemplar 2 case 6 posterior draw CI}%
\end{figure}

\noindent Posterior uncertainty was propagated to the model output in Case~6 from Exemplar~2 and 95\% CIs are plotted in Fig.~\ref{fig: exemplar 2 case 6 posterior draw CI}. The left column of plots show CIs for the trial in each design setting with the lowest amount of total variance, and the right column of plots show the trial with the highest level of total variance. For the adaptive design, the minimum total variance came from a trial with an optimal design of $\des^{*} = [\varepsilon_{11}, \varepsilon_{22}, \varepsilon_{11}, \varepsilon_{22}, \varepsilon_{11}, \varepsilon_{22}, \varepsilon_{11}]$, and the maximum total variance came from a trial with an optimal design of $\des^{*} = [\varepsilon_{22}, \varepsilon_{11}, \varepsilon_{22}, \varepsilon_{11}, \varepsilon_{22}, \varepsilon_{11}, \varepsilon_{11}]$. The narrow CIs in the adaptive design for both the minimum and maximum case indicate that the adaptive design consistently resulted in parameter inference with a low degree of uncertainty across the 100 trials. In constrast, the CIs in the static settings varied greatly from the minimum to the maximum total variance trials, indicating that reliable parameter inference was not consistent across the 100 trials.

